\documentclass[12pt,emulateapj]{article}
\usepackage[onecolumn]{emulateapj}
\tightenlines

\newcommand{\be}{\begin{equation}}
\newcommand{\ee}{\end{equation}}
\newcommand{\ba}{\begin{eqnarray}}
\newcommand{\ea}{\end{eqnarray}}

\slugcomment{Astrophysical Journal}
\begin{document}
\today
\title{Velocity Modification of HI Power Spectrum}
\author{A. Lazarian\altaffilmark{1,2,3} D. Pogosyan\altaffilmark{2} }
\altaffiltext{1}{Department of Astrophysical Sciences, Princeton University,
Princeton}
\altaffiltext{2}{Canadian Institute for Theoretical Astrophysics, University
of Toronto}
\altaffiltext{3}{Present address: Dept. of Astronomy, University of Wisconsin,
Madison}
%\begin{abstract}
The distribution of atomic hydrogen in the Galactic plane 
is usually mapped using the Doppler shift
of 21~cm emission line and this causes the modification of the
observed emission spectrum.
We calculate the emission
 spectrum in velocity slices of data (channel maps)
and derive its dependence on 
the statistics of velocity and density fields. We find
that (a) if the density spectrum is steep, i.e. $n<-3$, the
short-wave asymptotics of the emissivity spectrum is dominated by
the velocity fluctuations; (b) the velocity fluctuations make
the emission spectra shallower, provided that the data slices
are sufficiently thin. In other words, turbulent velocity creates 
small scale structure that can erroneously be identified as clouds.
The effect of thermal velocity is very similar to the change of
the effective slice thickness, but the difference is that while an
increase of the slice thickness increases the amplitude of the signal
the increase of the thermal velocity leaves the measured intensities
intact while washing out fluctuations. The contribution of fluctuations in
warm HI is suppressed relative to cold component 
when velocity channels used are narrower than warm HI thermal velocity
and small angular scale fluctuations are measured.
We calculate
how the spectra vary with the change of  velocity slice thickness and show 
that the observational 21~cm data 
is consistent with the explanation that the intensity fluctuations
within individual channel maps
are generated by a turbulent velocity field. As the thickness of velocity
slices increases density fluctuations get to dominate emissivity.
This allows to disentangle velocity and density statistics. 
The application of our technique to the Galactic and SMC data
reveals spectra of density and velocity with the power law index
close to $-11/3$. This is a Kolmogorov index, but the explanation
of the spectrum appealing to the Kolmogorov-type cascade faces substantial
difficulties.  
We generalize our treatment for the case of a
statistical study of turbulence inside individual clouds.
The mathematical machinery developed is applicable to other emission lines.

%\end{abstract}


\section{Introduction}

Galactic HI is an important component of the interstellar media 
(McKee \& Ostriker 1977) and much
efforts have been devoted to its studies (see Burton 1992). 
In particular,
Crovisier \& Dickey (1983) and Green (1993) measured the spectrum of 21 cm
intensity fluctuations 
in order to get a handle on the statistical properties of
this media. Lazarian (1995, henceforth L95) suggested a statistical
  inversion technique
for HI 
interferometric data (see also Lazarian 1994 a,b).
Note that statistical description is a nearly indispensable strategy while
dealing with turbulence\footnote{Speaking about interstellar turbulence we 
understand it in terms of unpredictable
spatio-temporal behavior of nonlinear systems (Scalo 1985, 1987) and accept
that Kolmogorov picture may be too simplistic while dealing with
such complex media (Lazarian 1999).}. Indeterminism of turbulence implies
that two virtually identical systems develop very different patterns
of turbulent motions. The advantage of statistical techniques is
that they extract underlying regularities of the flow and reject
incidental details.

Attempts to study interstellar turbulence date as far back as 50s
(Horner 1951, Munch 1958, Wilson et al. 1959) and various directions
of research achieved various degree of success (see reviews by
Kaplan \& Pickelner 1970, Dickman 1985, Lazarian 1992, Armstrong, Rickett
\& Spangler 1995). 
Studies of turbulence statistics of ionized media were successful
(see Spangler \& Gwinn 1990, Narayan 1992) and provided the information of
the statistics of plasma density\footnote{Incidentally the found
spectrum was Kolmogorov one.} at scales $10^{8}$-$10^{15}$~cm. 
This research profited
a lot from clear understanding of processes of scintillations and scattering
achieved by theorists (see Goodman \& Narayan 1985, Narayan \& Goodman 
1989). 

Deficiencies in the theoretical description have been, to our mind, the
major impediments to studies of turbulence in the neutral part of 
the interstellar
medium. For instance, important statistical studies of molecular clouds 
(Dickman \& Kleiner 1985, Miesch \& Bally 1994) has not
achieved the success parallel to the one in scintillation studies.
These problems were addressed in L95, where the
quantitative study of deconvolution of the 3D statistics from
the data averaged along the lines of sight was presented (see also
 Lazarian 1999a).

However, the issue of velocity distortions of the observed spectrum
was treated only qualitatively in L95, and
this motivates our present
study. Indeed, even in the absence of density fluctuations
in Galactic coordinates velocity fluctuations would make the
distribution of emission inhomogeneous in  velocity data cubes.
The release of the Canadian Galactic Plane survey
data makes this work very timely.

The present paper concentrates on the studies
of HI in Galactic plane. In also addresses the issue of studies of
individual clouds. We deliberately restrict our attention
to HI, although the  presented technique is potentially applicable
to other spectral lines. The pervasive distribution of neutral hydrogen
presents a sharp contrast to the localized distribution of
molecular species and this alleviates problems with averaging\footnote{ A
study by Stutzki et al (1998) shows that the worries about the turbulence
homogeneity are probably exaggerated even in the case of molecular clouds.
At least the wavelet analysis that they used revealed a well-defined spectrum
of turbulence.} 
(cf. Houligan \& Scalo 1990). The remarkable constancy of the
slope of the intensity power spectrum measured by Green (1993)
and Stanimirovic et al (1998)
indicates that the HI sample is homogeneous in
a statistical sense. However, most of our results are applicable to other
emission lines. 

It may seem that the Galactic plane presents simultaneously the best and the
worst
case for obtaining HI statistics. On one side, the Galactic
rotation curve presents a natural distance indicator, which
allows to roughly segregate emission from different parts of Galactic
disk. On the other side, the random velocity is being confused
with the one arising from regular rotation resulting in distance errors. 

To what extent
the statistics of the underlying density can be recovered is
the issue this paper deals with. The effects of projection and velocity 
distortions plague the
statistical studies of turbulence in interstellar clouds. This paper 
provides a guide on how to deal with such problems. We aimed our paper
at providing a general mathematical formalism necessary for dealing
with
velocity distortions of the observed intensity fluctuations. Modifications
of the technique and its application to various emission/absorption
 lines will be done
elsewhere.

Another motivation for our study stems from recent attempts
to describe the structures in the Galactic hydrogen in order
to estimate the fluctuations of microwave polarization arising from
interstellar dust. This contribution is extremely important in view of
present-day efforts in the CMB research (see Prunet \& Lazarian 1999,
Draine \& Lazarian 1999). Some of the studies, for instance, 
one by Sethi, Prunet and Bouchet (1998)
attempts to relate the statistics of density observed in the velocity
space and the statistics of magnetic field fluctuations. Our present
work and a subsequent paper show that the relation between the two
statistics is far from trivial and velocity fluctuations may create
a lot of emissivity structures in the channel maps.

The distortion of the density statistics in the
presence of velocity fluctuations is important for understanding
of the large-scale distribution of galaxies. 
The corresponding literature is vast
and we just mention a pioneering study by Kaiser (1987) and 
a recent review by Hamilton (1998). In cosmology the problem is known
as the problem of 'redshift-space' corrections to the statistics of galaxy 
distribution and has been addressed predominantly under the 
linear regime of the gravitational instability
or at the regime when the velocity contribution to the Fourier spectrum
can be factorized, e.g. by a Maxwellian factor (see Hamilton (1998)).
The problem that we study is much richer.
We have to deal with the case of general, nonlinear, density fields
transformed by coherent turbulent velocities.
Development of the formalism for HI studies requires to consider a variety
of spectra, including steep
(e.g. Kolmogorov) power spectra which have never been 
addressed in cosmological
literature.

Our treatment of the problem is based on the description of HI statistics
in terms of power spectra. Such a description is traditional in hydrodynamics
and MHD theory and it allows to relate the statistics of interstellar
media to the statistics of idealized fluids. Moreover, spectra provide a 
useful intuitive insight to the properties of turbulence. For instance,
shallow spectra correspond to more structure at a small scale, while
steep spectra correspond to a random field dominated by
large-scale fluctuations. Other descriptors are reviewed in
Lazarian (1999b). 

An additional complication arises from variations of gas temperature.
Gas thermal velocity smears up the turbulent velocity effects and
this smearing effect depends on the temperature. In the paper 
we show that in most circumstances the intensity fluctuations are
due to cold HI. 
 
In what follows, we formulate the problem of velocity mapping,
discuss the 3D power spectrum of intensity that can be measured
in velocity data cubes
(section 2). Then we
 show how the 2D spectra of channel maps (those spectra can be directly
obtained using an interferometer) depend on the thickness of
the velocity slice (section 3). For power-law density and velocity
statistics we
calculate analytically and numerically the 2D spectra when the 
density is dominated by small-scale
and large-scale inhomogeneities (section 4). We compare our predictions
with the available Galactic and Small Magellanic Cloud (SMC) observational
data in section 5, while section~6 contains a qualitative
discussion of our results and 
presents an 
outline of the outstanding problems. We provide the 
summary of our results in section~7. To simplify our presentation we moved
derivation of many results into Appendixes. They, however, have the
value on their own. For instance, Appendix~B presents asymptotics of
3D spectra in velocity space, while
Appendix~C contains derivations
for correlation functions in velocity space. The  
modification of our analysis when velocity and density fields are correlated
is presented in Appendix~D. Appendix~E describes a statistical
treatment of emission from individual HI clouds and establishes the
connection between this treatment and the treatment of fluctuations
of intensity from HI Galactic disc that the main body of the paper
deals with. Note, that it is this modification of the technique that
is most important for studying turbulence in molecular clouds.
Appendix F describes additional statistical tools, namely,
1D radial spectra which can be useful measures of
density and velocity.

\section{The Problem.}

\subsection{The Model.}

In studying Galactic HI observers use a model 
Galactic rotation curve (see Kerr et al 1986) to identify
points along the line of sight in their data cubes, in which one
of the coordinates is the Doppler shifted 
velocity. The model curves reflect the
averaged motion of HI in the Galactic potential or bulk deviations
from circular motions and therefore are
accurate up to turbulent velocities.

It is  easy to understand that the random velocity field
makes statistics of emissivity different from the statistics of
density. Indeed, the velocity fluctuations can make
two emitting 
elements at different distances from the observer to overlap in the velocity
space producing an element of doubled emissivity.

In the present paper we address an important issue of relating the the
spectrum of HI emissivity
 in {\it velocity} space $xyv$ ($v$ is the $z$-component
of velocity\footnote{These coordinates are also called PPV coordinates which
abbreviates Position-Position-Velocity (see Vazquez-Semadeni 1999).}
to density fluctuations
in real space $xyz$ coordinates, where $z$ is directed along the line of sight
and $x$ \& $y$ span the sky plane. The statistics in $xyz$ coordinates
is essential for 
understanding HI
structure and processes of fragmentation (see Scalo 1985), while observations
provide us with the statistics of velocity data cubes.

Although the problem of relating statistics in velocity space and
Galactic coordinates is quite general, in what follows, we limit
ourselves to the case of 21~cm emission.
HI observations include high precision measurements of
Doppler shifted emission lines. For the sake
of simplicity, further on we  shall disregard self-absorption of the emitted 
radiation. 
This looks as a reasonable assumption as 
numerous studies (see Braun 1997, 1999, Higgs 1999) confirm that 
for many HI regions self-absorption
is negligible\footnote{In general, if absorption is localized in the form
of individual blobs of dense cold material, those only 
alter marginally statistical measures on the sizes
less than the size of blobs (L95).}. 

\subsection{Space-Velocity Mapping.}

The data of HI surveys is stored as data cubes with velocity
$v=|\vec{v}_z|$ constituting the third, $z$, dimension. 
We analyze what effect the mapping from the $xyz$ space
to the $xyv$ space
has on the power spectrum of HI density. First,  HI gas
experience the regular motion induced by galactic rotation. 
A model of Galactic rotation
provides a relation between the $z$ projection (the only projection
available through Doppler measurements) of regular
velocity $v_z^{reg}$  and the distance along the line of sight, i.e.
$v_z^{reg}=\phi({\bf x})$.  
If observations are made in the slices of the finite
thickness the function $\phi$ can be expanded into Taylor series around
a particular point (e.g. the center of data slice with $v_{z0}^{reg}=
\phi(z_0)$),
$v_z^{reg}=\phi({\bf x_0})+ \phi_{,{\bf x}} ({\bf x_0})
({\bf x} - {\bf x_0})+1/2\phi_{,{\bf x},{\bf x^\prime}}({\bf z_0})
({\bf x} - {\bf x_0}) ({\bf x^\prime} - {\bf x_0}) + \ldots $.

In addition to the regular part ${\bf v}^{reg}$ arising from Galactic rotation
the total observed velocity of a gas element
${\bf v}^{obs}$, also contains  a random, turbulent, part ${\bf u}$, so that
${\bf v}^{obs}= {\bf v}^{reg} + {\bf u}$. Therefore the mapping from
velocity space to Galactic coordinates becomes more involved. 
Most interesting effects on HI statistics in velocity space arise from the random ${\bf u}$ part.
We simplify their treatment by approximating the regular $z$-component of velocity
by the linear, z only dependent term: 
\be
z_s=z_{s0}+A \left[ \phi_{,z}(z_0)(z-z_o)+u_z \right]
\label{initial}
\ee
where $z_s$ is velocity coordinate in spectral-line data cubes, $A$
is an arbitrary scaling coefficient that relates velocity and data
cube length units.
Naturally, a linear mapping given by eq.~(\ref{initial}) is only correct if 
the nonlinear terms in the expansion of $\phi$ are small (
in other words, we consider the region of space where coherent flow
can be described as flow of a constant shear) and  the scales of interest in image plane
are sufficiently short that x,y dependence of the shear can be ignored 
% As we are interested
% in the effects of $u_z$ we should compare it with $1/2\phi''(z_0)(z-z_0)^2$.
\footnote{It is obvious that if the velocity of Galactic rotation is
$V_{max}$ then $\phi_{,zz}(z_0) (z-z_0)/\phi_{,z}(z_0) \sim (\delta V/V_{max}) $,
where $\delta V $ is equal or less than the thickness of a velocity 
slice. For sufficiently small slice thickness the 
factor $(\delta V/V_{max})$
is small and the contribution of the nonlinear term is negligible.
It will be clear from the discussion below that studying turbulence
at a particular scale one should account for the non-linearity of
$\phi$ on this scale. 
Similarly, fluctuations are observed over a limited area of 
$\delta x \delta y$ plane and one can use $\phi(z_0, x_0, y_0)$
to characterize the regular velocity 
if variations of $\phi$ on the distances studied are
small. This variation decreases with a distance to the slice.
It is easy to see that this requirements for the map to be
linear are easy to fulfill.}.

To describe turbulence in Galactic disc we find advantageous to use parameter
$f(z_0)=\phi_{,z}(z_0)^{-1}$ and put $A=f$. Using 
vector notations we describe the effect of a
random velocity ${\bf u}$ as a {\it linear} map from
Galactic frame ${\bf x}$ to the velocity-space coordinates ${\bf x}_s$
\begin{equation}
{\bf x_s} = {\bf x} - f({\bf x_0}) \, ({\bf u} \cdot {\bf \hat z} ) 
\; {\bf \hat z}~~~, 
\label{eq:map}
\end{equation}
Here ${\bf \hat{z}}$ is a unit vector along the line of sight,
${\bf x}$ is the actual Galactic coordinate of the emitting
HI element and ${\bf x}_s$ is the position calculated using the Galactic
rotation curve  (see Blitz \& Spergel 1991).
We shall note that the values of $f$ vary from slice to slice 
in accordance with the Galactic rotation curve. In our Galaxy $f$ is of
order of $100$~pc/(km/s), although it depends on the direction and depth of
observations.

A limiting case of the mapping corresponds to the study of 
individual HI clouds. We discuss this case in Appendix E and present a 
generalized form of the map applicable for the zero shear, i.e. $f^{-1} \to 0$.

\subsection{Spectrum in the Velocity Space}

The relevant expression for the spectrum of density
{\it in velocity space} has been derived by
Scoccimarro, Couchman \& Frieman (1999). Let us write the
Fourier component  of the density $\rho({\bf x}_s)$ in velocity space
as the sum over positions ${\bf x}_s^a$ of all hydrogen atoms
\begin{equation}
\rho_s({\bf k}) = \sum _{atoms} e^{i {\bf k} \cdot {\bf x}_s^a}
= \sum _{atoms} e^{i \left[ {\bf k} \cdot {\bf x}^a - f ({\bf k} \cdot {\bf \hat z})
({\bf u}^a \cdot {\bf \hat z}) \right]}~~~,
\end{equation}
In a coarse-grained fluid description of HI atom density in real space,
the summation over all atoms is replaced by a continuous
integral in the $xyz$ (real space) coordinates:  $\sum_{atoms} \Rightarrow \int d^3{\bf x} 
\rho({\bf x}) $ with gas density $\rho({\bf x})$. Velocities of individual
atoms are averaged\footnote{The averaging is similar to that presented by
eq.~(\ref{eq:gauss}).} over the thermal distribution 
$\langle \exp[- i f ({\bf k} \cdot {\bf \hat z})
({\bf u}^a \cdot {\bf \hat z})] \rangle_{thermal} = 
\exp[{- i f k_z u_z({\bf x})}] \exp[-f^2 k_z^2 v_T^2/2] $,
where ${\bf u} ({\bf x})$ is now the peculiar turbulent velocity of the
fluid element at position ${\bf x}$. Thermal velocity with one
dimensional mean-square $v_T=0.75 \left(\frac{T}{100{\rm K}}\right)^{1/2}$ 
km/s is incoherent and
acts as a smoothing along the velocity coordinate, limiting the resolution
in velocity space. 
Thus we obtain for density Fourier amplitude 
\begin{equation}
\rho_s({\bf k}) = e^{-f^2 k_z^2 v_T^2/2} \int d^3 {\bf x} \rho({\bf x}) 
e^{-i f k_z u_z({\bf x})}
e^{i {\bf k} \cdot {\bf x}}~~~.
\label{eq:rkintegral}
\end{equation}

The variance of Fourier amplitudes in velocity space is
\begin{equation}
\langle \rho_s({\bf k}) \rho_s^{*}({\bf k^{\prime}}) \rangle = e^{-f^2
(k_z^2+k_z^{\prime 2}) v_T^2/2}
\int d^3 {\bf x} \int d^3 {\bf x^\prime}
e^{i {\bf k} \cdot {\bf x}- {\bf k^\prime} \cdot {\bf x^\prime}}
\; \langle \rho({\bf x}) \rho({\bf x^\prime}) 
e^{-i f \left[ k_z u_z({\bf x}) -
k_z^{\prime} u_z({\bf x^{\prime}})\right]}
\rangle~.
\label{eq:preroman}
\end{equation}
Henceforth
we use angular brackets $\langle \ldots \rangle$ to denote ensemble
averaging over turbulence realizations. We assume
statistical homogeneity of the ensemble in the Galactic frame, i.e.
we assume that the 
average of any expression with two spatial coordinates depends only
on the vector separation between these two points. The density Fourier modes in
velocity space are then uncorrelated and can be 
described by the power spectrum\footnote{Here we deal with small scale
coherent structures and do not deal with the large scale ones, e.g.
spiral arms.}  $P_s({\bf k})$
%AL 25. /2 is added
\begin{eqnarray}
\langle \rho_s({\bf k}) \rho_s^{*}({\bf k}^{\prime}) \rangle &=& P_s({\bf k}) 
\delta({\bf k - k^{\prime}}) e^{-f^2 k_z^2 v_T^2}\\
P_s({\bf k}) &=& 
\int d^3 {\bf r} \, e^{i {\bf k}  \cdot {\bf r}} 
\Xi({\bf k}, {\bf r}),
~~~~~~~ {\bf r}={\bf x} - {\bf x \prime}~~,
\label{eq:roman}
\end{eqnarray}
where the kernel is
\be
\Xi({\bf k}, {\bf r})=\langle e^{i f 
k_z (u_z({\bf x})- u_z({\bf x \prime}))}
\rho ({\bf x}) \rho ({\bf x \prime}) \rangle~~~.
\label{kernelXi}
\ee

Looking at eq.~(\ref{kernelXi}) one easily notices that although both
velocity and density contribute to $P_s({\bf k})$ their functional
dependences are different. This opens a perspective of separating the velocity
and density contributions to the emissivity statistics that we explore
in sections 4 and 5.

Eq.~(\ref{eq:preroman}) assumes a plane-parallel approximation when
one does not distinguish between the radial nature of the line of sight and
the Cartesian $z$-direction. The formula is accurate when the HI region is
observed from a distance that is large compared
to its thickness ${\cal L}_{sl}$. The statistical analysis when the 
converging geometry of
the lines of sight is accounted for is more involved (see Lazarian 1994b)
and will be discussed elsewhere.

In what follows we limit ourselves with the discussion of basics of the
space-velocity mapping and the role of coherent turbulent velocity.
It is easy to see that the thermal velocity effect in $z_s$ direction is similar to the 
effect of the finite velocity resolution of the telescope
in the $x$ and $y$ directions (see L95).
To make our formulae more comprehensible we defined the spectrum $P_s({\bf k})$
to contain only correlated turbulent contribution while keeping thermal
velocity factor separately. We can  neglect thermal effect altogether if we
study supersonic turbulence
in velocity slices $\delta V > v_T$. We, however, will remember about
the high $k_z$ truncation of our spectrum by thermal motions when necessary (see
Chapter 3).  

To proceed at this stage we make the following assumptions
about the properties of the random fields $\rho ({\bf x})$ and ${\bf u} 
({\bf x})$ . 
%Assuming statistical homogeneity of the random fields we make use of
%two point descriptors of turbulence, namely, the structure and correlation
%functions. 
First of all, we assume that the 
turbulent velocity is uncorrelated with the density
in Galactic coordinates, and therefore
$\langle e^{i f \ldots} \rho ({\bf x}) \rho ({\bf x \prime}) \rangle =\langle e^{i f \ldots} \rangle \xi ({\bf r})$, where the 
density correlation function  is assumed to be isotropic in $xyz$-space
\be
\xi(r)=\xi({\bf r}) = \langle \rho ({\bf x}) \rho ({\bf x}+{\bf r}) \rangle~~
~.
\label{xifirst}
\ee

We discuss the corrections due to velocity-density 
correlations in Appendix~D and show that they are of limited importance for
Lognormal model of density fluctuations. Future research will test our
assumption for the distributions produced by numerical simulations.
Interstellar velocity and density fields can be anisotropic and we discuss
this in sections 6.2 and 6.3. 

We do not have to make any further assumptions about density statistics.
The turbulent velocity we assume to be Gaussian, then
\begin{eqnarray}
\langle e^{i f \ldots} \rangle =\int dy \exp({ify})\exp{\left(-\frac{y^2}
{2 \langle \Delta u_i  \Delta u_j  \rangle \hat z_i\hat z_i}\right)}=
e^{-f^2 k_z^2  \langle \Delta u_i \Delta u_j \rangle \hat z_i 
\hat z_j /2}~,\nonumber\\
\Delta u_i =u_i({\bf x})-u_i({\bf x \prime})~.
\label{eq:gauss}
\end{eqnarray}

To simplify our analysis we also restrict ourselves to the case 
of locally isotropic
turbulent velocity fields\footnote{The issue of density anisotropy
is analyzed in section~6.2 and the data reduction technique 
dealing with anisotropies
is discussed in L95.}.
Thus the structure tensor $\langle \Delta u_i \Delta u_j \rangle$,
which fully describes an
isotropic vector field can be expressed via longitudinal
$D_{LL}$ and transverse $D_{NN}$ components
(Monin \& Yaglom 1972)
\begin{equation}
\langle \Delta u_i \Delta u_j \rangle = \left( D_{LL}(r)-D_{NN}(r) \right) {r_i r_j \over r^2}
+D_{NN}(r) \delta_{ik}~~~,
\label{struc}
\end{equation}
where $\delta_{ik}$ equals 1 for $i=k$ and zero otherwise. 
We define z-projection of the velocity structure function as
\begin{equation}
D_z({\bf r}) \equiv \langle \Delta u_i \Delta u_j \rangle \hat z_i 
\hat z_j = D_{NN}(r) + [ D_{LL}(r)-D_{NN}(r)] \cos^2\theta~,
~~~ \cos\theta \equiv {\bf \hat r \cdot \hat z}
\label{eq:Dz}
\end{equation}
Substituting eqs.~(\ref{struc}) and (\ref{eq:gauss})
into eq.(\ref{eq:roman}) we obtain:
\begin{equation}
P_s({\bf k}) = \int d^3 {\bf r} \, e^{i {\bf k}  \cdot {\bf r}} \xi(r) 
\exp\left[-{f^2 k_z^2 D_z({\bf r})\over 2} \right]~~~.
\label{eq:main}
\end{equation} 
It is easy to see that the power spectrum above 
is anisotropic, with velocity mapping
having most effect on the modes parallel to the line of sight ($k_z=k$) and 
not affecting
the perpendicular ones ($k_z=0$).

\subsection{Velocity field}

If the velocity field is solenoidal\footnote{Solenoidality of
the velocity field follows for the incompressible fluid from the
continuity equation.} then according to Monin \& Yaglom (1972)
 $D_{LL}$ and $D_{NN}$
are related via a simple relation:
\be
D_{NN}(r) = D_{LL}(r) + \frac{r}{2} \; \frac{d}{dr} D_{LL}(r)~~~.
\label{sol}
\ee
Observational data that we discuss in section~5 corresponds to a power
law. This is suggestive that the underlying statistics is also 
a power-law. 
For the power-law statistics velocity eq.~(\ref{sol}) provides 
$D_{LL}=2/(m+2) D_{NN} =  C  r^m $. 

For potential fields Monin \& Yaglom (1972) show that
\be
D_{LL}(r) = D_{NN}(r) + \frac{r}{2} \; \frac{d}{dr} D_{NN}(r)~~~,
\ee
thus for power-law structure functions $D_{LL}=(1+m/2)D_{NN}$ and
again $D_{LL}\sim r^m$.

In general, the velocity field will have both potential and solenoidal
components and this introduces an uncertainty of the 
order unity in the coefficients
relating $D_{LL}$ and $D_{NN}$. This
does not change much in our results, and therefore, for the sake of
simplicity, doing calculations we shall assume that the velocity
field is solenoidal, which probably is not too far from the truth.

The convergence 
of integrals restricts
$m$ to the range $0 < m < 2$.
The value $m=2/3$ is distinct as it corresponds to the Kolmogorov
turbulence, but in Appendix~B we consider also cases $m=1/2,1$. The 
 power spectrum in velocity space is then
\begin{eqnarray}
P_s({\bf k}) &=& \int d^3 {\bf r} \, e^{i {\bf k}  \cdot {\bf r}} \xi(r)
\exp\left[-\frac{(k_z\lambda)^2 \tilde D_z({\bf r}/\lambda)}{2} \right] ~~~,
\label{eq:kolmogorov}\\
\tilde D_z({\bf r}/\lambda)&=&(r/\lambda)^m \left(1+m/2 (1-\cos^2
\theta)\right) ~~~,
\label{Dzz}
\end{eqnarray}
where the turbulence scale in {\it velocity} space
 is 
\be
\lambda = {\left[f^2 C \right]}^{1 \over 2-m} ~~~.
\label{fbar}
\ee

	This scale
	depends on the Galactic rotation curve, direction of observations and
	turbulence intensity. The
	scale $\lambda$ corresponds to the scale over which random 
	velocities map density fluctuations. Physically this is the
scale at which the velocity dispersion $\sim C \lambda^{m}$
becomes equal to
the squared difference of the regular velocities determined by Galactic
rotation (i.e. $f^{-2}\lambda^2$).

While we discuss the properties of the 3D spectrum in velocity space 
 in Appendix B, the 2D spectrum of HI intensity fluctuations
available via observations is discussed in the next section.

\section{Thick and Thin Slicing of Data Cubes}

It is easy to see that a radiointerferometer gets the 2D spectrum
of 21~cm intensity (see L95). Therefore it is very important to relate this
2D spectrum to the underlying 3D statistics of turbulence.
The velocity resolution of an individual radiointerferometer
channel represents the thinnest slicing of the data available. In
practice, several channels can be combined to provide a thicker slice
(see Green 1993).

Henceforth we use capital letters ${\bf R}$ and ${\bf K}$ to 
denote 2D quantities and reserve small letters ${\bf r}$ and
${\bf k}$ for 3D quantities. In particular,
$\bf R$ is the vector in two dimensional plane
orthogonal to the line of sight, and the distance in $xyv$ space is
${\bf r}_s^2={\bf R}^2+z_s^2$, where $z_s$ is the z-coordinate in  velocity
 space.

An important property of 21~cm emission is that it does not depend
on gas temperature (see Spitzer 1978). For negligible  self-absorption
(see discussion in section 2.1)
the HI intensity at the point ${\bf R}$, observed in velocity slice
$\delta V$, is proportional to the integral of
the density over the slice thickness
\be
I({\bf R}) \sim \int d z_s^\prime \rho \left({\bf R},z_s^\prime\right)
W_e\left[\frac{z_s-z_s^\prime}{\cal L}_{\rm sl}\right]~~~.
\ee
The width of the normalized ($\int dx_s W_e = 1$) experimental window function $W_e$, 
${\cal L}_{\rm sl}$, represents the velocity 
slice thickness; $z_s$ is the value of the slice central velocity.
We expressed both quantities in length units, i.e. ${\cal L}_{\rm sl} \approx f \delta V$, 
keeping in mind the coordinate transformation (\ref{eq:map}).
The shape of the window function depends on the sensitivity profile and
the width of the individual interferometer channel and on how the channels are combined.
In idealized case of uniform sensitivity one can consider step-like window 
$W_e(\Delta z/{\cal L}_{\rm sl}) = 1/{\cal L}_{\rm sl} $ if $ |\Delta z| \le {\cal L}_{\rm sl}/2$ and $W_e=0$ otherwise.
Intensity $I({\bf R})$ implicitly carries dependence on $z_s$, which we omit for brevity.
 
The normalized correlation
function of intensity fluctuations is
\begin{eqnarray}
\langle \delta I({\bf R_1}) \delta I({\bf R_2})\rangle & \sim &
\int \int dz_{s2}^\prime dz_{s1}^\prime
\langle \delta \rho({\bf R_1}, z_{s1}^\prime) \delta \rho({\bf R_2}, z_{s2}^\prime)
\rangle W_e\left[\frac{z_{s1}-z_{s1}^\prime}{\cal L}_{\rm sl}\right] 
W_e\left[\frac{z_{s2}-z_{s2}^\prime}{\cal L}_{\rm sl}\right] \nonumber \\ 
&=& \int \int dz_{s2}^\prime dz_{s1}^\prime
\xi_s \left({\bf R_1}-{\bf R_2},z_{s1}^\prime-z_{s2}^\prime\right)
W_e\left[\frac{z_{s1}-z_{s1}^\prime}{\cal L_{\rm sl}}\right] 
W_e\left[\frac{z_{s2}-z_{s2}^\prime}{\cal L_{\rm sl}}\right]\label{eq:inten}
\end{eqnarray}
where $ \delta I$ and $ \delta \rho$ mean, respectively,
variations of intensity and density. The density correlation function
in the velocity space $\xi_s$ is the Fourier transform of the turbulence spectrum 
$P_s({\bf K},k_z)$ (see eq.~(\ref{eq:main})) smoothed by the thermal factor
\be
\xi_s({\bf r}_s)=(2 \pi)^{-3} \int d^3{\bf k}
e^{-i {\bf k}\cdot{\bf r}_s}P_s({\bf k}) e^{-f^2 k_z^2 v_T^2}~~~
\label{xi_s}
\ee 
The properties of the correlation function are discussed in Appendix \ref{app:3Dcorr}. 

The sum of the squared imaginary and real parts of the interferometer
visibility function is proportional to the two dimensional spectrum
of  intensity fluctuations (see L95), which is
\be
P_2({\bf K}) \equiv \int d^2{\bf R}\, e^{i {\bf K R}} 
\langle \delta I({\bf R_1}) \delta I({\bf R_1}+{\bf R})\rangle~~~.
\label{ppp}
\ee
Substituting eqs.~(\ref{eq:inten}) and (\ref{xi_s}) into eq.~(\ref{ppp})
after simple calculations similar to those in Appendix~E we derive
the expression for $P_2({\bf K})$ 
\be
P_2({\bf K})|_{\cal L} \sim
{1 \over 2 \pi} \int_{-\infty}^{\infty} dk_z  P_s({\bf K}, k_z)\,
\tilde W_e^2(k_z {\cal L}_{\rm sl}) e^{-f^2 k_z^2 v_T^2}~~~,
\label{eq:slice}
\ee
where $|_{\cal L}$ reflects the dependence on the slice thickness.
$\tilde W_e(k_z {\cal L}_{\rm sl})$ is the Fourier transform of the experimental window function.
For instance 
$\tilde W_e^2(k_z {\cal L}_{\rm sl})=2 \left[1-\cos(k_z {\cal L}_{\rm sl}) \right] / (k_z {\cal L}_{\rm sl})^2$ 
for the step-like window.

The effective filter $W(k_z {\cal L}_{\rm sl}, k_z v_T)$ acting on 3D turbulence spectrum in
eq. (\ref{eq:slice}) is the product of experimental window function and
the thermal term
\be
W(k_z {\cal L}_{\rm sl}, k_z v_T)=\tilde W_e^2(k_z {\cal L}_{\rm sl}) e^{-f^2 k_z^2 v_T^2} ~~~.
\label{window}
\ee
This demonstrates that thermal dispersion can be treated as part of the 
experimental velocity window.
The effect of velocity slicing $\delta v$ is similar to the effect of
$v_T$ in the sense that they both limit the $k_z$ resolution. 
Increase of $v_T$ smears the fluctuations. Thus
we may expect the fluctuations from colder, $v_T < \delta v$, parts of the medium
to dominate over the contribution from warmer, $v_T > \delta v$, parts.
The exact criterion when this is true is given below.

At small $k_z$ we have $W \propto 1 -  k_z^2 ( {\tilde W_e}^{2\prime\prime}(0)
{\cal L}_{\rm sl}^2 + 2 f^2 v_T^2)/2 $,
thus the characteristic scale of the window can be defined by an effective
slice thickness 
\begin{equation}
{\cal L} = 2 ({\tilde W_e}^{2\prime\prime}(0)
{\cal L}^2_{\rm sl} + 2 f^2 v_T^2)^{1/2}=
2 f({\tilde W_e}^{2\prime\prime}(0) \delta v^2+2 v_T^2)^{1/2}~~.
\label{newL}
\end{equation} 
This is the customary definition of a characteristic scale
as twice the value of $k_z^{-1}$ at which the filter value drops by a factor of
two of its maximum. This approximately corresponds to the full width at 
half maximum
of the window in velocity space and is the effective width of the velocity 
slice. In what follows we denote the window function given by 
eq.~(\ref{window}) by $W(k_z {\cal L})$ and disregard its dependence on
$v_T$ when it is possible. In section~4.3, we, however, will have 
to treat the dependence of ${\cal L}$ on $v_T$ in order
to determine the relative
contribution of the warm and cold HI.

For step-like experimental window,  ${\tilde W_e}^{2\prime\prime}(0)=1/6$.
In this case thermal velocity  
dispersion has much more profound effect than the averaging over
the velocity channel of similar thickness. This is due to relatively weak 
cutoff in Fourier
space provided by step-like averaging in velocity slice.

As long as the exact
shape of the high $k_z$ cutoff by the window function is not important,
using effective slice thickness fully accounts for the effect of finite gas 
temperature.
From now on we shall mean `effective' everywhere we speak about slice 
thickness.

%he window function has the following properties: $W(0)=1$, 
%$W(k_z {\cal L} \gg 1) \sim (k_z {\cal L})^{-2} $ and
%$\int_{-\infty}^{\infty} d k_z W(k_z {\cal L}) = 2 \pi /{\cal L}$.

The thinner is 
the velocity slice ${\cal L}$, the closer to unity is the window $W$ over
larger range of $k_z$ and more 3D modes contribute to 2D spectrum. 

We shall call a slice {\it thin},
when the $W(k_z{\cal L})$ can be approximated by unity over all $k_z$ of interest
and the one of {\it thick}, when it is important that at high
$k_z$ the function $W(k_z {\cal L})$ decreases . Therefore when the slice is {\it thin}
eq.~(\ref{eq:slice}) gives\footnote{The intuitive notion
of the thin slice is hardly associated with the integral over spectrum.
Note, however, that the integral is taken in $k$-space and, for instance, a
thin sheet in a real space corresponds to an integral in $k$-space. A
different way of understanding eq.~(\ref{eq:thin}) is to consider randomly
moving 3D waves that produce a pattern corresponding to the wavevector
${\bf K}$ on their intersection of the plane. Obviously enough, various
3D waves with ${\bf k}=({\bf K}, k_z)$ contribute to the pattern. Integration
over $k_z$ in  eq.~(\ref{eq:thin}) reflects this.}   
\be
P_2({\bf K})|_{t} \approx
{1 \over 2 \pi} \int dk_z P_s({\bf K}, k_z)~~~,
\label{eq:thin}
\ee
where $|_{t}$ denotes that the slice is thin
while in the limit of a {\it thick} slice the exact expression for
the window function $W(k{\cal L})$ (i.e. eq.~(\ref{window})) 
should be used. Obviously enough,
whether the slice is thin or thick depends not only on ${\cal L}$
but also on the properties of $ P_s({\bf K}, k_z)$.
The exact criteria for 
slices to be thin and thick are discussed in the next section.

One expects different behavior of resulting 2D spectrum $P_2$ depending 
on a slice being {\it thin} or {\it thick}.
To clarify this distinction,
let us consider the density power-law 3D spectrum
$P_s({\bf K}, k_z)\propto(|{\bf K}|^2+k_z^2)^{n/2}, ~n < -1$ that is 
{\it not} subject to a velocity mapping. For this simple
case $P_s({\bf K}, k_z)$ is nearly constant
if $k_z < |{\bf K}|$ and  $\propto k_z^n$ if $k_z > |{\bf K}|$. 
This fall-off at high $k_z$ ($n$ is negative) amounts to a cut-off of the integral
$\int_0^\infty dk_z P_s({\bf K}, k_z) \approx \int_0^{|{\bf K}|}
dk_z P_s({\bf K}, k_z) $
at $k_z \approx |{\bf K}|$ (assuming, as we did, $n < -1$).
Correspondingly, if ${\cal L}|{\bf K}| \ll 1$ 
the window function is unimportant and the slice is {\it thin}, while
the slice is {\it thick} when ${\cal L}|{\bf K}| \gg 1$. In other words, 
the transition between 
{\it thick} and {\it thin} slices occurs when we
consider wavelengths longer or shorter than slice thickness. 
For the thick slice, the contribution
to the integral (\ref{eq:slice}) for the case considered
mostly comes from the $k_z \sim 0$ modes i.e.
\be
P_2({\bf K})|_{\cal T} \approx P_s({\bf K}, \bar k_z) 
\int d k_z W(k_z {\cal L})/2 \pi
\approx {\cal L}^{-1}  P_s({\bf K}, 0)~~~, 0\le \bar k_z < 1/{\cal L},
\label{eq:thick}
\ee
where $|_{\cal T}$ denotes that the slice is thick,
and this result
corresponds to the finding in L95. The result can be easily understood.
For sufficiently steep spectra
($n<-1$) integration over $z$ corresponds to choosing $k_z=0$ mode in 
the Fourier space.
Thus the 2D intensity spectrum and the underlying 3D density spectrum with 
$k_z=0$ should be identical. 

Thus, the asymptotics
of the 2D power spectrum produced by integrating  density fluctuations
along a line of sight are
\begin{equation}
P_2({\bf K}) \propto \left\{ 
\begin{array}{l}
|{\bf K}|^n, ~~~~~~~ {\cal L} |{\bf K}| \gg 1 \\
|{\bf K}|^{n+1}, ~~~~~ {\cal L} |{\bf K}| \ll 1
\end{array}
\right.
\label{eq:purpower}
\end{equation}

However, the rule of transition  between {\it thin} and {\it thick} 
slices gets modified
when we consider the emissivity power spectrum in velocity space.
In the presence of velocity mapping (see eq.~(\ref{eq:map})) an additional
scale\footnote{If the underlying 3D spectrum has several built-in scales,
detailed analysis can distinguish between several intermediate asymptotics.}
$\lambda$ (see eq.~(\ref{eq:roman})) appears. This alters the 
wavenumber $|{\bf K}|$ at which the transition between the {\it thin}
and {\it thick} slices occurs.

%It is easy to understand how 
% the modification of $P_s({\bf K}, k_z)$ in the presence of velocity 
%fluctuations entail the change of the criterion for {\it thin} and
%{\it thick} slicing. Recall, that we call the slice {\it thin}
%if the window function $W( k_z{\cal L})$ can be approximated by
%unity. This is possible, for instance, if  velocity mapping results
%in $P_s({\bf K}, k_z)$ which is cut-offed at $k_z<{\cal L}^{-1}$. In the
%next section we shall show that this is
%exactly what happens over a substantial range of ${\cal L}$
%when velocity fluctuations are present.

The {\it thin}-slice expression for 2D spectrum in the velocity space 
follows from
eq.~(\ref{eq:thin}) with 3D kernel given by eq.~(\ref{eq:kolmogorov})
%AL 25, changed
\begin{equation}
P_2({\bf K})|_{t} \approx \frac{1}{(2\pi)^{1/2}\lambda}
\int d^3 {\bf r} \, e^{i {\bf K}  \cdot {\bf R}}~ 
\frac{\xi(r)}{\tilde D_z({\bf r}/\lambda)^{1/2}}~
\exp\left[-\frac{z^2}{2 \lambda^2 \tilde D_z({\bf r}/\lambda)}\right]
\label{eq:thinv}
\end{equation}

The exact relation between $P_2({\bf K})$ and the underlying velocity and density statistics
depends on whether the density spectrum of HI is shallow or steep. In the
next section we discuss these two cases separately. 
%AL 25
%It is also obvious that
%the very importance of the velocity fluctuations depends on the relation
%between the turbulent and thermal velocity dispersions. Obviously enough
%the contribution of velocity fluctuations is marginally important for subsonic
%turbulence.

\section{ Two Dimensional Spectra }

Below we consider the distribution of density 
dominated (a) by fluctuations at small scales and (b) by
fluctuations at large scales. As we mentioned earlier, we limit 
our discussion to the power-law spectra
both because observations suggest power-law dependences (see Green 1993,
Stanimirocic et al 1999). If future research suggests a 
different dependences for
interstellar statistics, our formulae can be used to fit the data.
We show that in terms of power-law spectra case (a) 
corresponds to a power-law
index $n>-3$ and case (b) to $n<-3$.

\subsection{Short-wave dominated spectrum of density field}

To describe the statistical properties of the density fluctuations  
dominated by short wavelengths, we use  power-law correlation functions
of {\it over-density}:
\begin{equation}
\xi(r)=\langle \rho \rangle^2 
\left(1 + \left( {r_0 \over r} \right)^\gamma\right), ~~~~~ \gamma > 0~~~.
\label{eq:xi}
\end{equation}
The power-law part of the correlation corresponds to the 3D power-law spectrum 
\begin{equation}
P(k) \propto k^n, ~~~~n=\gamma-3 > -3~~~.
\end{equation}
The unity term in eq.~(\ref{eq:xi})
reflects the fact that {\it density} has non-zero mean value.
After Fourier transformation this term leads to a delta function
 term at ${\bf k}=0$ added to the power spectrum.
The correlation scale $r_0$ gives us the second
parameter of the problem (the first one being $\lambda$).

Two terms in the expression for correlation function lead to the correspondent
split of the power spectrum in velocity space into two parts. We show in the
Appendix B that the 3D power spectrum in velocity space  can be presented as
\be
P_s(|{\bf K}|,k_z)=\langle \rho \rangle^2
\left[P_{v}(|{\bf K}|,k_z)+P_{\rho}(|{\bf K}|,k_z)\right] ~~~, 
\label{eq:split}
\ee
where we introduced a notation
\begin{eqnarray}
P_v(|{\bf K}|,k_z) &=&  \int d^3 {\bf r} \, e^{i {\bf k}  \cdot {\bf r}} 
\exp\left[-\frac{(k_z\lambda)^2 \tilde D_z({\bf r}/\lambda)}{2} \right] ~~~,
\label{eq:pv}\\
P_{\rho}(|{\bf K}|,k_z) &=& \int d^3 {\bf r} \, e^{i {\bf k}  \cdot {\bf r}} 
\left( {r_0 / r} \right)^\gamma
\exp\left[-\frac{(k_z\lambda)^2 \tilde D_z({\bf r}/\lambda)}{2} \right] ~~~.
\label{eq:prho}
\end{eqnarray}
In eq.~(\ref{eq:split}) $P_v(\bf k)$ 
is the emissivity spectrum in velocity space arising from random velocity 
fluctuations in an incompressible fluid and in terms of mathematics
arises from the unity term (mean density) in eq.~(\ref{eq:xi}), while 
$P_{\rho}(\bf k)$
describes the spectrum of  emissivity fluctuations
arising from both velocity and 
density fluctuations. Formally $P_{\rho}(\bf k)$ arises from velocity
mapping acting upon the density fluctuations given by
$(r/r_0)^{\gamma}$ term in eq.~(\ref{eq:xi}).
As expected, the $P_v$ term has a $\delta$-function
behavior at $k=0$, but in velocity space
it is also non-vanishing for $k > 0$. As we mentioned earlier
even uniformly distributed in space,
but turbulently moving incompressible
fluid produce fluctuating emission (given by  $P_v(\bf k)$) when observed
in velocity slice of finite thickness.

We derive the asymptotics for $P_v$ and $P_{\rho}$ in 
Appendix~B. The main result is that the velocity mapping leads to the 
rapid cutoff of modes with large $k_z$ component of the wave vector.
Both $P_{\rho}$ and $P_v$ falls sharply ($\propto k_z^{-2\alpha/m}$,
where $\alpha$ is 3 for $P_v$ or $-n$ for $P_\rho$)
for $k_z  \gg  \lambda ^{-1} (k\lambda)^{m/2}, ~k^2={\bf K}^2+k_z^2$.
This implicitly defines the cutoff scale $k_z^c$ which can be approximated as 
$k_z^c = \lambda ^{-1} (1+|{\bf K}|\lambda)^{m/2}$.

We will see that $P_v$ and $P_{\rho}$ emerge, although in
another combination, when the density spectrum is long-wave dominated.
Thus, below, we formulate a {\it general criterion} for the slice
to be {\it thin} or {\it thick} in the presence of velocity mapping.
 
As we discuss in section~5, for  scales of observational interests
$|{\bf K}| \lambda \gg 1 $. For these modes
$k_z^c \approx \lambda ^{-1} (|{\bf K}|\lambda)^{m/2} < |{\bf K}| $, 
since $m < 2$.
This strongly affects the integration in eq.~(\ref{eq:slice}).
Indeed, one can assume that the window function
$W({\cal L} k_z)\approx 1$ as long as ${\cal L} k_z^c < 1$. 
In particular, the 
slice remains {\it thin} for much larger $|{\bf K}|$
than one would expect
from the example of power-law spectrum unmodified by the velocity mapping
(see eq.~(\ref{eq:purpower})). 
Now, given $|{\bf K}| \lambda \gg 1 $,
the transition between the {\it thin} to {\it thick} slice regimes
is 
\begin{eqnarray}
thin~~slice: ~~~~~~~~ {\cal L}/\lambda \cdot (|{\bf K}|\lambda)^{m/2} \ll 1 
\nonumber \\
thick~slice: ~~~~~~~~ {\cal L}/\lambda \cdot (|{\bf K}|\lambda)^{m/2} \gg 1
\label{eq:transition}
\end{eqnarray}
The {\it thin}-slice asymptotics is achieved when the first condition
is satisfied due to very steep cutoff of 3D spectrum at high $k_z$.
In other words, due to a rapid decrease of 3D spectrum $P_s$ 
the window function can be ignored over the whole range
of $k_z$.
The last inequality essentially gives us the {\it necessary} condition
for the variations of the window function to be important. 
To appreciate the effect of the velocity mapping let us rewrite it
as ${\cal L} |{\bf K}| > (|{\bf K}|\lambda)^{1-m/2} \gg 1$. This is to be
compared with {\it thick slice} condition ${\cal L} |{\bf K}| > 1$ which 
would take effect if there is no velocity term, $\lambda \to 0$ (combined
rule ${\cal L} |{\bf K}| > max(1, (|{\bf K}|\lambda)^{1-m/2})$ is valid for
any $\lambda$).
For the wavenumber $|{\bf K}|=(10~{\rm pc})^{-1}$, typical velocity turbulence
scale $\lambda=3.5$~kpc (see Section 5) and $m=2/3$ we thus need 
${\cal L}  > 500$~pc for slice to be {\it thick}, which is fifty times
larger than Eq.~(\ref{eq:purpower}) suggests. 
Indeed, according to  Eq.~(\ref{eq:purpower}) ${\cal L} > |{\bf K}|^{-1}=10$~pc
is sufficient for the slice to be thick 
 if the velocity field is absent. 
%However, if $L$ gets larger than
%$\lambda$ the criterion ${\cal L}|{\bf K}|$ $\gg 1$ or $\ll 1$ determines
%whether the slice is, respectively, thick or thin.

The separation between {\it thin} and {\it thick} regimes is intuitively clear
if rewritten in physical
units. Substituting $\lambda$ from eq.~(\ref{fbar}) and expressing slice
thickness through the width of window function,
which for $v_T=0$ is proportional to the width
of the interferometer channel $\delta V$, 
${\cal L} = f \delta V$ we obtain 
\begin{eqnarray}
thin~~slice: ~~~~~~~~ C  |{\bf K}|^{-m} \gg \delta V^2 \nonumber\\
thick~slice: ~~~~~~~~ C  |{\bf K}|^{-m} \ll \delta V^2
\label{eq:phystrans}
\end{eqnarray}
In other words, if the velocity dispersion $Cr^m$ on the scale $|{\bf K}|^{-1}$
is larger than the squared width of the channel\footnote{Here, for the
sake of simplicity, we talk about measurements for which 
the thickness of the slice is determined by the velocity channel of
the telescope. In reality, data from many channels can be added to
produce a thicker slice (see section 6). 
Moreover, in most cases
$v_T$ contribution is important in eq.~(\ref{newL})
and in more precisely $\delta V=0.5({\tilde W_e}^{2\prime\prime}(0) 
\delta v^2+2 v_T^2)^{1/2}$. Generalizations of
our criterion to these cases are trivial.} (in velocity units)
 the slice is {\it thin}. If
the opposite is true -- the slice is {\it thick}.
We stress, that the critical wavelength depends only on the width of the 
interferometer channel
and the amplitude of velocity turbulence, but not on the direction 
of observations or the actual physical thickness of the slice involved
(al long as $|{\bf K}| \lambda \gg 1 $).
To observe the transitions between the two regimes one can either vary
$\delta V$ or $|{\bf K}|$.
We show in Appendix E that the criterion (\ref{eq:phystrans}) is
quite general and applicable not only to the Galaxy but to individual
clouds and external galaxies.

The 2D spectrum of intensity fluctuations 
in a velocity slice of data can be obtained using eq.~(\ref{eq:slice}).
It follows from eq.~(\ref{eq:split}) that
\begin{equation}
P_2(|{\bf K}|)=\langle \rho \rangle^2
\left[P_{2v}(|{\bf K}|)+P_{2\rho}(|{\bf K}|)\right]~~~,
\label{2Dspec}
\end{equation}
where 
\be
P_{2v}({\bf K})|_{\cal L} \sim
{1 \over 2 \pi} \int_{-\infty}^{\infty} dk_z  P_v({\bf K}, k_z)\,
W(k_z{\cal L})~~~,
\label{slicev}
\ee
and
\be
P_{2\rho}({\bf K})|_{\cal L} \sim
{1 \over 2 \pi} \int_{-\infty}^{\infty} dk_z  P_{\rho}({\bf K}, k_z)\,
W(k_z{\cal L}) ~~~,
\label{slicerho}
\ee
The calculation of appropriate integrals is straightforward using the
asymptotics  for 3D spectra $P_v$ and $P_{\rho}$ obtained in 
Appendix~B. In Table~1
we present asymptotical formulae for $P_{2\rho}$ and $P_{2v}$
 spectra in a {\it thin} and a 
{\it thick} 
slice regimes under the assumption $|{\bf K}| \lambda \gg 1 $. 
$P_{2\rho}$ and $P_{2v}$ also appear in the case of long-wave dominated density
fluctuations and therefore our results are quite general.

\begin{table}[h]
\begin{displaymath}
\begin{array}{lrr} \hline\hline\\
& \multicolumn{1}{c}{\rm thick~~slice}  &
\multicolumn{1}{c}{\rm thin~~slice}
\\[2mm]
& \multicolumn{1}{c}{{\cal L}/\lambda \cdot (|{\bf K}|\lambda)^{m/2} \gg 1};  &
\multicolumn{1}{c}{{\cal L}/\lambda \cdot (|{\bf K}|\lambda)^{m/2} \ll 1}
\\[2mm] \hline \\
\lambda^{-2} P_{2\rho}({\bf K}): &
A_n \cdot (\lambda/{\cal L})\cdot (r_0/\lambda)^{3+n}\cdot (|{\bf K}|\lambda)^n;
&S_{nm} (r_0/\lambda)^{3+n} \cdot (|{\bf K}|\lambda)^{n+m/2} \\[2mm] 
\hline \\[3mm]
\lambda^{-2} P_{2v}({\bf K}): & B_m \cdot (\lambda/{\cal L})^3 \cdot 
(|{\bf K}|\lambda)^{-3-m}; &
S_{-3m} \cdot (|{\bf K}|\lambda)^{-3+m/2} \\[3mm] \hline
\end{array}
\end{displaymath}
\caption{Asymptotics of the  2D spectrum in the {\it thin}
and {\it thick} velocity slices for 
$|{\bf K}|\lambda \gg 1$. Numerical constants $A_n$, $B_m$ and $S_{nm}$
are given in Appendix A. 
In Appendix~E we show that
these results are applicable to individual clouds provided that the cloud size is used instead of
$\lambda$.} 
\label{tab:2Dspk_asymp}
\end{table}

In the regime of a {\it thick} slice Table~\ref{tab:2Dspk_asymp} shows that 
the power-law indexes of the 2D spectrum and the underlying 3D density 
spectrum coincide.
This is the regime considered in L95. If the slice is {\it thin}, the 
2D spectrum 
index becomes shallower, i.e. $n+m/2$. However this spectrum is still
steeper than the spectrum of density in a {\it thin} slice
in the absence of velocity modulations. Indeed, in the latter case  
eq.~(\ref{eq:purpower}) predicts  the index $n+1>n+m/2$ for $m<2$.
This {\it thin}- slice behavior can be immediately found from the integral
representation given by eq.~(\ref{eq:thinv}). As can be easily shown, 
at high $|{\bf K}|$
the exponential factor can be set to unity, so
\begin{equation}
P_2({\bf K})|_{t} \approx \frac{1}{(2\pi)^{1/2}\lambda}
\int d^3 {\bf r} \, e^{i {\bf K}  \cdot {\bf R}}~ 
\frac{\xi(r)}{\tilde D_z^{1/2}({\bf r}/\lambda)}
\label{eq:thinhighK}
\end{equation}
An extra factor $D_z^{1/2}({\bf r}) \propto C^{1/2}r^{m/2}$ is responsible
for the change
of slope of the spectrum due to velocity effect.

Note that in a {\it thin} slice $P_{2v}$ behaves as a density part 
$P_{2\rho}$ taken
with $n=-3$. Comparing  expression for $P_{2\rho}$ rewritten as
$P_{2\rho} = S_{nm} (|{\bf K}| r_0)^{3+n} (|{\bf K}|\lambda)^{-3+m/2}$
with the formula for $P_{2v}$ shows that for short-wave
dominated spectra $n > -3$ the
density part is always dominant over $P_{2v}$ at wavelength
shorter than the density correlation length $|{\bf K}| r_0 > 1$.
The density term $P_{2\rho}$ is also dominant for {\it thick} slices.
Indeed,
\be
\frac{P_{2\rho}}{P_{2v}}\approx ( |{\bf K}| r_0)^{3+n}\left[\left(\frac{\cal L}{\lambda}\right)^2
( |{\bf K}| \lambda)^{m}\right]\gg 1~~~,
\ee
as shallow density with 
$n>-3$ and the terms in the second bracket $\gg 1$ for a thick slice.  
In this regime the velocity part
has an extra $(\lambda/ {\cal L})^2$ factor and 
extra damping $(|{\bf K}|\lambda)^{-m}$.

The magnitude of $P_{2v}({\bf K},k_z)$ is more tangible than one might
suggest in an analogy with eq.~(\ref{eq:thick}). 
The reason is that $P_{2v}({\bf K},k_z)$ is a growing
function of $k_z$ up to  $k_z'=\lambda^{-1}{\rm max}\left[1,(|{\bf K}|\lambda)^{m/2}\right]$
(see Table 2 in Appendix B). This is the case where the exact shape of the fall-off
of the window $W({\cal L} k_z)$ is important. Typically if $W({\cal L} k_z)$ 
falls fast at $ k_z > {\cal L}^{-1}$ we have 
\begin{equation}
\lambda^{-2} P_{2v}(|{\bf K}|) \propto \lambda \int_0^{1/\cal L}\!\!
dk_z \frac{(k_z \lambda)^2}{(k \lambda)^{3+m}} 
\propto  (\lambda/{\cal L})^3 (|{\bf K}|\lambda)^{-3-m}
\label{eq:pvthick}
\end{equation}
Peculiar is the case of step-like uniform window, which falls off only as $1/(k_z {\cal L})^2$,
then
\begin{equation}
\lambda^{-2} P_{2v}(|{\bf K}|) \propto \int_0^{(|{\bf K}|\lambda)^{m/2}}
d (\lambda k_z) \frac{(k_z \lambda)^2}{(k \lambda)^{3+m}} \frac{1}{(k_z {\cal L})^2}
\propto (\lambda/{\cal L})^2 (|{\bf K}|\lambda)^{-3-m/2}
\label{eq:thickv2}
\end{equation}
Our numerical results have been obtained for this case.

Details of the transition from  {\it thin} to {\it thick} slicing
as ${\cal L}$ increases require numerical study. The results,
plotted in Figure~1, confirm the accuracy of our analytical estimates.
\begin{figure}[h]
{\centering \leavevmode
\epsfxsize=.33\columnwidth \epsfysize=.45\columnwidth \epsfbox{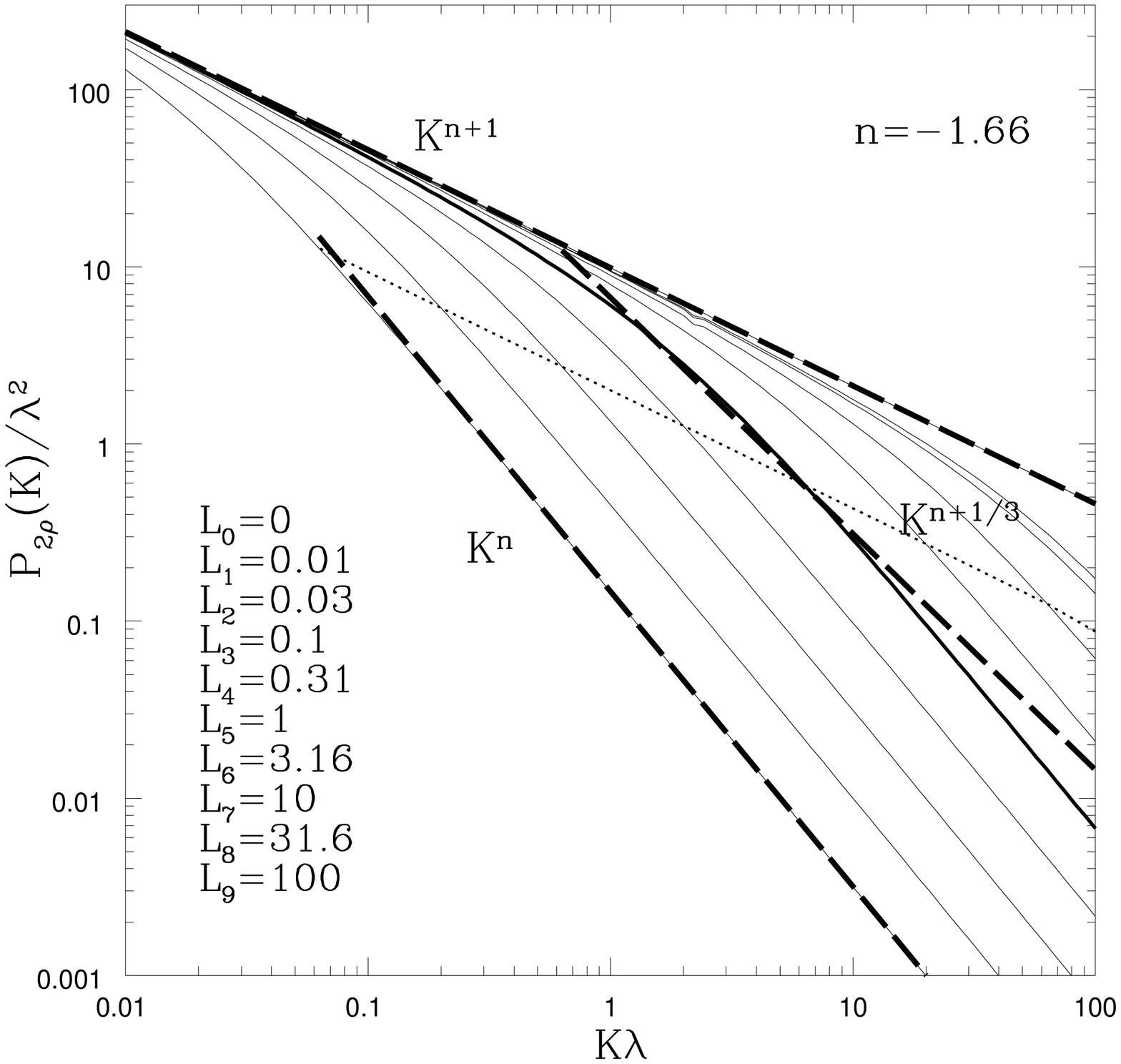} \hfil
\epsfxsize=.33\columnwidth \epsfysize=.45\columnwidth \epsfbox{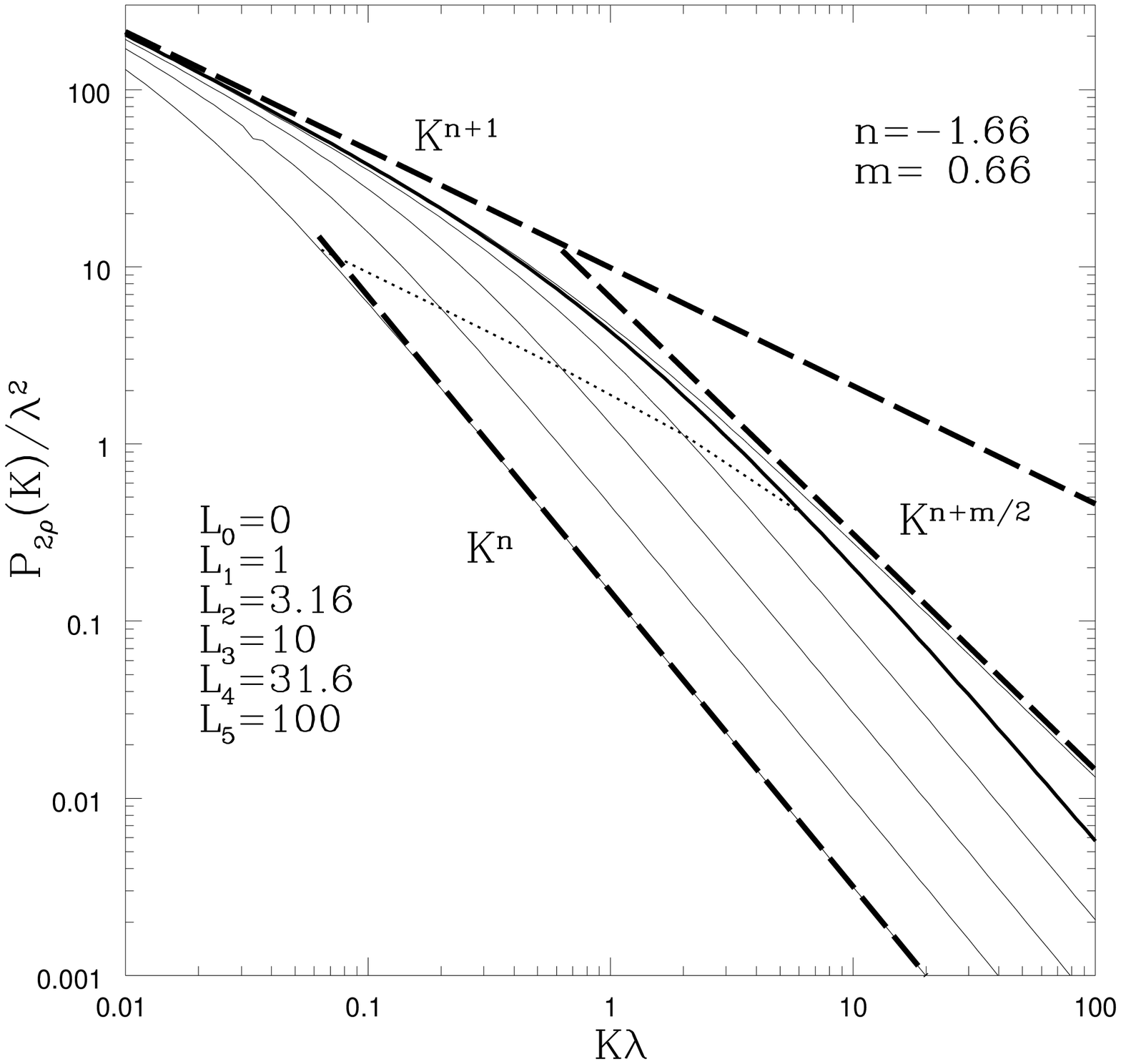} \hfil
\epsfxsize=.33\columnwidth \epsfysize=.45\columnwidth \epsfbox{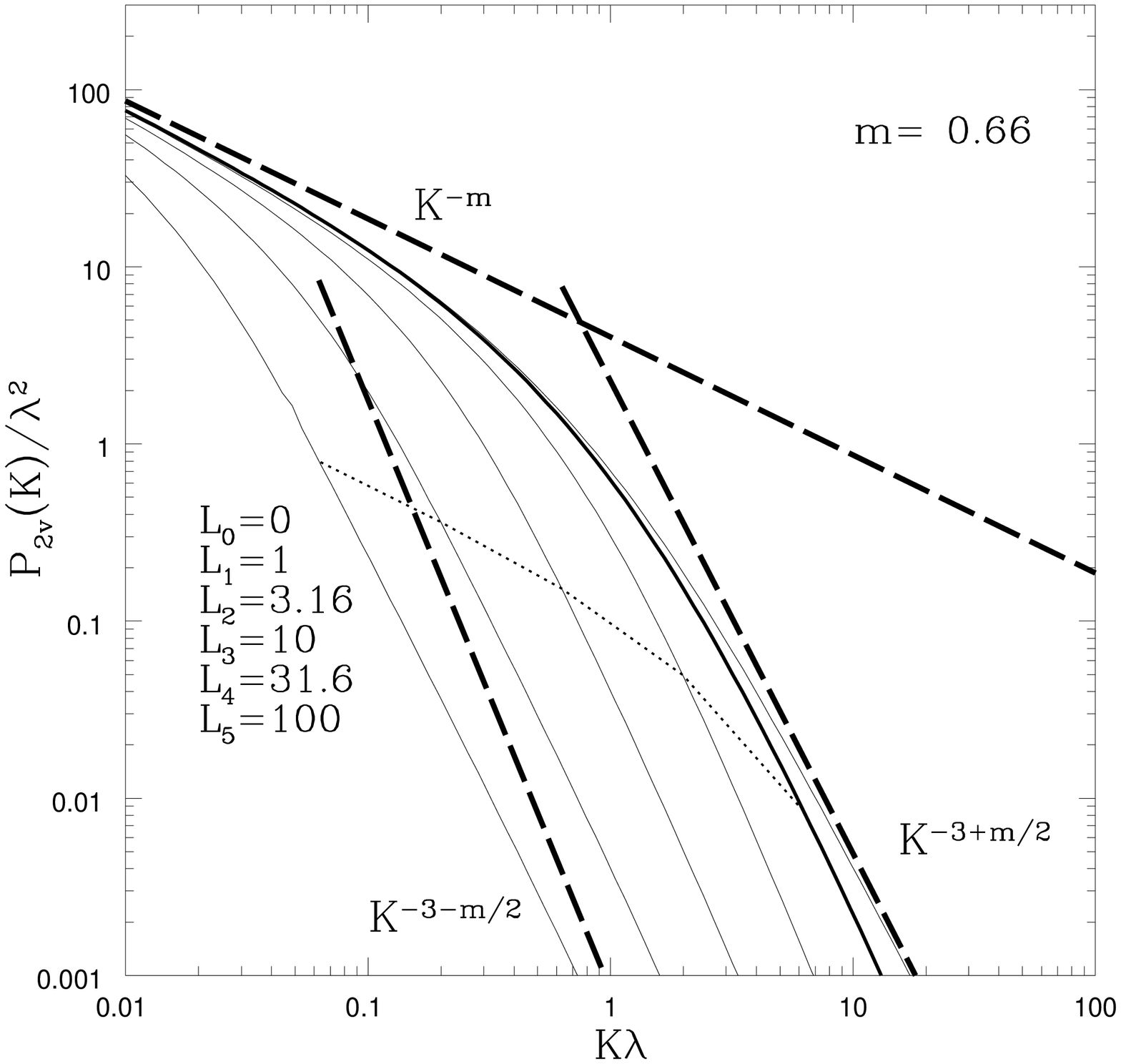} 
}
%\centerline{\epsfxsize=5.in\epsfbox{spect2Dn53m23.eps}}
\caption{Two dimensional power spectrum measured in a slice of variable 
thickness
$L={\cal L}/\lambda$ (upper curves correspond to smaller $L_i$). Left panel shows $P_2(|{\bf K}|)$ in the absence of velocity mapping. 
The middle and the right panels correspond to density 
$P_{2\rho}$
and velocity $P_{2v}$ terms.
The upper dashed lines show the {\it thin} slice power-law asymptotics
described in Table~\ref{tab:xi_asymp}; for comparison the short-wave 
$P_{2\rho}$
slope is also shown in the left panel.
The curve of enhanced weight corresponds to $L=1$.
The {\it thick} slice asymptotics are accurate below the dotted line.} 
\label{fig:2Dspek}
\end{figure}
Our calculations show how with the increase of slice thickness the 
spectral index of $P_{2\rho}$ steepens from
the value $n+m/2$, characteristic to thin slice, towards the value $n$ measured in thick
slices. The transition scale between these two regimes is significantly
increased as the result of velocity fluctuations.
For thin slices and large   ${\bf K}$ the $P_{2\rho}$ spectrum 
depends on both velocity and density fluctuations. The pure velocity term
$P_{2v}$ is always subdominant at short wavelengths.

All in all, 
on short wavelengths $|{\bf K}|\lambda \gg 1$ thick velocity 
slicing of 21~cm data cubes
provides the spectrum $|{\bf K}|^n$, while thin slicing provides 
$|{\bf K}|^{n+m/2}$.
Therefore if density statistics is dominated by small scale fluctuations,
interferometers can provide us both with density and velocity
statistics; the density spectral index being obtained from thick slices,
and combining this with the thin slice data one can obtain the 
velocity spectral index $m$.

\subsection{Long-wave dominated spectrum of density field}

Our analysis above does not cover $\gamma<0$ and $n<-3$. This is the regime
dominated by large-scale density fluctuations that we consider in this
section.

Kolmogorov turbulence is often assumed to be valid
not only for the velocity but also for the density\footnote{This is exactly
true for the density of a passive contaminant transported by the Kolmogorov
velocity field. It is not {\it a priori} clear that such a treatment is
applicable to HI.}. This type of turbulence
corresponds to $\gamma=-2/3=-m$, $n=-11/3$. For $\gamma<0$ 
we formally substitute
 in Eq.~(\ref{eq:kolmogorov})
\be
\xi(r)=d(\infty)[1-d(r)/d(\infty)]~~~,
\ee
where $d(r)$ is a structure function\footnote{We could start from
the structure functions from the very beginning but we preferred to
use correlation functions to make our presentation more uniform.} of density
\be
d(r)=
\langle(\rho({\bf x}+{\bf r})-\rho({\bf x}))^2 \rangle~~~.
\ee

Real world structure functions do not grow infinitely and therefore
we have to introduce a cutoff at some large scale. If the 
cut-off happens at $r_c$ then for the long-wave dominated turbulence 
($\gamma<0$)
\be
d(r)=d(\infty){r^{-\gamma} \over r^{-\gamma}+r_c^{-\gamma}}~~~,
\label{structure}
\ee
and the correlation function is
\be
\xi(r)=\frac{d(\infty)}{2}{r_c^{-\gamma} \over r^{-\gamma}+r_c^{-\gamma}}~~~.
\label{cor-structure}
\ee

For simplicity\footnote{This is a consistent assumption for the Kolmogorov
density turbulence.}, further on we assume that the structure functions of
velocity and density have the cut-off at the same value of $r_c$. Below we
show that our final results marginally
 depend on a particular
value of $r_c$ as long as $r_c$ is much larger than the turbulence
scales under study, i.e. $|{\bf K}|r_c \gg 1 $. Similarly, it is possible to
show that in the presence of two different cut-off scales $r_{1c}<r_{2c}$
the results do not depend on either scale provided that $|{\bf K}|r_{1c}
 \gg 1 $.

Observations indicate that the amplitude of random velocity is high
enough, so that $\lambda > r_c$ (see section~5). Thus we are 
predominantly interested in the short-wave
regime $ |{\bf K}|^{-1} < r_c < \lambda$.

For sufficiently small $r\ll r_c$ eq.~(\ref{cor-structure}) gives
\be
\xi(r)\approx \frac{1}{2} d(\infty) (1-[r/r_c]^{-\gamma})
\label{xinew}
\ee
Note that in eq.~(\ref{xinew}) $r_c$ plays the role of $r_0$ in
eq.~(\ref{eq:xi}). Both $r_c$ and $r_0$ are the scales at which
most of the energy of the field is stored.
It is straightforward to show that at short waves $|{\bf K}|r_c > 1 $
eq.~(\ref{xinew}) entails
\begin{equation}
P_2(|{\bf K}|) \approx \frac{1}{2} d(\infty) 
\left[ P_{2v} +  P_{2\rho} 
\right]~~~,
\label{eq:kolmolimit}
\end{equation}
where $P_{2\rho}$ coincides with the definition in section~4.1 if
a change $r_c\rightarrow r_0$ is made. $P_{2v}$ arises from the
unity term in the definition of correlation functions and
therefore is identical to that in section~4.1.
Eq.~(\ref{eq:kolmolimit}) is an analog of eq.~(\ref{2Dspec}) for
the case of a density field dominated by large-scale fluctuations.
The comparison between the two reveals a substantial difference between
the two regimes.
Note, that by construction $P_{2\rho} $ is the only part which carries the 
information about
the index of underlying density statistics\footnote{The amplitude of 
$P_{2\rho}$ for $n<-3$
is negative and this compensates for a minus sign in 
eq.(\ref{eq:kolmolimit}).}. 
The other term, namely, $P_{2v}$ arises purely from velocity.
Notice, that in this case the variation of the amplitude $d(\infty)$ 
of the density fluctuations
in $xyz$ space affects  $P_{2v}$ and $P_{2\rho}$ 
(see eq.~(\ref{eq:kolmolimit}))
in the same way.
Asymptotics of $P_{2v}$ and $P_{2\rho}$ are given by Table~1
provided that $r_c$ is used instead of $r_0$ in the expression
for $P_{2\rho}$.  

The feasibility
of  extracting the density statistics from the observable
emissivity spectrum 
depends on the thickness ${\cal L}$ of the data slice, which governs the 
relative magnitude
of $P_{2v}$ and $P_{2\rho}$ terms. In a {\it thin} slice,
i.e when $ ({\cal L}/\lambda) \cdot (|{\bf K}|\lambda)^{m/2} > 1$, 
a comparison between $P_{2v}$ and $P_{2\rho}$ is similar to that
 carried out in the
previous section. It shows that for $n<-3$  $P_{2v}$
always dominates if $|{\bf K}| r_c > 1$. 
In {\it thick} slice, eq.(\ref{eq:pvthick},  $P_{2v}$
has the same spectral index as $P_{2\rho}$ but its amplitude differs
by $(\lambda/{\cal L})^2$.  $P_{2\rho}$ dominates when
\be
{\cal L}/\lambda > (r_0/\lambda)^{m/2}
\label{very thick}
\ee
This condition does not qualitatively change when thermal line broadening 
is taken into account. In general, for ${\cal L}\gg \lambda$ the 
density spectrum should be observable.

It is clear that sufficiently weak velocity turbulence
cannot distort the density statistics. Our formulae
confirm this intuitive prediction. As the level of velocity 
turbulence decreases so does the scale $\lambda$ given by
eq.~(\ref{fbar}). As the result ${\cal L}$ for very weak turbulence
becomes much greater than $\lambda$ and therefore the density
statistics reveals itself. 

Thus we can summarize that when ${\cal L} < \lambda$
the velocity term dominates, producing universal spectral slope independent
of the underlying density slope $n$. This means that
for the {\it thick} slices the expected slope is $-3-m/2$ and for
{\it thin} slices the slope is $-3+m/2$. 
If the turbulent velocity as index $m=2/3$ ( the Kolmogorov index), 
the slope is equal to $-8/3$ if the slice
is {\it thin} and steepens to $-10/3$ for thicker slices. 

\begin{figure}[ht]
{\centering \leavevmode
\epsfxsize=3in\epsfbox{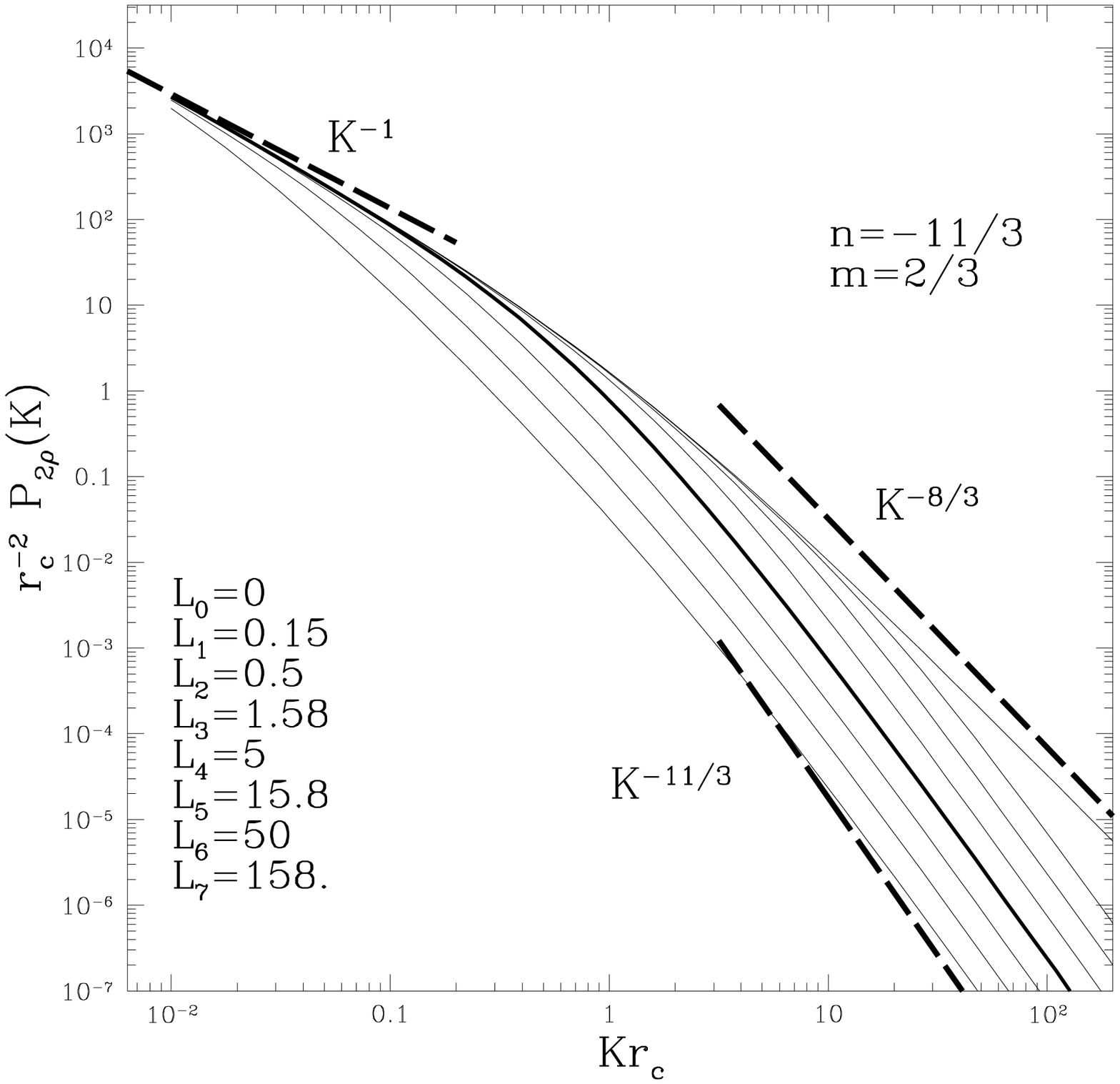}
\epsfxsize=3in\epsfbox{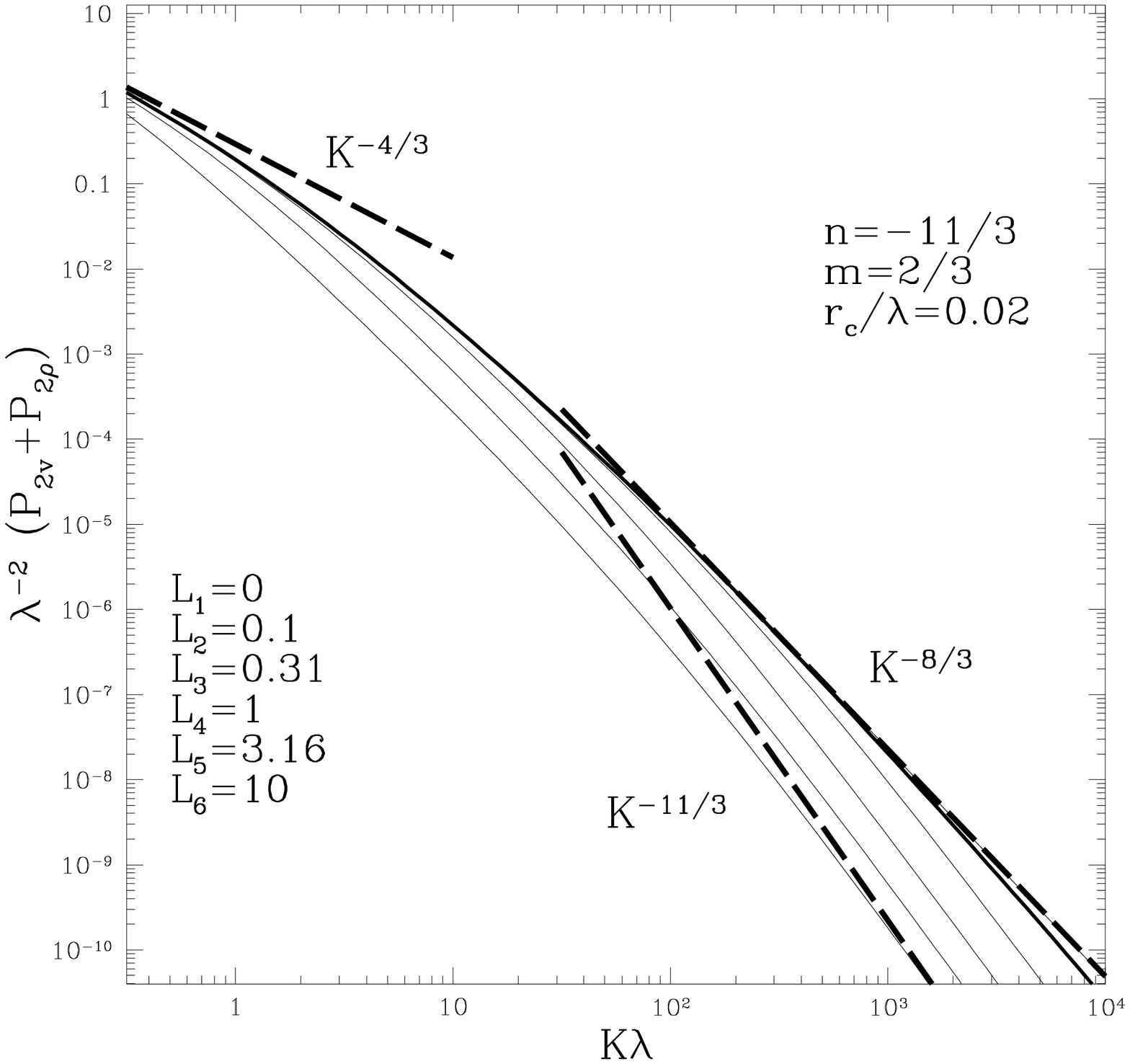} 
}
{\centering \leavevmode
\epsfxsize=3in\epsfbox{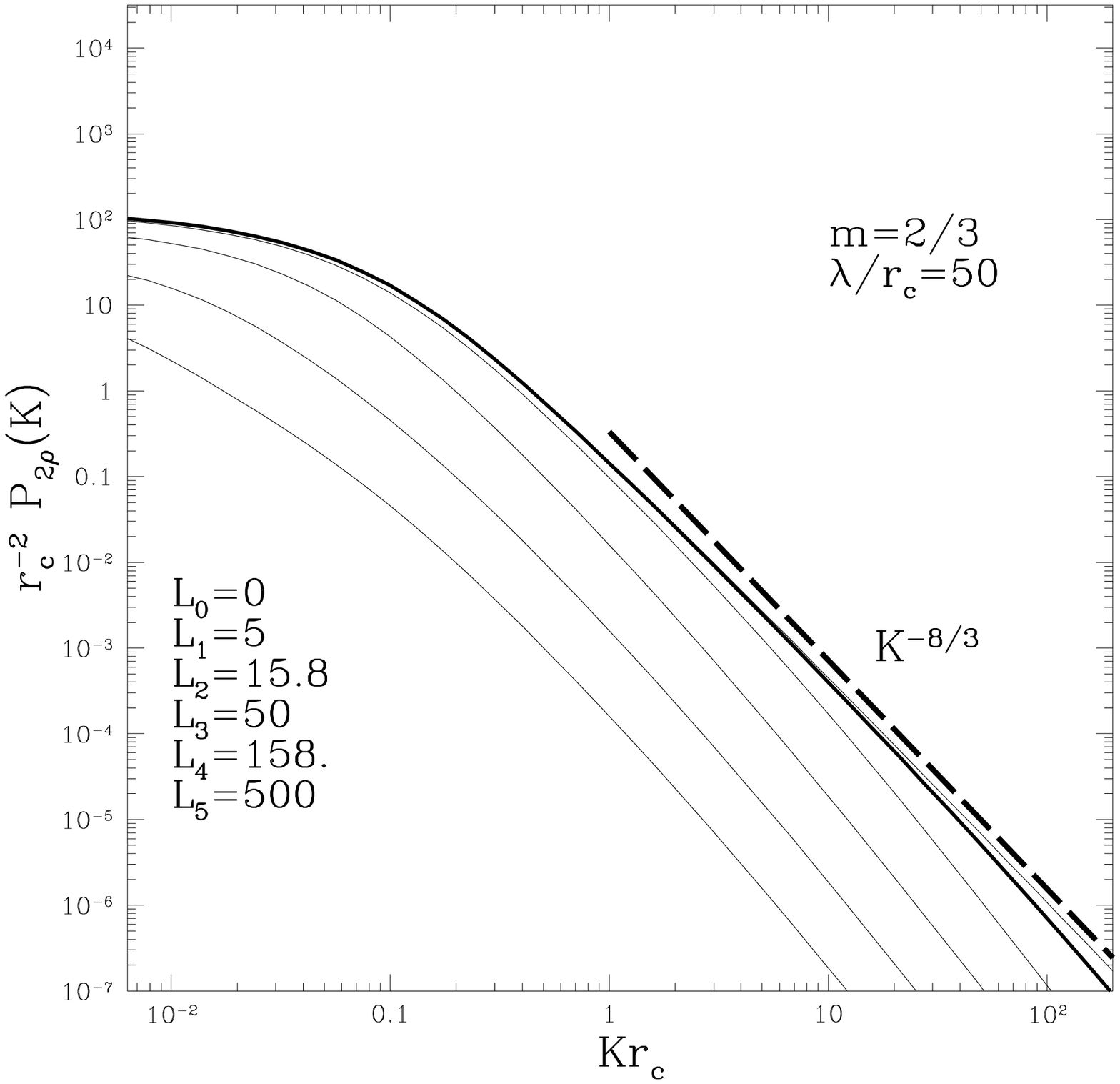}
\epsfxsize=3in\epsfbox{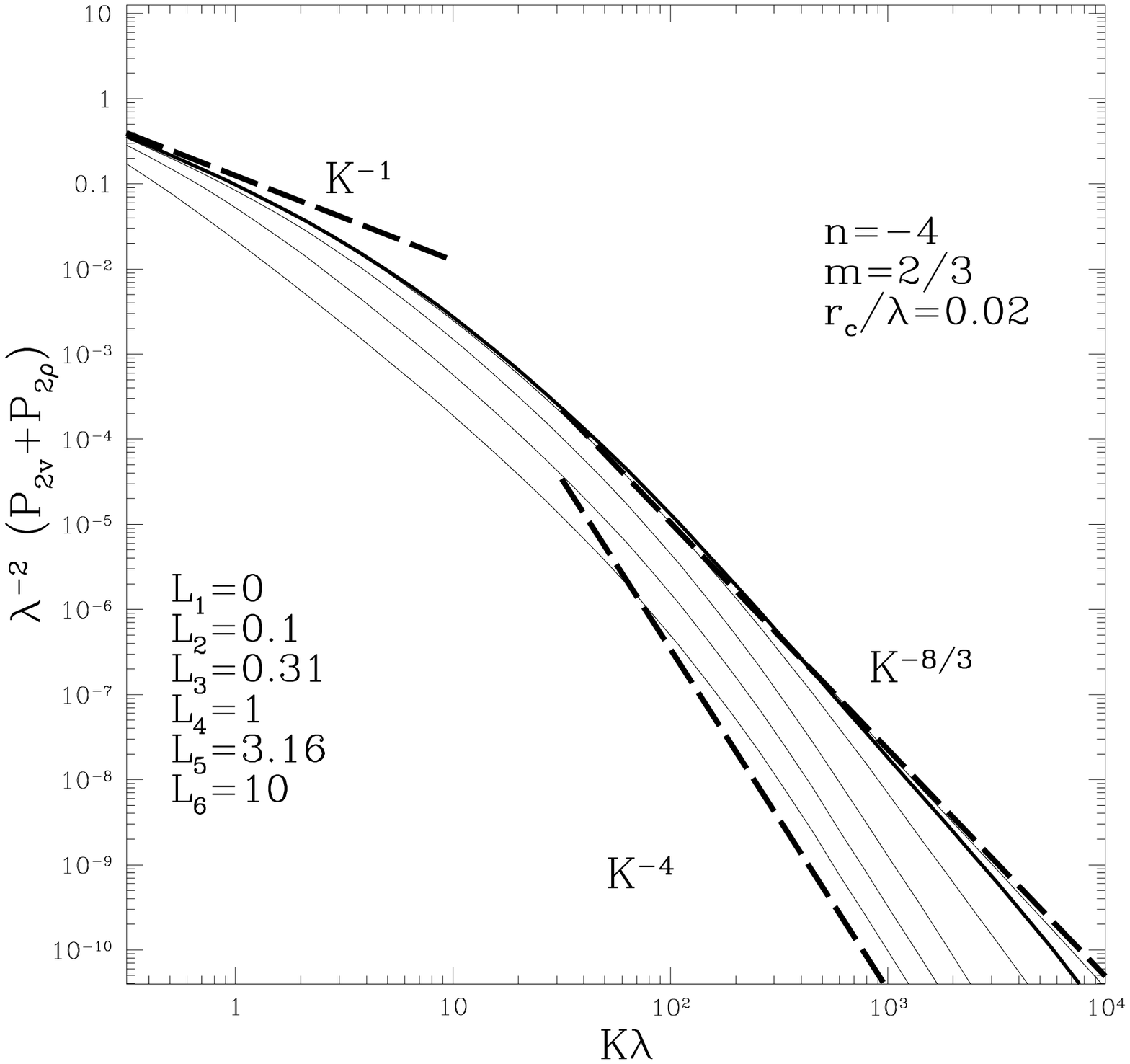} 
}
%\centerline{\epsfxsize=5.in\epsfbox{spect2Dn53m23.eps}}
\caption{Two dimensional power spectrum for ``Kolmogorov'' turbulence
(which does not mean more that the statistics of density and velocity
have the Kolmogorov indexes $m=-\gamma=2/3$) 
measured in a slice of variable thickness $L={\cal L}/\lambda$ (upper curves
correspond to smaller $L_i$).  
The ratio $\lambda/r_c = 50$ is fixed. 
Upper left panel shows $P_2(|{\bf K}|)$ when the velocity mapping
is absent. All quantities are expressed in units of $r_c$ for this panel.
The bottom left panel presents the density $P_{2 v}$ term 
arising from velocity fluctuations only.
The upper right panel shows $P_{2 v}+P_{2\rho}$
when both
velocity and density are ``Kolmogorov''.
The dashed lines show  power-law asymptotics described in detail in Section~4.
Heavy line corresponds to the slice thickness ${\cal L} = 0.1 \lambda$. 
The bottom right panel repeats the previous calculations  but for $n=-4$. 
This confirms that in the relatively {\it thin} slice ${\cal L} < \lambda $
 both $n=-11/3$ and $n=-4$ cases,
irrespectively of $n$ have slope and amplitude determined by velocity 
fluctuations,
as comparison with bottom left panel reveals. The slope is equal to 
$-8/3$, if the
slice is {\it thin} i.e. 
$({\cal L}/\lambda)| ({\bf K}|\lambda)^{m/2} < 1$, and 
tends to $-10/3$ for thicker slices.
Only slices much thicker than the velocity scale, i.e.
$ {\cal L} \gg \lambda $, 
reveal the underlying density spectrum.}
\label{fig:2Dspekkolm}
\end{figure}

The effect of velocity mapping on the density power spectrum is 
illustrated
in Figure \ref{fig:2Dspekkolm}. An additional asymptotics, namely
$K^{-4/3}$, arises from the evident expansion of eq.~(\ref{cor-structure})
for $r\gg r_c$. In this regime density fluctuations dominate those of
intensity.  

There may be an element of confusion since  
``Kolmogorov''\footnote{We emphasize that the name does not imply the
actual connection to the picture of the Kolmogorov energy cascade, but
only to the particular power index.} density 
fluctuations in the absence of velocity mapping will also produce 
$n=-8/3$
spectrum in a thin slice. However, as Figure~\ref{fig:2Dspekkolm} demonstrates, in 
the regime when random velocities are large, it is the velocity term 
that dominates for large $|{\bf K}|$. 
The fact that changes of underlying density 
spectrum do not alter the observed spectrum 
for $|{\bf K} r_c| \gg 1$ and ${\cal L}/r_c <1$
confirms our claim (see Figure~2).

Now it is time to address the question whether our results
depend on the cut-off size $r_c$. For this purpose we plot the
results of calculations for very different values of $r_c/\lambda$
in Figure 3. It is easy to see that only marginal dependence
of the spectral slope on $r_c$ exists for sufficiently large values 
of $|{\bf K}|$.

All in all, in the case of long-wave dominated density statistics both
thin and thick slices provide velocity spectral index.
To determine the density spectral index one should use
{\it very 
thick} slices with
${\cal L} > \lambda$. Only then the emissivity spectrum 
begins to reflect the index of the actual HI spectrum of density (i.e. the
density spectrum in Galactic coordinates). The regimes of long and
short-wave dominated spectra may be distinguished by varying the thickness
of velocity slices. 

\begin{figure}[h]
\centerline{\epsfxsize=5.in\epsfbox{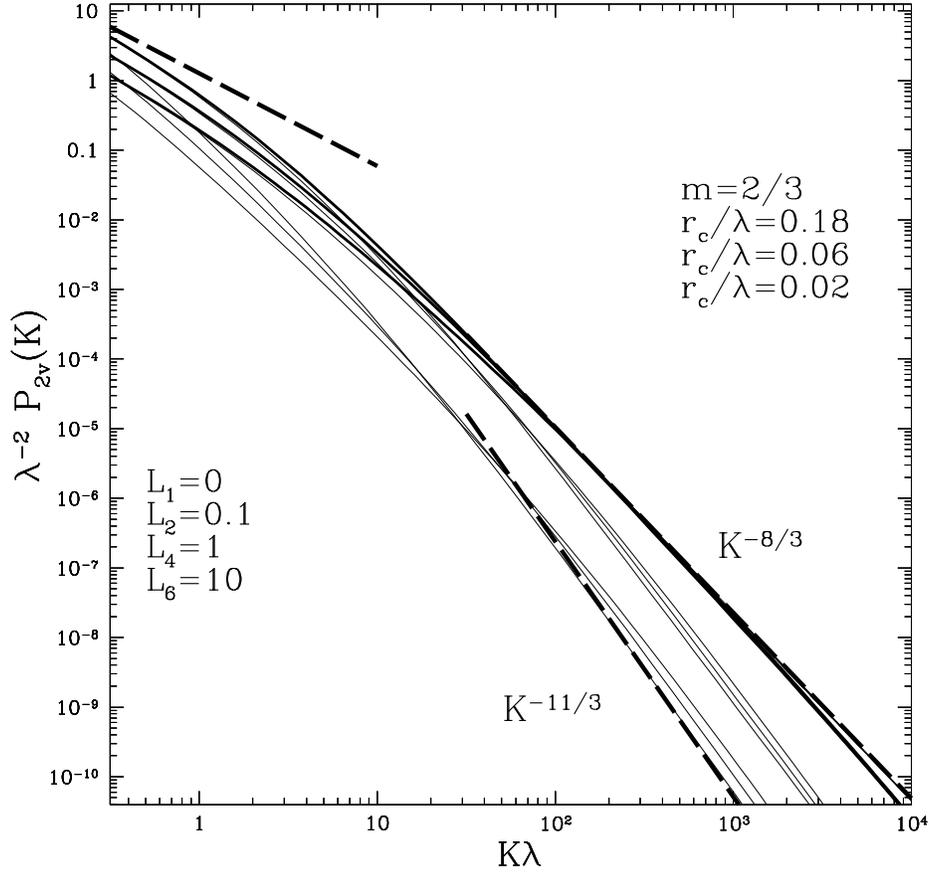}}
\caption{Dependence of the two dimensional power spectrum
on the structure function cutoff $r_c$. The results for four values of the 
slice
thickness $L={\cal L}/\lambda$ and three values of $r_c$ are shown.
In shortwave regime $|{\bf K}|r_c > 1$ 2D spectrum is completely insensitive to
$r_c$ for {\it thin} slice and exhibit only minor dependence on $r_c$
if slice is {\it thick}. The latter dependence reflects the fact that 
{\it thick}
asymptotics is achieved a bit later for smaller value of $r_c$.}
\label{fig:rc}
\end{figure}

\subsection{Warm and Cold HI}

When we deal with cold ($T \approx 100K$) HI,
the turbulent velocity is supersonic, while
for warm ($T\approx 6000$~K) HI the turbulent velocity is comparable
with the thermal velocity. Our studies above dealt with supersonic 
turbulence and the effect of the thermal velocities was shown to 
be equivalent to increasing the thickness of the slice from
$\sim \delta V$ to $\sim 2*(\delta v^2/6+2 v_T^2)^{1/2}$ (see section~3,
we use $W_e^{\prime\prime}=1/6$ assuming step-like experimental window).
As a consequence
there is no point of getting experimental velocity slices narrower than 
$\sqrt{12} v_{T,cold} \approx 2.6km/s $. For warm HI
the turbulence corresponds to the Mach number
of the order of unity and the minimal thickness of the slice is
$ \ga 17$~km/s. What are the observational consequences of this?

First of all, we may attempt to estimate the actual temperature of HI.
We calculated how spectra should change with the thickness of the 
velocity slice. When $\delta v\gg v_T$ the change of the $\delta v$
should entail the corresponding change in the effective slice thickness.
However, as $\delta v$ approaches $v_T$ a further increase of the instrument
velocity resolution (i.e. decreasing $\delta v$) will not result in the change of the
power slope. In the medium with a few HI phases having different temperatures
$T_i$ the transitions will happen for a number of $\delta_i V=3.5 v_{Ti}$.
Modulations of power-law spectral index as a function of
$\delta V$ should reflect  the distribution of temperatures
of HI components along the line of sight. We shall discuss the particular
technique of probing HI temperature distribution elsewhere.

For the generally accepted picture of
the mixture of warm and cold components 
the power spectrum can be presented as the sum of a three:
\be
P_2({\bf K})=P_{2~(cold)}({\bf K})+P_{2~(warm)}({\bf K})+P_{2~(warm-cold)}({\bf K})
\label{cold_warm}
\ee
where the part $P_{2~(warm-cold)}({\bf K})$ arises from correlations
between warm and cold components. Note that 
the minimal effective thickness of the slice for the warm
component is $\approx 20$~km/s. When $\delta v$ approaches this
thickness we expect to see variations of the spectral index if the
warm gas constitutes a substantial fraction of the total HI. As
mentioned above, this potentially
may be used to determine the relative abundance of the components. 

Assume, for the sake of simplicity, that the mass fraction of the
cold and warm media are comparable and the same is true for
the amplitudes of the
three dimensional spectra $P_{s({\rm warm})}$ and $P_{s({\rm cold})}$.
When the experimental slice width $\delta V$ is smaller than 
$\sqrt{12}v_{T(warm)}$
the the ratio of cold to warm contributions 
$P_{s({\rm warm})}/P_{s({\rm cold})}$
as a function of wavenumber ${\bf K}$ will be constant 
of order unity at long wavelength
${\bf K} \ll (C^2/3.5 v_{T(warm)})^{1/m}$ (see criterion 
(\ref{eq:phystrans})), then fall off as ${\rm K}^{-2/m}$
until  ${\bf K} \approx (C^2/\delta V)^{1/m}$ and be constant again 
but suppressed as $\sim \delta V/3.5 v_{T(warm)}$  at shorter wavelength (if
$\delta V< 3.5 v_{T(cold)}$), until 
the maximum suppression $\sim v_{T(cold)}/v_{T(warm)}$ is reached)
\footnote{Our earlier statement of the possible change of the spectral
index can be understood as through this suppression. Indeed, for
$\delta V/3.5 v_{T(warm)}\ll 1$ the contribution of the warm media is
negligible, while it increases with $\delta V$.}.
These three regimes correspond to wavelengths for which effective 
slice is, first,
{\it thin} for both cold and warm component, then still {\it thin} 
for cold, but already
{\it thick} for the warm one and, finally, {\it thick} for both contributions.
Addressing the actual Galactic HI data in
Fig.~4 we may see that the slicing is frequently thin for cold 
HI and thick for warm HI.
As $\delta V$ increases above
thermal velocity of warm component, shortwave relative suppression 
of contribution 
from warm HI disappears.

In reality   $P_{s({\rm warm})}$ and $P_{s({\rm cold})}$ may have different
amplitudes and even different spectral slopes. Moreover, the mass fraction
in those component may vary. The variations $P_2$ with the slice thickness
may allow to determine the characteristics of the turbulence in the multiphase
media. 

The $P_{2~(warm-cold)}$ contribution arises from the correlations between 
fluctuations in warm and cold media. The density fluctuations are spatially
separated in two media and therefore their correlation is likely to be 
negligible. We may
not be so sure about the velocity fluctuations as cold and warm gas may
participate in a coherent motions. It may be shown, however, that the 
difference
between $v_{T(warm)}$ and $v_{T(cold)}$ will suppress a possible contribution
from $P_{2~(warm-cold)}$ in a manner similar to the suppression of 
$P_{2~(warm)}$.

\section{Observed Spectra}

{\bf Revisiting L95}\\
In L95 the relations between 2D statistics available via observations
and the underlying 3D 
statistics were obtained on the assumption that the observed statistics
is determined by density fluctuations. Below we use our
results on velocity mapping to interpret power spectra of intensity obtained
by Green (1993) in terms of HI density and velocity fluctuations.  

Green's  observations
of the HI emission were accomplished with the Synthesis
Telescope of the Dominion Radio Astrophysical Observatory (DRAO)
towards $l= 140^{\circ}, b=0^{\circ}$ ($03^{h} 03^{m} 23^{s},
+58^{\circ} 06' 20'$, epoch 1950.0) and they revealed a power law spectrum of
2D intensity. This spectrum is proportional to $P_2({\bf K})|_{\cal L}$
and its interpretation depends on whether the slicing is thick or thin.

To answer this question we should estimate both $\lambda$ and $|{\bf K}|$.
To estimate $\lambda$ we use a crude estimate of $f$, namely,
we describe ordered motion of atomic hydrogen in the outer parts of the Galaxy
by a simple flat rotation curve without any distorting radial motions.
In this case 
the parameter $f$ which describes line-of-sight projection of the shear in
a rotational flow is
\be
f\approx - \frac{1}{2 A}
\frac{\sin l}{(\alpha + \sin l)^2\sqrt{1-(\alpha + \sin l)^2}}
\label{f_approx}
\ee
where $A=14~$km/s/kpc is the Oort's constant and $\alpha = V_z/V_0$, 
where $V_0\approx 220$~km s$^{-1}$
 and $V_z$ is the relative velocity of
HI parcel. This relative velocity varies from 0 at $z=0$ to $-V_0 \sin(l)$
for the gas parcel at $z=\infty$. Correspondingly, the parameter $f$,
which is negative, grows in magnitude from the familiar $ 1/A \sin(2l)$
in the Sun vicinity  to larger negative values for distant HI regions.
In particular, Green's data covers the velocities
$V_z$ up to $\sim -100$ km/s, thus $f$ ranges from $-72.5$ pc/(km/s) 
to $-460$ pc/(km/s). 

If we assume that velocity variations at the scale $30$~pc amount
to $10$~km/s and arise from the Kolmogorov turbulence, the structure
functions of velocity are
\be
D_{LL}(r)\approx 100
\left(\frac{r}{30 {\rm pc}}\right)^{2/3}~{\rm km}^2{\rm s}^{-2}~~~,
\ee
where the cut-off of the function at large scales is disregarded
on the assumption
that turbulence is being studied at the scales much less than $r_c$.
Thus $\lambda$ given by eq.~(\ref{fbar}) is $\approx 3.5$~kpc for
the closest slice and $\approx 56$~kpc for the most distant slice.

The width of the interferometer channels combined to give
a single data point in Green's dataset is $\delta V =5.94~\mathrm{km/s}$.
The slice thickness in parsec is ${\cal L}\approx \delta V f$~pc, 
and varies from
$\approx 600$~pc for the closest slices to  $\approx 2200$~pc for the 
distant ones\footnote{Note, that the cut-off due to thermal velocity (see
section~2.1) in Warm Neutral Medium (see table of idealized phases
in Draine \& Lazarian (1999) ) is $\sim 6$~km/s. If the WNM constitutes
the dominant fraction of the neutral phase (Dickey 1995) then the
velocity resolution above is optimal and no further decrease in $\delta V$
will result in getting new information. However, if close to Galactic
plane Cold Neutral Media constitutes a substantial portion of mass, the
increase of velocity resolution up to 1~km/s is desirable.}. The wavenumber of transition from {\it thin} to {\it think}
slice given by eq.~(\ref{eq:phystrans}) is equal $0.16\,pc^{-1}$. 

\begin{figure}[ht]
\centerline{\epsfxsize=3.in\epsfbox{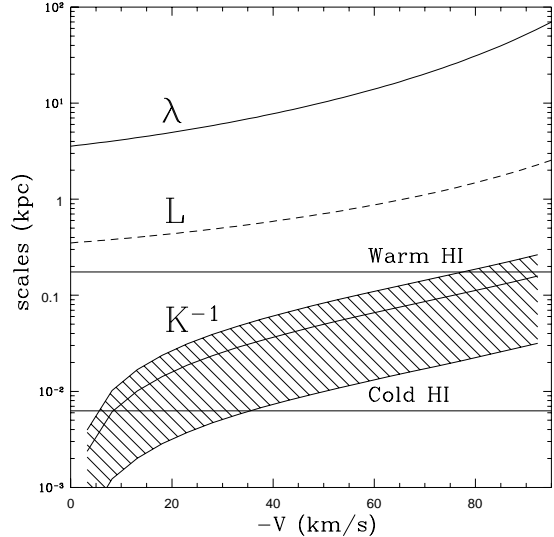}}
\caption{The variations of geometric scales with the sampling velocity are
shown. The upper curve corresponds to the variations of the correlation
scale $\lambda$ 
in the velocity space. The middle curve corresponds to the variations
of the slice thickness $\cal L$. The darkened area in the Figure depicts
to the range of the turbulence scales under study in Green (1993). 
The solid curve within the darkened area corresponds to the interferometric
measurements with the baseline 21~m. The lower horizontal line denotes value of
$|{\bf K}|^{-1}$ which separates {\it thin} (above) and {\it thick} slice
regimes for cold HI and the upper line denotes a similar value for the warm HI.}
\label{fig:flat_rot}
\end{figure}

Figure~\ref{fig:flat_rot}
shows the range of the scales involved (see Green (1993)). 
The smallest 
$|\bf K|$ span from $\sim 1/3$~pc$^{-1}$ for the closest slices
to $1/200$~pc$^{-1}$ for the distant ones\footnote{We stress that our analysis
should be modified 
for the largest scales which are larger that $r_c$.}, and therefore
$|{\bf K}| \lambda \gg 1$. The range of values corresponding
to $|{\bf K}|^{-1}$ is shown by the darkened region in the plot.
If we deal with cold HI, most of them, except the nearest slices 
correspond to the regime
where slice is {\it thin}. For warm HI the border line is different and
most of the scales correspond to {\it very thick} slice. 

We have shown earlier that the contribution of the warm component is
suppressed unless the slices are thin for both components. The latter
regime corresponds to the parameter space below the line ``cold HI'' in
Fig.~4. For such close slices the actual geometry of the diverging lines
of sight must be accounted for (see Lazarian 1994b)
and we consider the resulting complications
elsewhere. If the intensity contributions of the warm and cold HI are
comparable, it is possible to show (see section 4.3) that the cold HI
contribution will dominate the spectra above the ``cold HI'' line. 

Our analysis of Figure~\ref{fig:flat_rot} and Figures 1 and 2
shows that {\it if the spectrum of density
is shallow (i.e. $n>-3$)}, the observations by Green (1993) reveal 
the spectrum with index $n+m/2$. For $m=2/3$ the spectrum of emissivity
obtained by Green (1993), namely the emissivity with the index $\sim -2.7$,
corresponds to $n \sim -3$.
This conclusions is not  sensitive to the values
of density correlation scale $r_0$ (see Eq.~(\ref{eq:xi})), which
we estimated to be of order of $20$~pc.

{\it If, however, the density spectrum is steep (i.e. $n<-3$)}, the
fluctuations of 21~cm intensity observed by Green (1993) can arise
from velocity fluctuations. In this case the spectral index is
$-3+m/2$. For $m=2/3$ one gets the slope $-8/3\approx -2.7$ which
is exactly what is observed. 

\begin{figure}[ht]
\centerline{\epsfxsize=3.in\epsfbox{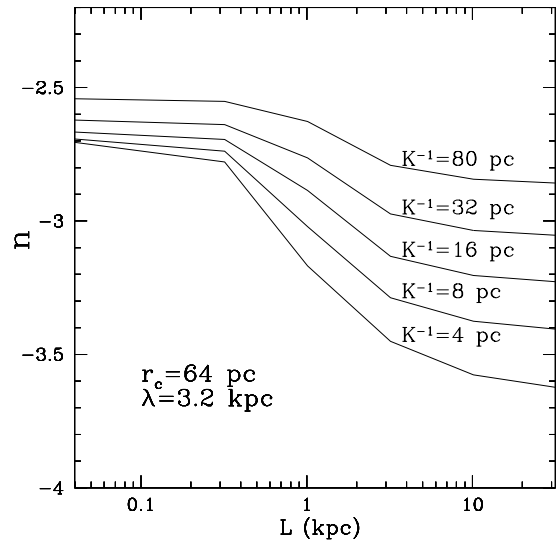}}
\caption{The variations of the spectral index $n$ of the 21~cm 
intensity spectrum
with the thickness of the slice ${\cal L}$ for the turbulence with Kolmogorov
velocity spectrum and long-wave dominated density spectrum. To 
provide the feelings of the scales involved we expressed our
results in parsecs.}
\label{change}
\end{figure}

In the case of the Galaxy, the steepening of the spectrum depends on
the scale $|{\bf K}|^{-1}$ as shown in Figure~5. This figure should
be not interpreted literally for large ${\cal L}$, however. First
of all, when the thickness of the region becomes comparable to the
distance to it, one should account for the divergence of lines of
sight (Lazarian 1994b). Moreover, our model that assumes infinitely
extended media does not hold and effects of finite size of the
emitting region become important when ${\cal L}$ becomes comparable with
the size of the Galactic disk (see Appendix~E).

{\bf SMC}\\
Small Magelanic Cloud (SMC) is rather irregular galaxy and the Earth observer
looks at it from outside. For its study the modification of our technique
described in the Appendix~E is applicable. As the result, the size
of the SMC plays the role of the parameter $\lambda$ and with this
modification our earlier formulae are applicable.

An advantage of using SMC is that the effects related to the divergence
of lines of sight are negligible for the study and this simplifies the
analysis for thick slices. Data on SMC in  Stanimirovic et al. (1999),
Stanimirovic et al. (2000) and Stanimirovic (private communication)
shows that for thin slices the slope
is $\approx -2.8$ but it steepens up to $\approx -3.4$ as the integration
over the whole emission line is performed. Note, that in the latter
case the intensity fluctuations arise from density fluctuations only 
 and the spectral index of $P_2$ corresponds to the 3D spectral
index of underlying density turbulence. This means that the density
spectrum has an index $-3.4$, which is close to the Kolmogorov $-11/3$
index. As this index is less than $-3$, the fluctuations 
should be ``velocity-dominant'' and the corresponding index 
should be $-3+m/2$, provided that the slice is ``thin''. 
The measured spectrum is a bit steeper than in the Galaxy. Is this
difference meaningful? Does it mean that the spectrum of
turbulence deviates from the Kolmogorov law? Is this a reflection of
a transition to thicker slicing?
A detailed study of the SMC data should answer these sort of questions.
At the current stage it looks to us that there exist a rough
agreement between the data and the assumption that both density and
velocity spectra have the Kolmogorov index of $-11/3$.

An additional interest to SMC data stems from the fact that variations of
the power law may shed light on the nature of turbulence in warm HI.
We discussed above that the fluctuations in sufficiently thin slices
are dominated by the cold component. As the thickness of slice increases
the relative contribution of the warm component increases. The fact
that the velocity-integrated data provides the index close to the 
Kolmogorov one either indicate that the fluctuations of warm HI density
are insubstantial or that they have the Kolmogorov index.

{\bf Dust and Molecular Data}\\
Dust and HI are well correlated (see Boularger et al. 1996). 
Both Galactic and SMC 
Far Infrared emission exhibits
shallow spectrum  with the index $-3$ (Wall \& Waller 1998, Waller et al 1998,
Stanimirovic et al. 2000). However Stanimirovic et al. 2000 combined
data at 100~$\mu$m and $60$~$\mu$m to show that the actual dust spectrum 
of dust column density in steeper. In fact
the spectrum of dust column density was found has index $\approx -3.6$
which is in  rough correspondence with the $\approx -3.4-- 3.5$ spectral index
for HI (see above). Thus both HI and dust supports the idea that
the spectrum of density is steep for diffuse media.

Are there any evidence that that the interstellar density
may be shallow? In the theory domain
Henriksen (1994) derived a quite shallow spectrum for compressible 
turbulence, but we do not know whether this result is applicable to HI.
Although molecular cloud statistics may be different from that of
diffuse HI, comparison of the two is worth doing.
 
Recently molecular cloud data was discussed in Stutzki et al. (1998). 
There  for both
 $^{12}$CO (data from Heithausen \& Thaddeus (1990) and
Falgarone et al (1998))
and $^{13}$CO (data from Heyer et al. (1997))
transitions the spectrum of intensity was observed
to have a power law index $\sim -2.8$. As the data is averaged over
velocity, the fluctuations of intensity are due to density fluctuations
and the spectrum of density should have the same
slope as the intensity (i.e. $n=2.8$, (see Eq.~(\ref{eq:thick})), provided
that the transitions are optically {\em thin}. 
The trouble is that the lines are optically {\em thick} and the interpretation
above is not applicable. Further research should determine what the
particular emissivity intex means in terms of underlying statistics
of velocity\footnote{Although the data is integrated over the emission
line, the random velocity is important as
self-adsorption depends on the gas velocity dispersion.}
and density. 

Strangely enough, the shallow
spectrum of molecular density fits well in the picture of
molecular cloud structure. Stutzki et al 1998 
derive the relation \footnote{This relation
can be applied to individual structures in velocity slices provided that
the index of the two dimensional spectrum of intensity fluctuations is used
instead of $n$. In the velocity dominated case the masses would reflect not
the masses of the real clumps but caustics produced via projection.} between 
the power spectral
index $n$, the spectral index $\varepsilon$ of clump 
mass-size relation   $M\propto r^{\varepsilon}$,
and the index  $\aleph$ of the clump mass spectrum $dN/dM\propto M^{\aleph}$:
$$n=(3-\aleph)\varepsilon$$
$\varepsilon=2$ corresponds to the Larson's relations (Larson 1992) and $\aleph$
in the range 1.6 to 1.8 corresponds to the 
analysis of clumps (Kramer et al. 1998). The trouble is that this relation
was derived to be true for optically thin clouds, while the clouds
in the analysis by Stutzki et al 1998 are optically thin makes one
worried. Further research should concentrate on providing the adequate
statistical description for optically thick tracers and comparing the
statistics for thin and thick tracers.

\section{Discussion}

\subsection{Our approach}

{\bf Velocity mapping}
In the paper thus far we address the problem of the velocity modification
of HI statistics. Although the paper deals with HI only, the
problem it addresses is quite general and therefore our results are applicable
to any optically thin lines. We plan to address the effects of finite optical
depth elsewhere.

We showed that intensity fluctuations can arise both from velocity and
density fluctuations and derived the 3D spectrum of intensity 
(eq.~(\ref{eq:main})) within velocity data cubes. A quick look
at eq.~(\ref{eq:main}) shows that the
density and velocity enter the expression in a different manner and this
eventually enables us to separate the velocity and density contributions.

{\bf Statistics used: 2D Spectra}\\
Channel maps and the related 2D spectra constitute the central topic
of this study. This is because the channel maps are readily available
via single dish observations, while channel 
map spectra $P_2({\bf K})$ can be directly obtained
with an interferometer. 
 
As we discussed in the previous section it is important to determine
whether 21~cm
intensity fluctuations are due to density or velocity fluctuations.
If the density spectrum is long-wave dominated, our results
in Figure~2 indicate that HI intensity spectrum does not reflect
the underlying HI density statistics. Therefore the spectrum of the density
fluctuations may have any value $n<-3$, but this
will not alter the intensity fluctuations unless a substantial portion of 
the 21~cm
line is integrated over and the slice become ``thick'' in accordance
with the criterion (\ref{eq:phystrans}). 
 
{\bf Additional statistics: 1D and 3D Spectra}\\
In view of the problems we face with the interpretation of present
HI results it is important to make use of other statistical tools. Indeed,
2D statistics of channel maps is not the only statistics that is
available. Using data cubes one can obtain the 3D spectrum $P_s$
which properties are described in the Appendix~B.
 Alternatively, an additional insight can be obtained if 1D spectra are
formed using velocity cubes. We discuss these new
statistical tools in Appendix~F and show that 1D spectra is complementary
to the tools we employed in the main text. In fact, using this
statistics one does not need to change the velocity slice thickness to
distinguish between long and short-wave dominated density spectra.
If the density field is long-wave dominated we expect
to see the power index $-2/m$, while if the density spectrum
is short-wave dominated we expect to get the index $2(n+2)/m $.
This test can be performed with the existing 21-cm data cubes.
For $n=-3$ the predictions for the two cases coincide.
Naturally the spectra are not independent, but they are produced
using different procedures of data handling and therefore provide
an additional test.

\subsection{Problems of Interpretation}

{\bf ``Big Power Law''}\\
The application of our technique to the available data (see section~5)
testifies that the density spectrum of HI in SMC
has a power-law index close to the Kolmogorov  value of $-11/3$
over the scales from $40$~pc to $4$~kpc. Assuming that density statistics
in our Galaxy is steep (as in SMC) we also get that the velocity 
has an index of $\sim -11/3$ from $10$ to $100$~pc.  
The proximity of indexes for both velocity and density spectra to the
Kolmogorov values is both remarkable and unexpected. First of all,
there are good reasons why we should not {\it a priori} expect to 
find Kolmogorov-type
turbulence in interstellar medium (Lazarian 1999a). Moreover, Kolmogorov
interpretation would entail problems with understanding of the nature of
energy injection. We also noted earlier that SMC possibly shows a density
spectrum with only marginal velocity contribution, which would mean
that the structures we observed are static. The last point requires a
further investigation, however.

It is very strange, but Astrophysical turbulence whenever it is measured
seem to exhibit Kolmogorov power slope.
The existence of the ``Big Power Law'' (see  Armstrong 
et al 1995, Spangler 1999) is
one of the great Astrophysical mysteries. Mesurements of turbulence
in ionized media established a spectrum  with $-11/3$
index at the scales from a few AU to a fraction of a parsec. 
Measurements at larger scales are much less reliable (Cordes 1999)
and although compatible with the $-11/3$ index allow other interpretations.
Contrary to our intentions (see L95, Lazarian 1999a) we contributed
to making the problem
even less tolerable. It looks now that the ``Big Power Law''
spreads up to scales of several kpc.

{\bf MHD turbulence}\\
Ignoring for the moment the issue of energy injection scale and associated
problems we may ask whether we expect to see the turbulence with the
index close to the Kolmogorov one when the fluid is strongly magnetized.

The Kolmogorov-type spectrum of plasma density fluctuations
observed via radio scintillations and scattering (see Armstrong 
et al 1995 and references therein) has been interpreted recently 
as the consequence of
a new type of MHD cascade by Goldreich \& Sridhar (1995). 
The Goldreich-Shridhar model of turbulence\footnote{A qualitative
discussion of the model and the role of reconnection for the
cascade can be found in Lazarian \& Vishniac (1999).} differs considerably
from the Kraichnan one (Iroshnikov 1963, Kraichnan 1964).
It accounts for the fact that hydrodynamic motions can easily mix up
magnetic field lines in the plane perpendicular to the direction of
the mean field. Such motions provide eddies elongated along the 
field direction and the velocity spectrum close to the Kolmogorov one.

The Goldreich-Shridhar turbulence is anisotropic with eddies 
stretched along magnetic
field direction. 
The wavevector component parallel to magnetic field $k_{\|}$
scales as $k_{\bot}^{2/3}$, where $k_{\bot}$ is a wavevector component 
perpendicular
to the magnetic field. Thus the degree of anisotropy increases with the
decrease of the scale.  The fact that during the observations various 
scales contribute
to the correlation functions of intensity 
(see discussion in L95) and the contribution of the
largest scales dominates the signal mitigate the anisotropy of the 
intensity statistics.
The component of magnetic field parallel to the line of sight
decreases anisotropy as well. 
Indeed, it is easy to see that only the component of magnetic field that 
is in the
plane of the sky will create the anisotropy, while the component of 
magnetic field
parallel to the line of sight ``mix up'' parallel and perpendicular 
direction\footnote{No
anisotropy is expected for the intensity correlations if the total 
magnetic field
is parallel to the line of sight.}. 
Therefore the anisotropy of the 21-cm statistics may
not be easy observable (see Green 1994). 
In our future papers we plan to discuss the observational
tests of the Goldreich-Shridhar picture for HI. A simultaneous
use of other techniques of turbulence study (see Lazarian 1992,
Lazarian 1993) should allow to have a more comprehensive picture
of MHD turbulence. 

Incompressibility is assumed within the Goldreich-Shridhar picture
of turbulence. In actual ISM turbulence the compressibility may
be important in view of the fact that the media may be pliable to
compressive forces (soft equation of state). It is unclear how
the picture of MHD cascade will be changed in this case. 

When magnetic field is intermittent, one may expect the thermal instability
to be more efficient in the regions with a smaller value of magnetic field.
Indeed, if varios regions of magnetized gas are at pressure equilibrium,
the gas pressure within the regions with lower magnetic field will
be higher. This shall increase the density of gas and may initiate thermal
instability. Indeed, the heating of gas is proportional to the gas density,
while cooling is proportional to the (density)$^2$. As the result dense
gas will tend to cool down further increasing its density (Spitzer, 1978)
making regions with lower magnetic field colder and denser. This would
qualitatively correspond to the picture of magnetic measurements in HI
by Carl Heiles (private communication). We will quantify the picture
and its implications on the small scale structure of HI elsewhere.

{\bf Superbubbles and Supershells}\\
One may speculate that the observed statistics is the result of 
the superbubbles and supershells. Indeed, potentially superbubbles
may inject energy on the kpc scale. The issue in this case is why we
do not see ``knees'' in the spectra at the 
intermediate scales when singular shells inject energy. Does this mean
that the supershells should carry an overwhelming amount of energy or the
available resolution is not sufficient to see peculiarities related to more
localized energy injection?

In general, energy injection scale is critical for understanding interstellar
turbulence. For decades it was generally believed that MHD turbulence damps
slowly (Iroshnikov 1963, Kraichnan 1964). In this situation supersonically
broadened linewidths of molecular clouds
could be explained as locally generated, e.g. due
to a cloud collapse. However, new understanding of MHD turbulence
(Goldreich \& Shridhar 1995) and numerical simulations 
(e.g. Vazquez-Semadeni 1999)
suggest that the turbulent damping in magnetized media takes place on the
scale of an eddy\footnote{It is shown in Lazarian \& Vishniac (1999) that
MHD turbulence allows equivalent descriptions via wave-wave, wave-eddy and
eddy-eddy interactions.}  turnover (similar to the case of unmagnetized media).
Therefore to allow starless cores to have linewidths similar to those
of star-forming cores (Benson \& Myers
1989)
the energy injection should happen on the scale much
larger than the cloud size (for larger eddies the timescale of turnover is
larger).
Nevertheless, scales of the order of several kpc look excessively large.

Energy injection to molecular clouds may arise 
directly from shocks
(Scalo \& Kornreich 1999),
and this also can be classified as injection from large scales. The
spectrum of shocks is steep ($\sim k^{-4}$) and does not correspond
to the observations.

One have to accept that power spectra may be rather rough statistical
tools and the index $-11/3$ may have nothing to do with a cascade.
For instance, such a density spectrum may emerge in a media with HI shells
having a distribution $P(x)\sim x^{8/3}dx$. However, the particular 
power law does
not seem to have particular physical explanation and moreover, small
shells are expected to have larger velocities which does not correspond
to the observed velocity spectrum.

\subsection{Model assumptions and future work}

This work is the first, as far as we know, that treats quantitatively
fluctuations arising from turbulent velocities. It is not surprising
therefore that there are many issues to be studied in depth in the future.

{\bf Density-Velocity Correlations}\\
Our derivations above assumed that the turbulent velocity
and density are not correlated in Galactic coordinates. This seems to
be true at large scales, but may fail on small scales where self
gravity may contract HI structures\footnote{Whether this important in
practical terms of HI studies is unclear, as a transition to higher
densities entails a transition to molecular hydrogen.}. Such a correlation may
enhance the amplitude of fluctuations observed if the local
divergence of the velocity field is negative in the regions of density
enhancement (i.e there is systematic inflow of matter into overdense
regions) and, {\it vice versa}, would decrease the amplitude 
of fluctuations if overdense regions are 
associated with outflows. However, interstellar medium is far
from being quiescent even on the small scales and the sense of velocity-density
correlation is unclear for the ISM turbulence. It is easy to see
from the equation of continuity that the logarithmic time derivative
of density equal to the divergence of the velocity field. This
means, that the instantaneous density itself depends on the integral of
the velocity divergence and in the complex medium with sequential expansions
and collapses the importance of this correlation is difficult to estimate.
In any case, we do not expect {\it a priori} any strong velocity-density
correlations.

To check whether the  assumption about density-velocity correlation
is important for our final results
one needs either to do numerical simulations or to assume the
statistics for joint density-velocity distribution. In Appendix~D
we check our results assuming Lognormal distribution
for random density\footnote{Lognormal distributions
naturally arise in numerical simulations of isothermal gas
(see Vazquez-Semadeni \& Passot 1999).} and the velocity and
logarithm of density are correlated Gaussian fields.

To estimate the upper limit of the effect of density-velocity correlation 
we assume the maximal possible value of this
 correlation. Naturally,
in all realistic situations we expect the value of this
correlation to be much smaller than this maximal value.
 Our results for this worst possible regime
can be summarized as follows. The criterion for
the slice to be thick is increased at most by a factor of a few and
all our findings for the thick\footnote{This includes both long-wave and
short-wave density field.} slices (see Table~1) hold. The correspondence
between our earlier results and ``extremely correlated'' results
holds for the thin slices in the long-wave regime, which is
favored by our analysis of data. For thin slices
and short-wave dominated density field (shallow density spectrum) 
the correlations
are shown to enhance the velocity effects. In fact, in the extreme
regime of correlations it is velocity, rather than density
that dominates the expected channel emissivity. However, we stress that
we used an upper limit for these correlations and 
for interstellar data sets we expect to observe density effects.
Changes of the spectrum within the thin slice regime when the density is
short-wave dominated can be a signature
of strong velocity-density correlations.
Using the thick slices it is possible to check whether the density
is short-wave or long-wave dominated. A more detailed study of 
the regime of velocity-density correlations can be done using
our expressions in Appendix~E and substituting a different law of
velocity-density correlations (rather than taking its maximal value),
but this study is beyond the scope of the
present paper.

{\bf Anisotropy and Linear Expansion}\\
It is both  challenging and important to determine
the degree of the HI spectrum anisotropy and its variation
from slice to slice. This information can provide an insight
to the nature of HI turbulence and may be used as a diagnostic for the
interstellar magnetic field. For instance, measuring the structure functions
of HI intensity as a function of a positional angle in an analogy with
the statistical treatment of synchrotron fluctuations (see Lazarian 1992)
may reveal magnetic field direction in various portions of the sky, provided
that the turbulence is indeed Goldreich-Shridhar type. So far, the
attempts to measure anisotropy in HI are limited to the Green (1994) study,
where no anisotropy was detected. Apparently a better analysis is
needed. For the slices with high degree of
anisotropy our statistical technique should be modified as suggested in L95.

Another issue is related to the linear expansion (\ref{eq:map}). If
the slice is sufficiently thick the map becomes non-linear. 
Potentially, knowing a detailed Galactic rotation curve one
can describe a non-linear mapping. We found, however, that the linear
mapping is OK for both thin and thick regime if we study sufficiently
small scales (see discussion after eq.~(1)). 
We also found that if ${\cal L}$ is smaller than the distance to the
slice, it is possible to ignore the complication related to the
radial nature of lines of sight. Thus the theory at its present stage
is widely applicable.

{\bf Applicability of the theory}\\
The theory presented in the paper is quite general. The mathematical
machinery developed here can be used for studying turbulence in
various emission/adsorption lines.
For instance, velocity effects and line of sight integration are the problems 
that are also encountered in  studies of turbulence inside 
molecular cloud. Unfortunately,
there are additional complications related to studying molecular clouds.
For instance, these studies face problems with samples being not 
statistically homogeneous (see Miesch \& Bally 1994).
Indeed, it is widely believed that for individual molecular complexes 
the statistics (especially at large separations)
is dominated by regular gradients rather than the random 
component. To eliminate the inhomogeneous component Zurfleh (1967)
or other types of spatial filtering (see Spicker \&
Feitzinger 1988a,b) can be used. However, Stutzki et al (1998) have shown that the
statistics can be obtained without such filtering if the particular
modification of wavelet analysis, namely, so-called Alan variances
are used. For power-law statistics
Stutzki et al. (1998) relate their statistical measures
to the spectrum and correlation functions that are used in the present
study. Therefore the translation of our results to the language of Alan-variables
is trivial. 

What is less trivial is to account for the finite 
optical depth effects for particular emission lines.
In treating this problem the adsorption by dust presents the
least of the evils. We expect more complex dependences treating
self adsorption within an emission line. The optical thickness
in this case will depend on the velocity adsorption.

Our present results are obtained for 21~cm emission transitions which intensity
is proportional to the integrated density of HI along the line of sight.
Other emissivities, e.g. those of H$_\alpha$ lines are proportional
to the squared density of species. It is possible to generalize our
results for those transitions and contribute to the studies of  ionized
emitting media, e.g. of HII regions (O'Dell 1986, O'Dell \& Castaneda 1987).

Other generalizations are also straightforward. Our present
research was motivated primarily
by the necessity to interpret the existing interferometric data and therefore
in the paper we mostly dealt with spectra. As the result,
correlation functions (see Appendix C) serve auxiliary purposes in our study.
With other sets of data, correlation functions may be more advantageous
and our analysis of their properties in Appendix C is useful.
 Moreover, modifying our technique it is possible
to study various moments of the line-shape (Lazarian 1992), 
e.g. velocity centroids 
(see Dickman 1985).
In our future paper we shall discuss all these possibilities.

{\bf HI data and simulations}\\
In the paper we predicted how the power slope will change with the
thickness of the velocity slice as a function of velocity and density
statistics and the ratio of the cold and warm HI filling factor.
It is important to test these our predictions against observations.
The SMC data is an example of the data set that can be used for the
purpose.

It is also important to verify our ideas using simulations. Our results
on velocity and density correlation were tested so far
only for Lognormal distribution where we managed to get an analytical
result. Using the statistics that arises from simulations is on the agenda.
We would stress that the only 3D simulations are useful. Due to a limited
dynamical range these simulations do not show power-law behavior and
additional care is required for comparing these results with our predictions.

{\bf Clouds and Filaments}\\
HI data cubes, exhibit a lot of small scale 
emissivity structure\footnote{It was noticed by Langer, 
Wilson \& Anderson (1993) that more structure
is seen in spectral line data cubes
 than in the integrated intensity maps.}. The question
is what part of them is real, i.e. is associated with density enhancements
in galactic coordinates and what part of them is produced by velocity 
fluctuations. A related question is whether the structures we see
 are produced dynamically, through
forces, e.g. self-gravity, acting on the media or they may be produced
statistically exhibiting the properties of random field. The second
question was partially answered in Lazarian \& Pogosyan (1997), where it 
was shown that density fluctuations with Gaussian distribution and
power spectra result in filamentary structures. The structures become
anisotropic and directed towards the observer when the velocity effects
are accounted for. 

The issue of density enhancements produced by velocity fluctuations is
closely related to the statistics of ``clouds'' observed in
velocity space. The results on velocity mapping that we discussed earlier
suggest that spectra of fluctuations observed in velocity slices
are more shallow than the underlying spectra. This means
more power on small scales 
or, in other words, more small scale structure (``clouds'')  
appears in the velocity slices due to velocity fluctuations.
Understanding of the velocity mapping
 may help to solve the problem of discrepancy between counts of 
clouds observed in emission and extinction (compare Scalo \& Lazarian 1996).

The integrated velocity maps are sensitive to density fluctuations
and may provide means of solving the problems of the statistics of actual
density inhomogeneities.

{\bf Beyond 2-point Statistics}\\
Is the power spectrum  the best one can get? Power spectrum does not
reflect the clustering properties of the medium. To find those
characteristics one should use multi point statistics, of which
the simplest would be the so-called bi-spectrum (Scoccimarro 1997).
This type of statistics is sensitive to clustering properties of the media
and we may hope to relate the bi-spectrum with the inflows observed
in molecular clouds (see
Myers \& Lazarian 1998) and the processes of star formation.
Other potentially useful tools, e.g. genus analysis are discussed in 
Lazarian (1999b).
A choice of a particular tool depends on the sort of questions one
attempts to answer. This paper is limited to making use of channel
map spectra. 

All in all, we believe that 
the machinery developed in the present paper opens 
avenues for both theoretical and observational research
and the appropriate studies should contribute to
unraveling the mystery of ISM turbulence. At the same time
a search for new statistical tools appropriate for diagnostic of
ISM turbulence should continue.

\section{Conclusions}

The results of the present work can be summarized as follows:

1. The emissivity spectrum arises both from velocity and density
fluctuations. The emissivity statistics depends on the thickness of
the velocity slice. The minimal value of the effective thickness
depends on the temperature of the gas. We found limiting regimes that we
termed ``thick'' and ``thin'' velocity slices (for criteria see 
Eq.~(\ref{eq:transition})) and obtained
asymptotics for these regimes. Thermal velocity
acts to increase the effective thickness of the slice without increasing
the amplitude of the signal. For a mixture of cold and warm HI the slice
may be thin for cold phase and thick for the warm phase. In this case the 
the expected relative contribution of warm medium is suppressed.

2. Two dimensional spectrum of 21~cm 
intensity (that can be measured by interferometer)
corresponds to the integral
of the underlying
3D spectrum if the slicing is thin while in the thick slice regime the
result depends on whether the density is short- or long-wave dominated
(see below for the criterium definition).
The 3D spectrum is anisotropic and can be expressed through the
velocity and density statistics.

3. For Galactic data we find that 
the slice is thick if the two dimensional wavenumber
$|{\bf K}| > (\delta V^2/C )^{1/m}\approx 0.16$~pc$^{-1}$, which means
that 
the present day radiointerferometers provide a thin slicing of Galactic
disk data. Density and velocity enter our formulae in a non-symmetric
fashion. Therefore the slope of the HI emissivity spectrum changes with the 
thickness of the slice and this allows to find the slopes of the
spectra of velocity and density fluctuations separately.

4. For thin slices the velocity fluctuations make spectra of emissivity
more shallow creating a lot of structures in PPV space that can
be erroneously identified as clouds. For long-wave dominated density
most of the small scale structures are due to velocity fluctuations
both in thin and thick slice regimes.

5. If the overdensity is dominated by
small scale inhomogeneities (i.e. the correlation functions scale
as $1/r^{\gamma}$, $\gamma >0$), the statistics of the channel maps
is dominated by density fluctuations. The HI observational data seem
not to support shallow density spectrum, however.

6. If the overdensity is dominated by 
large scale fluctuations (i.e. the structure functions scale
as $1/r^{\gamma}$, $\gamma < 0$), the statistics of the channel maps
is dominated by velocity fluctuations. 
In this case, to measure density one should use {\it very} thick slicing
${\cal L}\gg \lambda$, where $\lambda$ is given by Eq.~(\ref{fbar}). 
In thin slices one will find universal 2D spectral slope $\approx -8/3$
if velocity field is described by the Kolmogorov index $m=2/3$.
Data on the galactic HI statistics and that of SMC roughly agree with
this spectrum. 

7. The two cases above (i.e. $\gamma >0$ and $\gamma <0$)
can be distinguished by varying the thickness of the
velocity slice. The analysis of the observational data
suggests that the HI density spectrum has an index close to the
Kolmogorov one, namely, $-11/3$ and the statistics of 
the velocity field is also  Kolmogorov-type.  
Thus one may argue that the observed emissivity spectrum
 might be a signature of the Goldreich-Shridhar (1995)
MHD turbulence. 
The developed mathematical machinery is applicable to studies of turbulence
both in Galactic disc and individual clouds (see Appendix~E). 
Various emission lines can be
used and the technique can be modified to account for finite optical depth
effects.

\acknowledgements

The authors are grateful to Roman Scoccimarro for sharing his results
before publication and numerous useful discussions. The presentation
of the material has been improved following very helpful comments
by John Scalo. Discussions with Robert Brawn, Snezana Stanimirovic
and Steve Shore during the ``Cosmic Evolution and Galaxy Formation'' 
conference, Puebla, Mexico were helpful at the final stage
of our work and we would like to thank Jose Franco for inviting us.
We would like to thank Nick Kaiser for illuminating
exchanges. AL is grateful to Jay Gallagher,
Linda Spark whose suggestions helped to improve the presentation of
the material.
A.L. also acknowledges   NASA grant
NAG5 2858 and 
valuable suggestions by  Chris McKee.

\appendix

\section{List of Numerical Constants}\label{app:constants}
 Relevant to Table \ref{tab:3Dspk_asymp}
\begin{displaymath}
\begin{array}{clc}
\multicolumn{1}{l}{\mbox{{\it 3D power spectrum}
%, relevant to Table \ref{tab:3Dspk_asymp}:
}}\\[3mm]
a_n= & 2 \pi^2  \Gamma(-n) / \left|\Gamma\left({3+n \over 2}\right)\right|
\Gamma\left({1-n \over 2}\right) &
\mbox{eq.~(\ref{eq:f=0})}\\[3mm]
b_m=& \frac{m}{12}\pi^2 \Gamma(5+m)/\Gamma\left(1-m/2\right)
\Gamma\left(3+m/2\right) &
\mbox{eq.~(\ref{eq:I2lowk})}\\[3mm] 
s_{nm}= &
\begin{tabular}{|l|ccc|c|} \hline
m/n & -1 & -5/3 & -2 & -3 \\ \hline
1/2 & 74.8279 & 434.657 & 1361.75 & 84244.6 \\
2/3 & 35.4366 & 88.5916 & 172.584 & 2205.93 \\
1   & 19.1096 & 23.0004 & 29.4949 & 92.5001 \\ \hline
\end{tabular}
& \mbox{eq.~(\ref{eq:asymp})} \\[5mm]
\end{array}
\end{displaymath}
where the numbers at the right hand sight correspond to the
formulae in Appendix~B where the constants appear first.

Relevant to
Table \ref{tab:2Dspk_asymp} and Table \ref{tab:xi_asymp}
\begin{displaymath}
\begin{array}{clc}
\multicolumn{1}{l}{\mbox{{\it 2D power spectrum}, 
%relevant to Table \ref{tab:2Dspk_asymp}:
}}\\[3mm]
A_n= & a_n
& \\[3mm]
B_m= & (13/5 \pi)\, b_m  & \\[3mm]
S_{nm}= &  2^{-n-(m+1)/2} \pi \Gamma\left(-\frac{2n+m}{4}\right) /
\left|\Gamma\left(\frac{3+n+m/2}{2}\right)\right| & \\[3mm]\multicolumn{1}{l}
{\mbox{{\it Correlation function}, 
%relevant to Table \ref{tab:xi_asymp}:
}}\\[3mm]
c_{\gamma m}=&2^{(\gamma+m-1)/m}
\Gamma\left({2\gamma+m-2 \over 2m}\right) / m \pi^{1\over 2}& \\[3mm]
C_{\gamma m}= &\Gamma\left(\frac{2\gamma+m-2}{4}\right) /
2^{1\over 2} \Gamma\left(\frac{2\gamma+m}{4}\right) & 
\end{array}
\end{displaymath}

\section{Asymptotics and Special Regimes for 3D Power Spectrum}\label{app:asym}

We discuss asymptotics  and introduce useful approximations 
of the integrals involved in computation of 
the 3D power spectrum $P_s({\bf k})=P_{\rho}({\bf k})+
P_v({\bf k})$. If the density correlation function is given by 
eq.~(\ref{eq:xi})
Introducing new dimensionless variables  
\ba
y& =& kr\\ \nonumber
x&= &\cos\theta\\ \nonumber
\phi& =&(k \lambda)^{2-m}\mu^2=\lambda^{2-m} k_z^2 \, k^{-m}~~~,
\label{def}
\ea
we can present the 3D spectrum $P_s({\bf k})$  (see 
eq.~(\ref{eq:kolmogorov})) as
\begin{eqnarray}
P_s({\bf k}) & = & P_v({\bf k}) + P_{\rho}({\bf k})~~~, \\ \nonumber
P_v({\bf k}) & = & k^{-3} {\cal I}_3 (\mu,\phi)~~~, \\ \nonumber 
P_{\rho}({\bf k}) & = & k^{\gamma-3} r_0^{\gamma}
{\cal I}_{-n} (\mu,\phi)~~~.
\label{eq:insum}
\end{eqnarray}
\begin{equation}
{\cal I}_{\alpha}(\mu,\phi) = 2 \pi \int_0^{\infty} y^{\alpha-1} dy \int_{-1}^1 dx
J_0 \left(y \sqrt{(1-\mu^2)(1-x^2)} \right)
\cos(y \mu x) e^{-\phi y^m (2+m(1-x^2))/4},
\label{eq:Iterm}
\end{equation}
where $\alpha=3$ or $-n$ and $J_0$ is the Bessel function of the
zeroth order. We have to calculate $P_v$ and $P_\rho$ in another combination
when the density correlation function is given by eq.~(\ref{cor-structure})

 It is most illuminating to start by expanding the plane wave
in the eq.~(\ref{eq:main}) into multipoles 
\begin{equation}
e^{i {\bf k} \cdot {\bf r}} = 4 \pi \sum _{\mathrm{lm}} i^{\mathrm{l}}
 j_{\mathrm{l}}(kr) Y_{\mathrm{lm}}({\bf \hat k})
 Y_{\mathrm{lm}}^*({\bf \hat r})~~~,
\end{equation}
where $ j_{\mathrm{l}}(kr)$ is a spherical Bessel function and $Y_{\mathrm{lm}}
$ are spherical harmonics.
After performing trivial integration over the polar angle in  
eq.~(\ref{eq:main}) and
making the $dy$ integration the innermost one we get 
\begin{equation}
{\cal I}_{\alpha}({\bf k})=8 \pi^2 \sum_{\mathrm{l}} (i^{\mathrm{l}}) Y_{\mathrm{l0}}(\mu)
\int_{-1}^1 dx Y_{\mathrm{l0}}^*(x) \left(
\int _0^\infty y^{\alpha-1} dy j_{\mathrm{l}}(y) e^{-\phi y^m (2+m(1-x^2))/4} \right)
\label{eq:expansion}
\end{equation}

In this Appendix we consider two asymptotics, one ``long-wave'' 
corresponding to small $\phi$ and the ``short-wave'' corresponding 
to large $\phi$.
In the long-wave limit $\phi$ (see eq.~(\ref{def})) tends to zero and the
integration is trivial. In this case
${\cal I}_{\alpha} $ is just a constant
which coincides with the value of ${\cal I}_{\alpha}$ in the absence of random
velocity
\be
{\cal I}_{\alpha}(\mu,0) =  2 \pi^2 \frac{\Gamma (\alpha )}
{\Gamma \left({3-\alpha \over 2} \right)
\Gamma \left({1+\alpha \over 2} \right)} ~~~.
\label{eq:f=00}
\ee
The result is the same if one considers the wave modes orthogonal to the line
of sight ($\mu=0$) which are unaffected by velocity mapping
$ {\cal I}_{\alpha}(0,\phi) = {\cal I}_{\alpha}(\mu,0) $. 

For the term $P_{\rho}$ arising from density
perturbations $\alpha=-n < 3$ and ${\cal I}_{-n}(\mu,0)$ is
finite. For the velocity term $P_v$, $\alpha=3$ and eq.(\ref{eq:f=0})
gives zero value. The $\delta$-function at $k=0$ which is contained in $P_v$ and
describes the mean density in velocity space has been implicitly omitted
by the choice of a variable $y$.
To get more informative result,
we expand eq.~(\ref{eq:expansion}) up to terms with $\phi$ to the first power,
and after integration over angles obtain
%\be
%{\cal I}_3(\mu,\phi \to 0)\approx -\pi \phi \left[ {\scriptstyle\frac{6+2m}{3}}
%\int y^(2+m j_0(y)~ dy
%-m\pi \phi (\mu^2-{\scriptstyle\frac{1}{3}}) \int y^{2+m} j_2(y)~ dy \right]
%\label{I0}
%\ee
%Thus,
%"reqularized" integral approaches zero at $\phi=0$ linearly in $\phi$
\be
{\cal I}_3(\mu,\phi \to 0) \approx b_m \left[1 +
\frac{3}{5} \left(\frac{1}{2}-\mu^2 \right) \right] \cdot \phi,
\label{eq:I2lowk}
\ee
where
\be
b_m=\frac{m}{12} \pi^2 \frac{ \Gamma(5+m)}{\Gamma (1-m/2) \Gamma (3+m/2)}.
\label{eq:I2lowk1}
\ee
Note, when we expand the exponent in powers of $\phi$ 
each next term of the expansion subsequently
gives rise to higher
multipoles in $l$ series.
Namely, the zeroth-order term in $\phi$ leads only to a monopole
contribution, the first-order terms add quadrupole, the
second-order add an octupole etc.

Let us derive the short-wave asymptotics for eq.~(\ref{eq:expansion}).
We note that
the function $j_{\mathrm{l}}(y)$ is peaked, roughly, at $y={\mathrm{l}}+1/2$,
so if the exponential
cut-off occurs at lower $y$, the corresponding $l$-term can be neglected.
Thus only low multipoles remain in the sum for large $\phi$. Indeed, for 
a given $\phi$ high multipoles
$l+\frac{1}{2} > {\left(\frac{3+m}{6} \phi \right)}^{-1/m} $ give vanishing
contribution into eq.(\ref{eq:expansion}). Here we use an estimate for the
average value of the angle $<x^2>=1/3$ to get $(2+m-m<x^2>)/4=\frac{3+m}{6}$.

Short-wave (i.e. high $\phi$) asymptotics follows when only monopole is left
in the multipole expansion (\ref{eq:expansion}).
This is valid, formally, when $l=2$ quadrupole is cut out (since the integrals
of a dipole is zero). The criterion for
this is $\phi > \frac{6}{3+m} \left( {2 \over 5} \right) ^m \approx 1 $. 
Therefore 
\begin{equation}
{\cal I}_{\alpha}(\mu,\phi > 1) \approx 2 \pi \int _0^\infty y^{\alpha-1} dy j_0(y) \int_{-1}^1 dx e^{-\phi y^m (2+m(1-x^2))/4}
\label{eq:highk}
\end{equation}
For very short waves, i.e. $\phi \gg 1$  one can put
$j_0(y)=1$ in eq.(\ref{eq:highk}) to obtain the limit
\begin{equation}
{\cal I}_{\alpha}(\mu,\phi \to \infty) \to 
\left[ {\scriptstyle \frac{8 \pi}{m} \left(\frac{4}{m+2} \right)^{\frac{\alpha}{m} }}
\int _0^\infty y^{\frac{2\alpha-m}{m}} e^{-y^2}
D \left( y \sqrt{{\scriptstyle \frac{m}{m+2}}} \right) \right]
\phi^{-\frac{\alpha}{m}} 
\label{eq:asymp}
\end{equation}  
$D(y)$ is the Dawson integral. The factor in brackets, $s_{nm}$, is tabulated in
Appendix A.

We found that eq.~(\ref{eq:highk}) well approximates expression 
(\ref{eq:expansion}) for all $\phi$. Indeed, both long-wave asymptotics
eq.~(\ref{eq:f=00}) and the short-wave asymptotics eq.~(\ref{eq:highk})
are obtained with only $l=0$ term in eq.~(\ref{eq:expansion}).
We checked numerically that eq.(\ref{eq:highk}) provides an
accurate approximation for ${\cal I}_{-n}$, $n\neq -3$,
for all values of $\mu$ and $\phi$.
This is a very welcomed simplification, since eq.~(\ref{eq:highk})
depends on $\phi$ rather than on $k$ and $\mu$ separately. For 
$\alpha =3$ at small $\phi$ the
zeroth-order $\mu$-independent term (\ref{eq:f=00}) is equal to zero
and the in linear $\phi$ term  of ${\cal I}_3$ at low $\phi$
has the $\mu$-dependent amplitude (see \ref{eq:I2lowk}).
In this case
Eq.~(\ref{eq:highk}) provides an angle-average estimate.

The deviation of the expression (\ref{eq:highk}) from (\ref{eq:expansion}) 
is less than $3\%$ for $\alpha = 1$.
In fact, we found the following formula to be
even more accurate numerically while still depending only on $\phi$ variable
\begin{equation}
{\cal I}_{\alpha}(1,\phi) = 2 \pi \int_0^{\infty} y^{\alpha-1} dy \int_{-1}^1 dx \cos(yx) e^{-\phi y^m (2+m(1-x^2))/4}
\label{eq:m=1}
\end{equation}
This expression is exact for the modes parallel to the line of sight
($\mu = 1$) with $\phi=(k \lambda)^{2-m}$. Being extended to general
$\phi=(k \lambda)^{2-m} \mu ^2$
it gives an excellent approximation to the full result.

Above  we discussed the asymptotics of the 
integral
(\ref{eq:insum}) that can be found analytically. The 
Table~{\ref{tab:3Dspk_asymp} 
summarizes our results for the power spectrum in the velocity space. 
In velocity
space the spectrum is anisotropic, depending separately on 2D wavevector 
${\bf K}$, perpendicular to the line of site and $k_z$ component, parallel to
the line of site, $ k^2=|{\bf K}|^2+k_z^2$.  

\begin{table}[h]
\begin{displaymath}
\begin{array}{rcr} \hline\hline\\
&\multicolumn{1}{c}{(k_z \lambda)^2 \ll (k \lambda)^m} &
\multicolumn{1}{c}{(k_z \lambda)^2 \gg (k \lambda)^m}  \\[3mm] \hline\\
\lambda^{-3} P_\rho({\bf K},k_z): & a_n  (r_0/\lambda)^{n+3} \cdot (k\lambda)^n
& s_{nm}  (r_0/\lambda)^{n+3} \cdot (k_z \lambda)^{2 n/m}  \\[3mm] \hline\\ 
\lambda^{-3} P_v({\bf K},k_z): &   
b_m \left[1 + \frac{3}{5}\left(1/2-k_z^2/k^2\right)\right] \cdot
(k_z \lambda)^2 / (k \lambda)^{3+m}&
s_{-3m} \cdot  (k_z \lambda)^{-6/m} \\[3mm] \hline
\end{array}
\end{displaymath}
\caption{Asymptotical behavior of the 3D spectrum in velocity space.
Numerical constants $a_n,b_m$ and $s_{nm}$ are given in
Appendix A.}
\label{tab:3Dspk_asymp}
\end{table}
The first (``long-wave'') asymptotics is achieved for waves longer than 
a velocity correlation scale $\lambda$ {\it or} for the waves 
transverse to the line of sight, 
which
are unaffected by velocity mapping. The second (``short-wave'') asymptotics 
is reached
for waves that are shorter than the turbulence length {\it and} have 
non-negligible
component in z-direction.

In Fig.~\ref{fig:3DPr0L1} and Fig.~\ref{fig:3DPr0L03}
we show results of numerical calculations for the total spectrum in
velocity space $P_s=P_{\rho}+P_v$ obtained for
$r_0=\lambda$ and $r_0=0.02\lambda$ respectively.
In the former case the transition for
the short-wave asymptotics happens exactly at the same scale as
density correlation scale, while in the latter,
more realistic case the velocity scale significantly exceeds density
correlation scale. 
Naturally, when density in Galactic coordinates is less correlated
the caustics caused by velocity fluctuations become important.
The comparison of Fig.~5 and Fig.~6 confirms this tendency.
\begin{figure}[ht]
\centerline{\epsfxsize=5.in\epsfbox{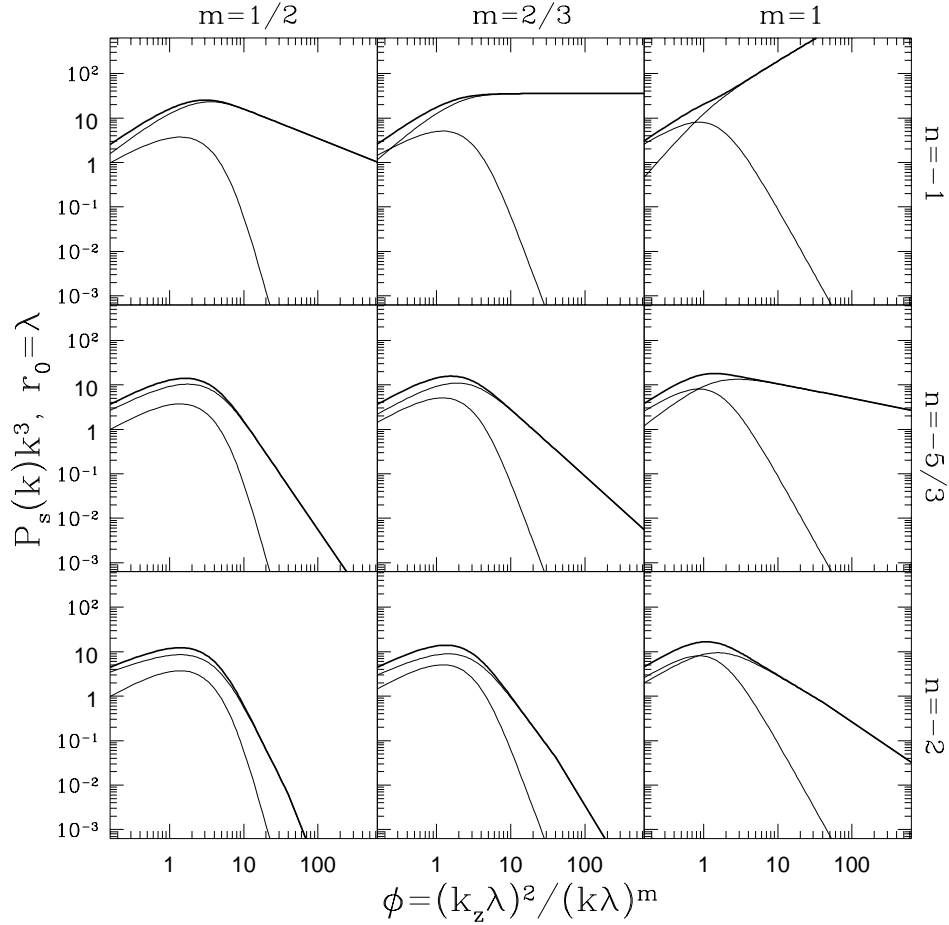}}
\caption{The quantity $P(k)k^3$  for various values of $m$ and
$n=\gamma-3$; $r_0=\lambda$. The short wave part of the spectrum is 
the function of the variable $\phi=(k_z\lambda)^2/(k\lambda)^m$ only. Although
the longwave spectrum depends separately on $k_z$ and $k$, we have shown the total
spectrum in $\phi$ variable, assuming the modes to be along the line of sight
$k_z \sim k$. In each plot the heavy line shows the total
spectrum $P_s=P_{\rho}+P_v$, while two light lines show $P_{\rho}$ and $P_v$
terms separately. The term originated from the underlying density perturbations
$P_{\rho}$ dominates at large wavenumbers. }
\label{fig:3DPr0L1}
\end{figure}
\begin{figure}[ht]
\centerline{\epsfxsize=5.in\epsfbox{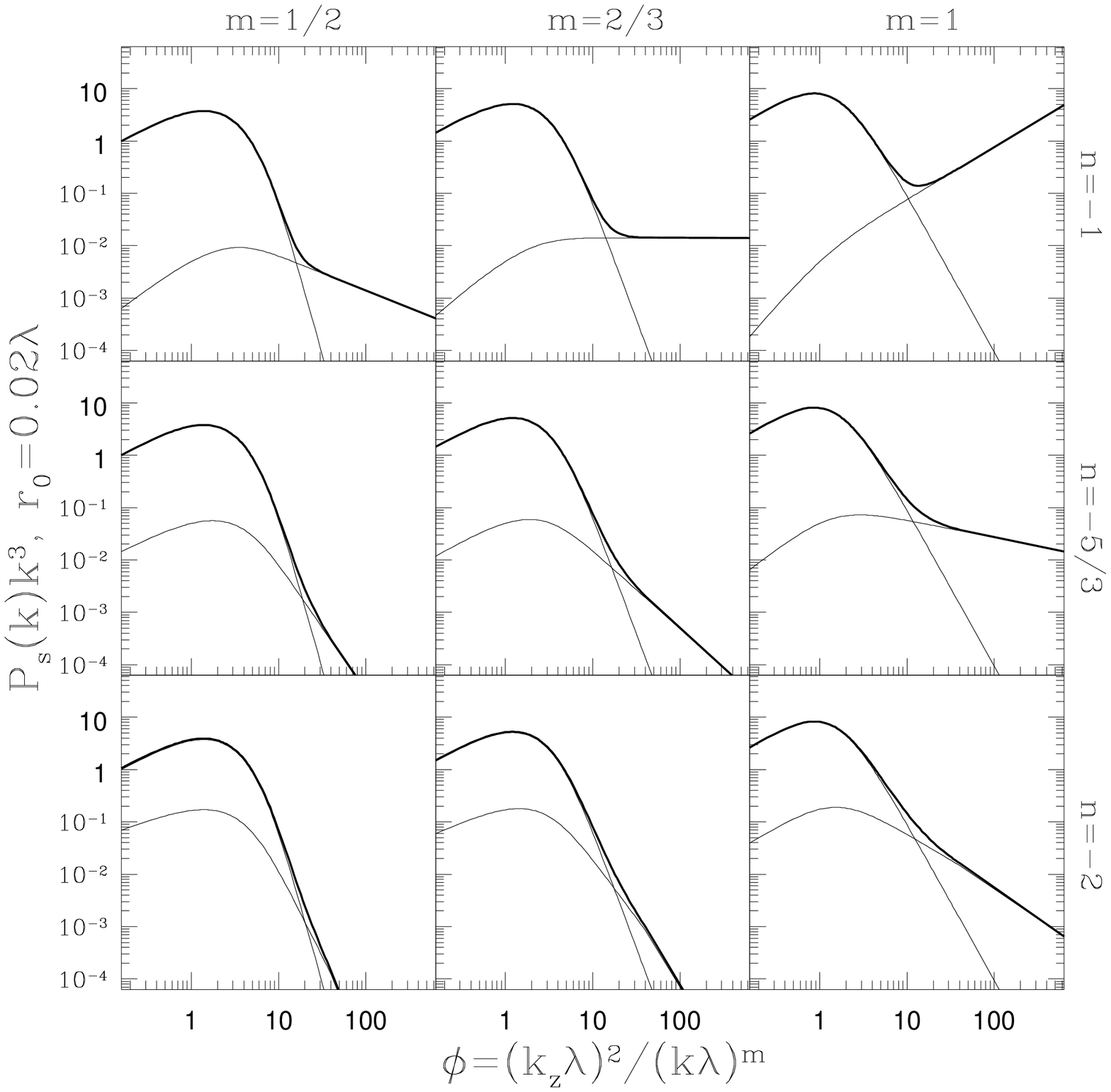}}
\caption{The same as in Fig.~\ref{fig:3DPr0L1} but for
$r_0=0.02\lambda$.  }
\label{fig:3DPr0L03}
\end{figure}
The results for $P_v$ and $P_\rho$ constitute the basis for our
computation of 2D spectra in the main text and 1D spectra in Appendix~F.

\section{3D Correlation Functions}\label{app:3Dcorr}

The two-point correlation function is an alternative to a
power-spectrum
description of the density field. In the velocity space the correlation
function $\xi_s({\bf R},z) $ is obtained by the inverse Fourier
transformation of the spectrum $P_s({\bf k})$,
$\xi_s({\bf R},z)=(2 \pi)^{-3} \int d^3{\bf k}
e^{-i {\bf k}\cdot{\bf r}_s}P_s({\bf k})$. 
Here, as in the main text, we use capital letters to denote vectors
perpendicualar to the line of sight and therefore the total vector
in the velocity space is ${\bf r}_s=({\bf R},z_s)$ and
$R\equiv|{\bf R}|$.
With an anisotoropic $P_s({\bf K},k_z)$ given by (\ref{eq:main}), the 
integration over ${\bf K}$ in eq.~(\ref{xi_s})
leads to delta function, which can be used
to eliminate the integral over $\bf R$.
The integral over $k_z$ in eq.~(\ref{xi_s}) 
can be done analytically to obtain
\begin{equation}
\xi_s(R,z_s)=\int \! dz \: \xi (R, z)
\left[
\left(2\pi \lambda^2 \tilde D_z(R,z)\right)^{-{1 \over 2}}
\exp\left(-{(z_s-z)^2 \over 2 \lambda^2 \tilde D_z (R,z)}\right)
\right]
\label{eq:corrs}
\end{equation}
Here $\tilde D_z(R,z)$ describes a velocity
structure tensor, i.e.
$\tilde D_z (R,z)=(r/\lambda)^m \left[1+\frac{m}{2}(1-z^2/r^2) \right]$
in the case of solenoidal turbulence,
and $r^2={\bf R}^2+z^2$.
In the absence of the random velocity field $\lambda \to 0$ the kernel
in the brackets transforms to a  delta function $\delta(z_s-z)$, and,
as one expects $ \xi_s(R,z_s) \to \xi(r) $. Similarly to the split of
power spectrum, the correlation function can be split in two parts
$\xi_s=1+\xi_v+\xi_{\rho}$, where unity describes mean density in
velocity space,
$\xi_v$ comes from the random velocity map of the mean density term and
corresponds to the first integral in (\ref{eq:corrs}) minus unity, while
$ \xi_{\rho}$ arises from the transformation of the underlying correlations
of density perturbations $\xi(r)$. It is useful to introduce 
dimensionless variables $\tilde r, \tilde {\bf R}, \tilde z$ which measure
distances in terms of the velocity correlation scale $\lambda $,
i.e $\tilde r =r/\lambda $, etc. Then, with $\gamma=3+n$ denoting a slope
of the correlation function $\xi(r)$
\begin{eqnarray}
\xi_v(\tilde R,\tilde z_s) &=& -1 + \int  \! d \tilde z \:
G(\tilde R,\tilde z_s, \tilde z) \\
\xi_{\rho}(\tilde R,\tilde z_s) &=& (r_0/\lambda)^{\gamma}
\int \! d \tilde z \:
\left({\tilde R}^2 +{\tilde z}^2 \right)^{-{\gamma \over 2}}
G(\tilde R,\tilde z_s, \tilde z) \\
G(\tilde R,\tilde z_s, \tilde z) &=&
\left(2 \pi \tilde D_z(\tilde R,\tilde z)\right)^{-{1 \over 2}}
\exp\left(-{(\tilde z_s-\tilde z)^2 \over
2 \tilde D_z (\tilde  R,\tilde z)}\right)
\end{eqnarray} 

Two specific cases are of particular interest, namely,
2D correlation function in a slice of fixed velocity
$ \xi_s(R,0) $ and 1D correlation along the line of sight $ \xi_s(0,z_s) $.
The first one corresponds to 2D correlation function measured in a {\it thin}
slice.
Asymptotical behavior of $\xi_s(R,z_s) $ is given in Table~\ref{tab:xi_asymp}.
\begin{table}[h]
\begin{displaymath}
\begin{array}{rcr} \hline\hline\\
& \multicolumn{1}{c}{(r/\lambda) \to 0}&\multicolumn{1}{c}{(r/\lambda) \to \infty}\\[5mm]
(\lambda/r_0)^{\gamma} \xi_{\rho}(R,0):
& C_{\gamma m} \cdot
{(R/\lambda)}^{-\gamma+(1-m/2)} &  {(R/\lambda)}^{-\gamma} \\[3mm]\hline\\
(\lambda/r_0)^{\gamma} \xi_{\rho}(0,z_s):
& c_{\gamma m} \cdot 
{(z_s/\lambda)}^{-\gamma-{(2-m)\over m}(\gamma-1)} &
{(z_s/\lambda)}^{-\gamma} \\[3mm]\hline\\
\xi_v(R,0): & m / (2-m) &
(m^2 / 4) \cdot {(R/\lambda)}^{\,m-2} \\[3mm]\hline\\
\xi_v(0,{z_s}): & m / (2-m) &
{m (m-1) \over 2} \cdot {(z_s/\lambda)}^ {\,m-2} \\[3mm] \hline
\end{array}
\end{displaymath}
\caption{Asymptotical behavior of the correlation function in velocity space.
Numerical constants $c_{\gamma m}$ and $C_{\gamma m}$ are given in Appendix A.}
\label{tab:xi_asymp}
\end{table}

The asymptotical formulae for $\xi_{\rho}$ are valid for $\gamma+m/2 > 1$.
As it should be, the correlation function is unchanged by velocity map
at large separations, having the same slope $\gamma$ as the underlying
$\xi(r)$. If $\gamma > 1$, the slope of correlation function at small
separations is
changed, becoming more shallow $-\gamma + (1-m/2)$ for correlation in
the slice of fixed velocity, but steeper $-\gamma-{(2-m)\over m}(\gamma-1)$
for the correlation along the line of sight, if $\gamma > 1$.
Interestingly, for $\gamma=1$ there is no change of correlation
slope along the line of sight. However the amplitude
of correlation at small distances is enhanced by the factor $1/m$.

The term $\xi_v$ has a finite value at a zero separation
and is not important at small distances. Correlation at the very long 
distances is, however, dominated by velocity term if $\gamma > 2-m$.
The above results are illustrated in Fig.~\ref{fig:xi}.
\begin{figure}[ht]
{\centering \leavevmode
\epsfxsize=.33\columnwidth \epsfbox{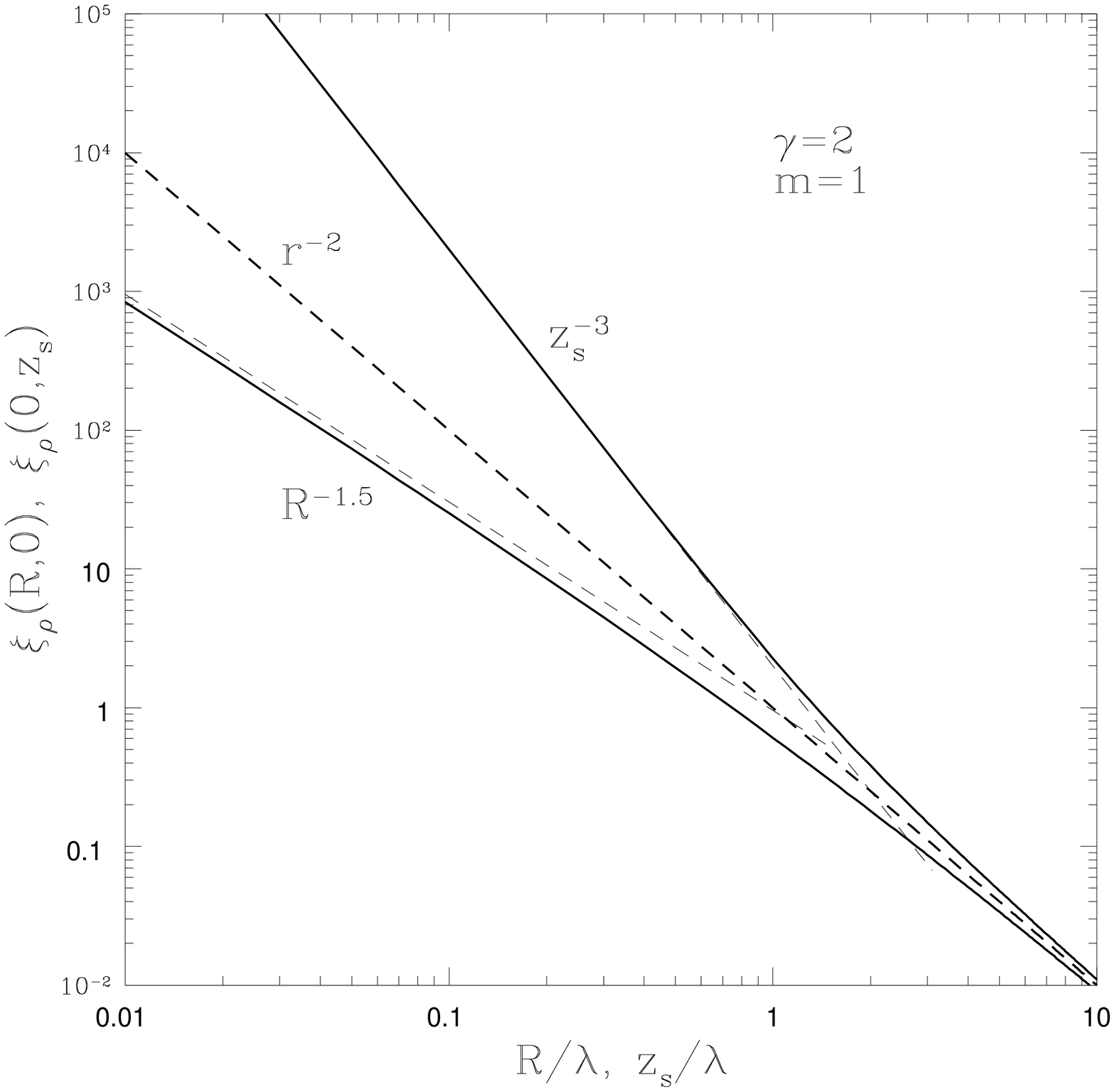} \hfil
\epsfxsize=.33\columnwidth \epsfbox{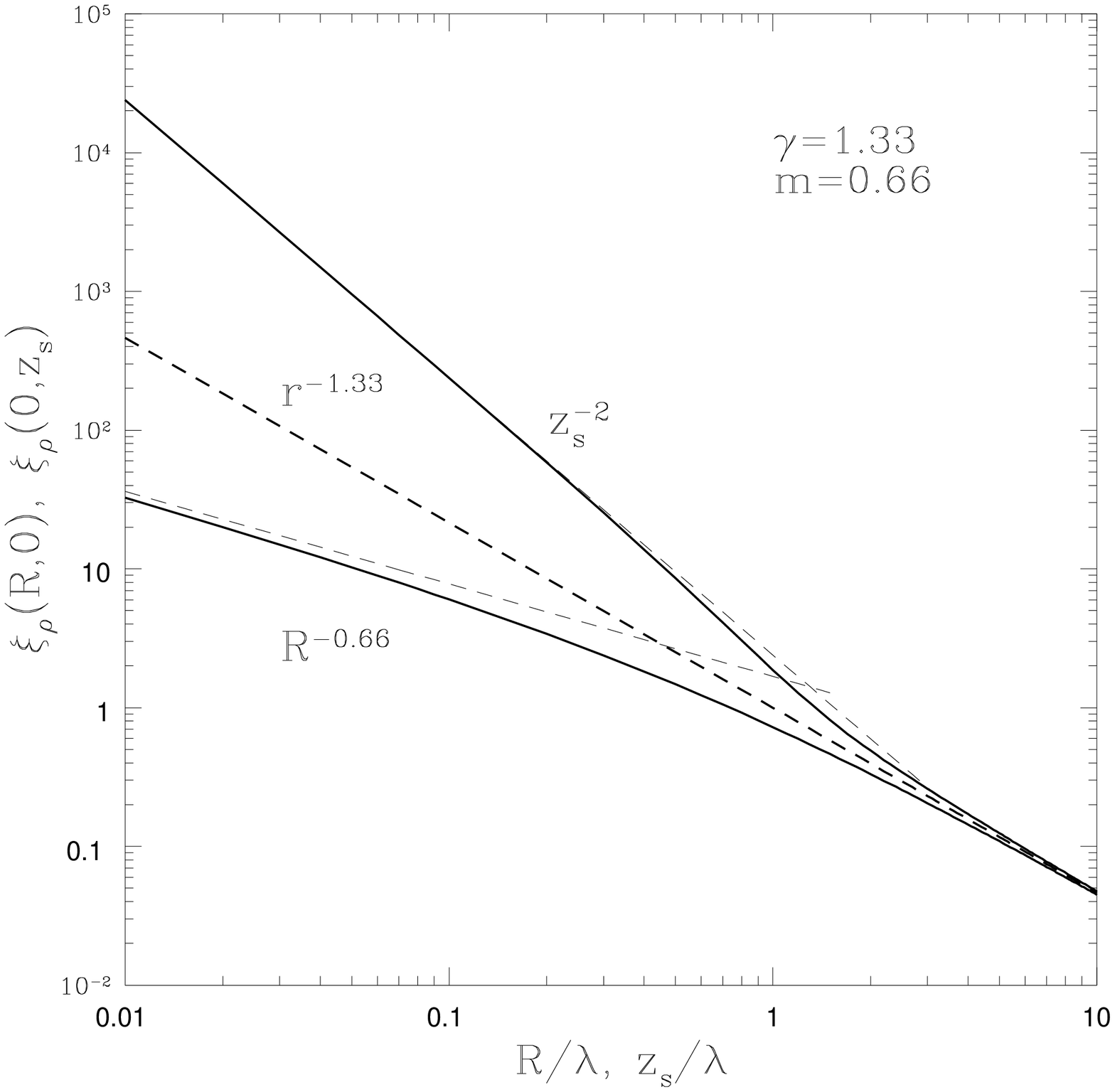} \hfil
\epsfxsize=.33\columnwidth \epsfbox{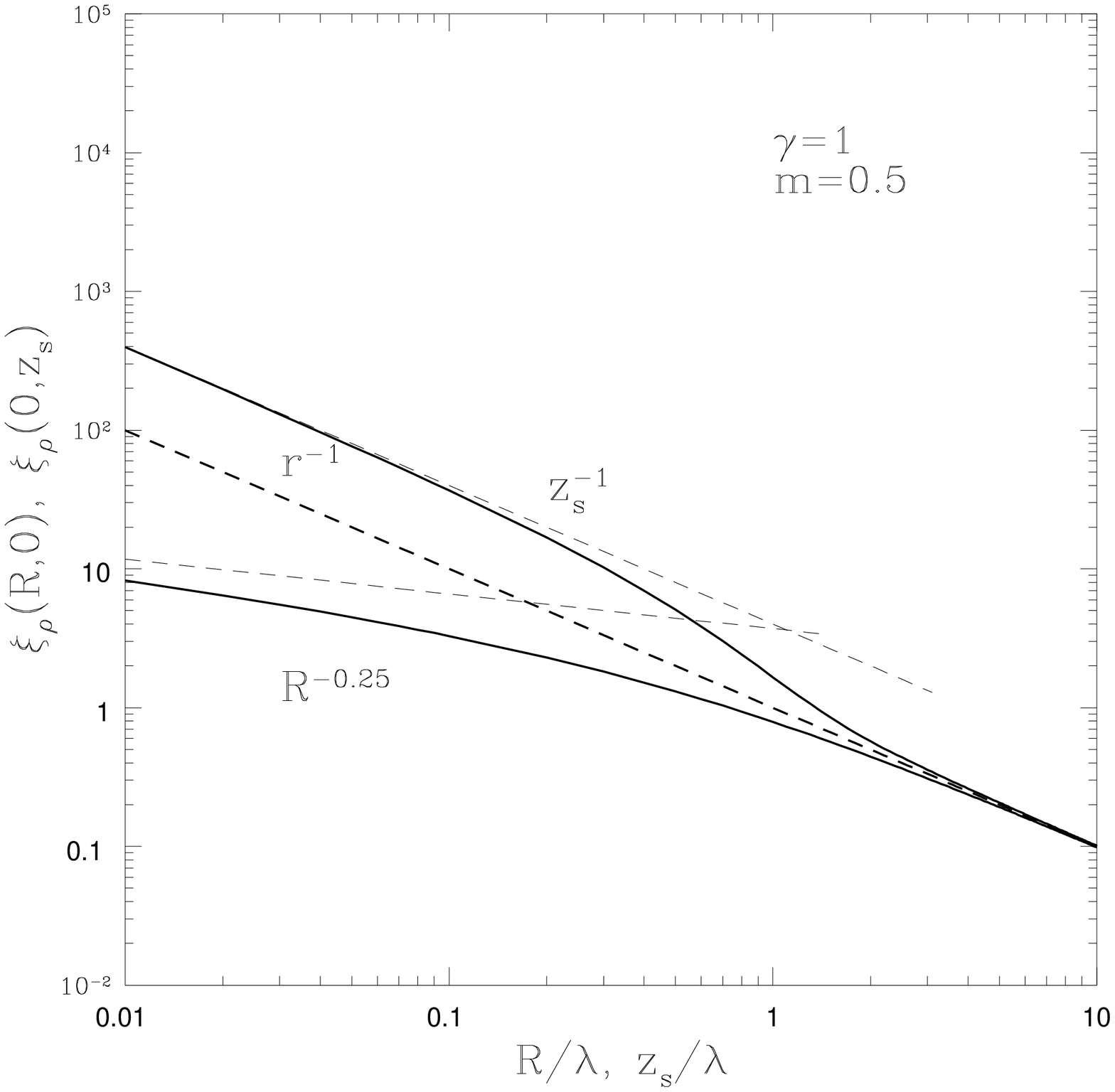} 
}
{\centering \leavevmode
\epsfxsize=.33\columnwidth \epsfbox{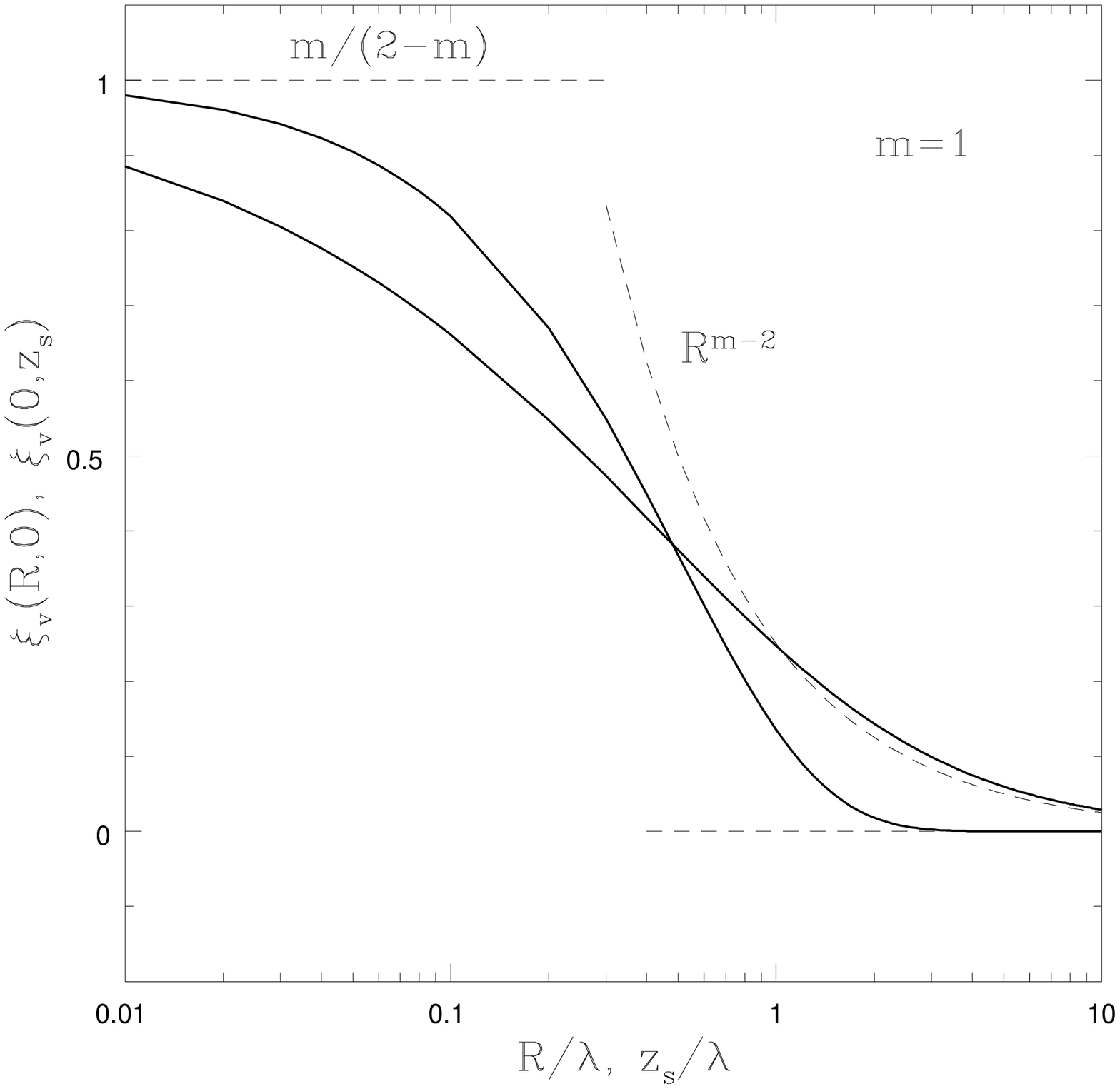} \hfil
\epsfxsize=.33\columnwidth \epsfbox{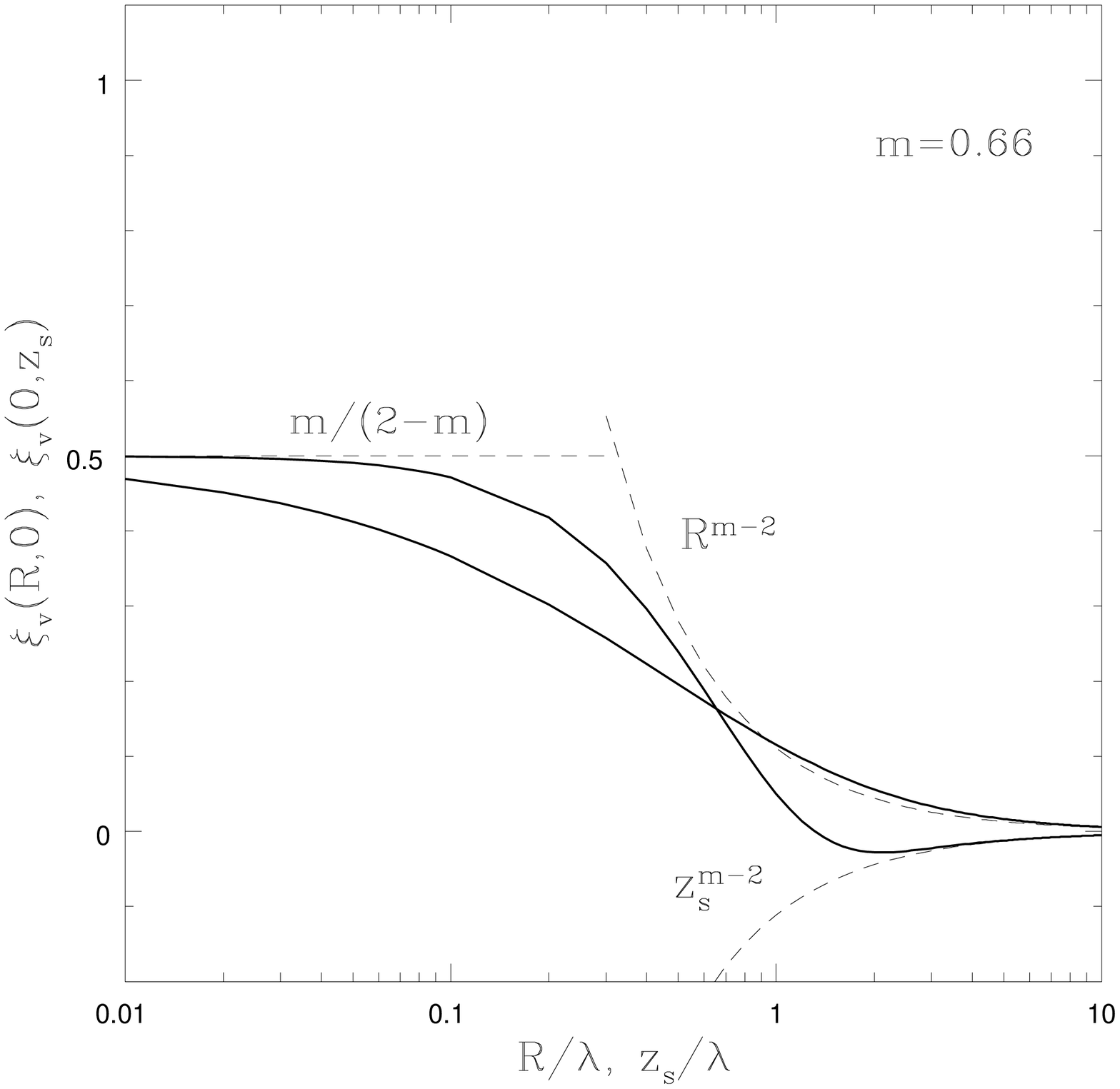} \hfil
\epsfxsize=.33\columnwidth \epsfbox{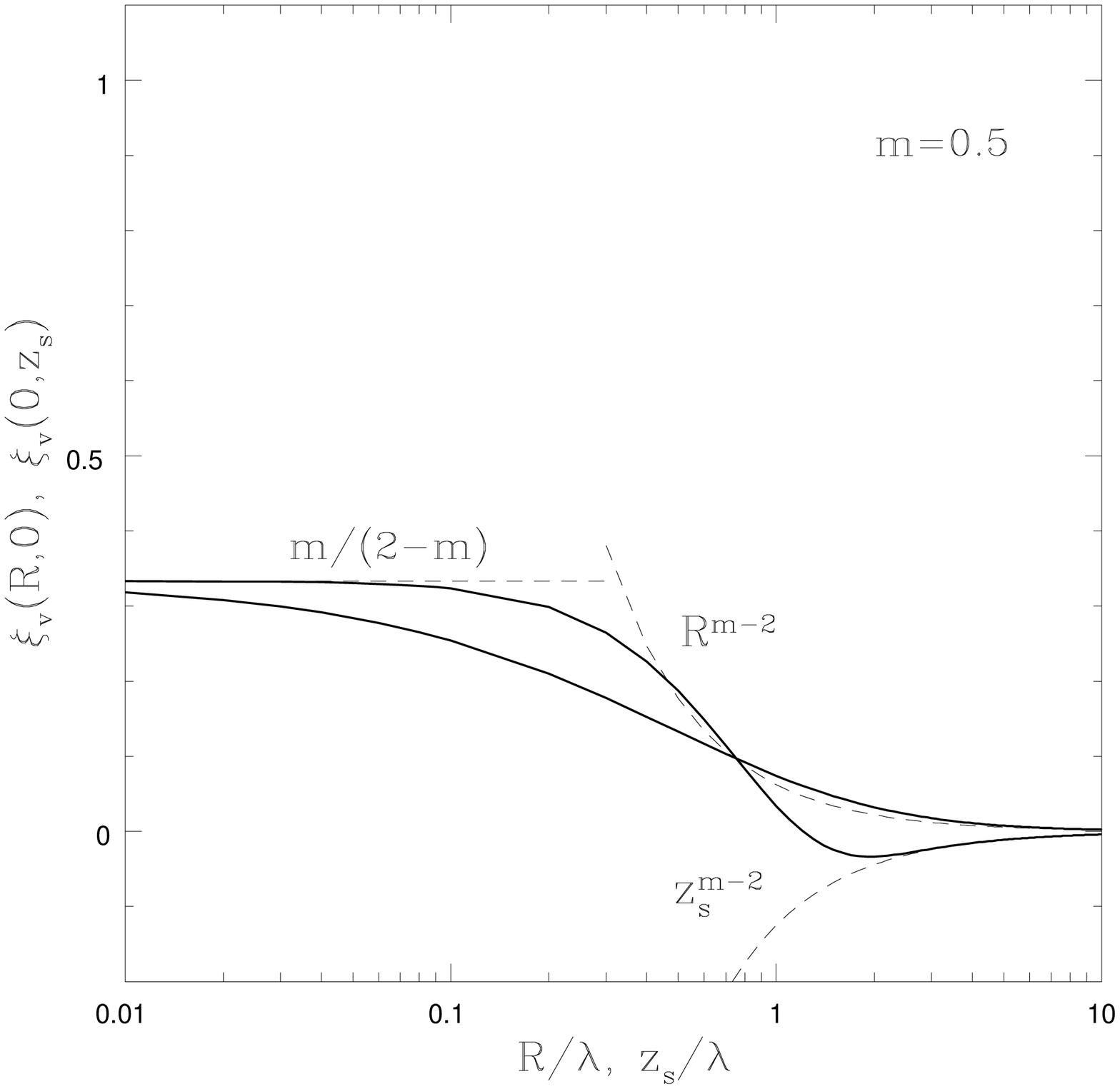} 
}
\caption{
Correlation function in velocity space for various choice of $\gamma$ and $m$.
Upper set of panels shown the density term $\xi_{\rho}$
the line of sight, $\xi_{\rho}(R,0)$, and in a transverse {\it thin} slice,
$\xi_{\rho}(0,z_s)$. Asymptotics also include underlying $r^{-\gamma}$
slope for density correlation function before velocity mapping.}
\label{fig:xi}
\end{figure}

In view of large values of $\lambda$ the regime $r/\lambda\rightarrow 0$
is the most important.

\section{Correlations between density and velocity}\label{app:kernel}

Our assumption in the body of the paper was that the density and
velocity are uncorrelated. Therefore it is interesting to study
whether possible velocity-density correlations alter our result.
To  do this we have to assume a statistical model for the density field.

In what follows we 
shall consider the Lognormal density distribution.
Particularly, we assume that the density can be expressed as
the exponent of the Gaussian field $z({\bf x})$, 
\begin{equation}
\rho({\bf x})=\rho_0 e^{z({\bf x})}.
\end{equation}
The advantage of this
model for density is that it can describe nonlinear density fluctuations
$|\delta| > 1$ while remaining simple\footnote{The Gaussian model
of the density field cannot describe large fluctuations and therefore
we do not consider it.}. 

We parametrize the Gaussian distribution for $z({\bf x})$ in the following
way 
\begin{equation}
P(z)=\frac{1}{(2 \pi)^{1/2} A^{1/2}} 
\exp \left[ -\frac{z^2}{2 A} \right]
\end{equation}
Then the average and the second moment of the density distribution, which can be
directly measured, are related to the parameters $\rho_0$ and $A$ as
\begin{eqnarray}
\bar \rho & \equiv & \langle \rho \rangle = \rho_0 e^{A/2}\nonumber \\
\sigma^2_{\rho} &\equiv &  \langle \rho^2 \rangle = \rho_0^2 e^{2A}
\label{eq:Bdef}
\end{eqnarray}
In the absence of density inhomogeneities $A \to 0$, $\bar \rho \to \rho_0$.

Power spectrum in velocity space is a Fourier transform of the
kernel (see eq.~({\ref{eq:roman}))
\begin{equation}
\langle e^{i f k_z [u_z(1)-u_z(2)]}\rho(1)\rho(2)\rangle
\label{eq:kernel} 
\end{equation}
where $u_z$ is z-component of velocity, $\rho$ 
and $1,2$ refer to two points in space, $k_z$ 
is a wavevector component in the velocity space.

To calculate the statistical average in eq.~(\ref{eq:kernel})
we use the variables
\begin{eqnarray}
y & \equiv & f/\lambda ~ [u_z(1)-u_z(2)] \\
z_+ & \equiv & z(1)+z(2)
\end{eqnarray}
which are Gaussian random quantities obeying 
bivariant joint Gaussian distribution. It is well known that a
 general Gaussian distribution
of $N$ correlated variables (with a zero mean values) described
by a vector ${\bf X} = \left(x_1,x_2,\ldots,x_N\right)$ 
is given by the probability function
\begin{eqnarray}
&&P\left(x_1,x_2,\ldots,x_N\right)=\frac{1}{ (2 \pi)^{N/2} \|{\bf C}\|^{1/2}}
\exp \left[-\frac{1}{2}{\bf X C^{-1} X} \right] \nonumber\\
&&{\bf C} \equiv
\left( \begin{array}{cccc}
\langle x_1^2 \rangle & \langle x_1 x_2 \rangle & \ldots & 
\langle x_1 x_N \rangle \\
\langle x_1 x_2 \rangle & \langle x_2^2 \rangle & \ldots & 
\langle x_2 x_N \rangle \\
\ldots & \ldots & \ldots & \ldots \\
\langle x_1 x_N \rangle & \langle x_2 x_N \rangle & \ldots & 
\langle x_N^2 \rangle~~~.
\end{array}
\right)
\label{C:dist}
\end{eqnarray}

We shall characterize the correlation properties of a pair $(y,z_+)$ 
by three functions
\begin{eqnarray}
\langle y^2 \rangle &=& D_z({\bf r}) \nonumber \\
\langle z_+^2 \rangle &=& B({\bf r}) \label{eq:dencor}\nonumber\\
\langle y z_+ \rangle &=& \langle u_z(1) z(2) \rangle
- \langle u_z(2) z(1) \rangle \equiv F ({\bf r})
\end{eqnarray}
where $F({\bf r})$ is a cross correlation
function\footnote{In the equation for $F({\bf r})$ we have used 
the fact that a random vector
quantity (velocity in our case) always has a zero correlation with any
scalar quantity (as density) at the same location in space, i.e,
$\langle u_z(1) \rho(1) \rangle=\langle u_z(2) \rho(2) \rangle=0$. Indeed,
the correlation functions of density-velocity in one point can be obtained
by correlating velocity and density in two different points 1 and 2 and
then bringing the points together. It is obvious that if points 1 and 2
are along z-axis and very close together, $\langle u_z(1) \rho(2) \rangle
\approx -\langle u_z(2) \rho(1)\rangle$. This proves that the correlations
are zero when the points coincide.} of velocity and density, 
$ D_z({\bf r}) $ is the z-projection
 of velocity structure function given by eq~({\ref{Dzz}) and
$ B({\bf r}) $ can be expressed through the density correlation function
$\xi({\bf r})$. Substituting those functions in eq.~(\ref{C:dist})
we get the joint distribution function
\begin{equation}
P(y,z_+) = \frac{1}{2 \pi (D_z B - F^2)^{1/2}}
\exp\left\{ -\frac{
B y^2 -2 F y z_+ + D_z z_+^2}{2 (D_z B - F^2)} \right\},
\label{eq:distr}
\end{equation} 
using which one can easily compute that
\begin{equation}
\xi({\bf r}) \equiv \langle \rho(1) \rho(2) \rangle / \rho_0^2 =
\langle e^{z_+} \rangle =   e^{B({\bf r})/2},
\end{equation}
and, thus, $B(0)=4 A=2 \ln \sigma_\rho^2 /\rho_0^2$.

Averaging kernel ({\ref{eq:kernel}) with the distribution function
given by eq.~(\ref{eq:distr}) yields 
\begin{equation}
P_s({\bf K},k_z)=
\int d^3{\bf r} e^{i {\bf k \cdot r}} \exp 
\left[ -\frac{(k_z\lambda)^2 D_z({\bf r})}{2} \right]
\exp \left[ i k_z \lambda F({\bf r}) \right] \xi ({\bf r})~~~,
\label{eez}
\end{equation}
which generalizes eq.(\ref{eq:kolmogorov}) for a non-zero correlation
of velocity and density.

Let us now project $P_3$ to $P_2$ in a 
{\it thin} slice and for high $|{\bf K}|$.
Using eq.~(\ref{eq:thin}) and $P_s$ given by eq.~(\ref{eez})
we obtain 
\begin{equation}
P_2({\bf K}) \approx \frac{1}{(2 \pi)^{1/2} \lambda}
\int d{\bf r} ~e^{i {\bf KR}} 
\frac{\xi({\bf r})}{D_z^{1/2}({\bf r})} 
\exp \left[-\frac{F^2({\bf r})}{2 D_z({\bf r})}\right]
\exp\left[-\frac{z^2}{2 \lambda^2 \tilde D_z({\bf r}/\lambda)}\right]
\label{eq:thinap}
\end{equation}
This expression is an analog of eq.~(\ref{eq:thinv}) in section~3
for $F\neq 0$. For high $|{\bf K}|$ the last exponent may be shown to 
be $\approx 1$ and then the equation become analogous to 
eq.~(\ref{eq:thinhighK}). 

An upper limit
for $F$ cross-correlation function follows from general Cauchy-Swartz
inequality (see Mathews \& Walker 1970),
$\langle y z_+ \rangle ^2 \le \langle y^2 \rangle
\langle z_+^2 \rangle $,
\begin{equation}
F^2({\bf r}) \le D_z({\bf r}) B({\bf r}) = 2 D_z({\bf r}) \ln \xi({\bf r}).
\label{eq:cauchylimit}
\end{equation}
and this limiting case
is achieved when density (logarithm) and velocity are perfectly
correlated and are not independent statistical quantities.
If we adopt this limit, $\xi({\bf r})$ cancels out and the
velocity fluctuations in the thin slice limit are
determined by the spectrum of random velocity only:
\begin{equation}
P_2({\bf K})|_t  \approx \frac{1}{(2 \pi)^{1/2} \lambda}
\int d{\bf r} e^{i {\bf KR}} D_z^{-1/2}
\label{p2app} 
\end{equation}

It is easy to see that equation eq.~(\ref{p2app}) coincides with
the expression for $P_{2v}$ and therefore in the case of of long-wave
dominated density field we reproduce the thin-slice asymptotics
found on the assumption of no correlation of velocity and density. 
Indeed, our analysis in section~4 showed that in this regime
the $P_{2v}$ part of the spectrum dominates the signal.
Formally, we do not reproduce the result for the short-wave dominated
density regime, where $P_{2\rho}$ part of $P_2$ is important. However,
we assumed the absolute maximum (unrealistic!) of the velocity-density
correlations. Therefore, for more realistic cases we do expect
density to reveal itself within the thin slices in the regime of
short-wave density fluctuations.

Eq.~(\ref{eez}) also indicate that the criterion for the velocity 
slice to be thick may be altered if the velocity-density correlations 
are present. Apart from the criterion (\ref{eq:transition}) one
need to require that  ${\cal L}/\lambda > (1/\lambda|{\bf K}|)F$.
For realistically small $F$ this condition is fulfilled when
(\ref{eq:transition}) is fulfilled. In the case of upper limit
of correlation given by eq.~(\ref{eq:cauchylimit}) the thickness
of the slice should be increased by a factor of the order of a few.
Asymptotics in the regime of thick slicing given by Table~1
stay the same when velocity-density are correlated.

All in all, thick slice regime does not depend on velocity and
density being correlated. This regime can provide us with the information
whether density is long or short wave correlated. The density dominates
for ``very thick'' slices (see criterion (\ref{very thick})) and for them it
is possible to determine the density spectrum irrespectively of
velocity-density correlations. The regime of thin slices depends on the 
velocity random field if the density field is long-wave
correlated and the spectra of both velocity and density field if the density
is short-wave correlated (see Table~1). In the case
of a thin slice we recover the velocity power spectrum
for the long-wave dominated regime. The only limiting case when 
velocity-density
correlations can matter is the thin slice for the short-wave dominated
density field. In the extreme case of a perfect velocity-density correlation
the emissivity depends only on the velocity spectrum, while the emissivity
does depend on density if the correlations are absent. This means
that if the analysis of the thick sliced data reveals short-wave correlated
density, an additional care for the analysis of the data in the
thin slice regime might be needed.

Our analysis of velocity-density correlations was performed for the
case of the HI turbulence study in Galactic disc. However, it is
also applicable to the studies of turbulence in individual clouds (see
Appendix~E).

\section{Application to individual clouds}
In the main body of the paper we were concerned with the studies of HI in the
Galactic disc when the Galactic rotation curve served as a
distance indicator. If studies of individual
clouds are concerned (e.g. high latitude clouds) no analog of
the distance-velocity relation exists. Similar problems arize when we
deal with external galaxies.

Here we derive the expression for 2D spectrum of HI emission
 applicable to studies of turbulence within
individual clouds. To do this we modify our formalism presented in the
main text to account for the finite size of the emitting region.
We will generalize map (\ref{eq:map}): 
\begin{eqnarray}
{\bf X_s} &=& {\bf X} \nonumber  \\ 
z_s &=& A \left[ f^{-1} z - {\bf u}({\bf x}) \cdot {\bf \hat z} \right]~~~, 
\label{eq:mapnoz}
\end{eqnarray}
 where the parameter $A$ is just a conversion factor which specifies the units
of $z_s$ coordinate. Our previous map (\ref{eq:map}) corresponds to the 
choice $A=f$. In the present form the map is applicable to a zero shear,
which corresponds to  $ f^{-1} \to 0$.

Our further treatment of the problem repeats the steps discussed in 
the main text but we take into account the finite extend of the cloud.
Since in the image plane  we consider  only short scales
relative to our object size, we can simplify our task by accounting for
only for a
finite thickness $S$ of the cloud along the line-of-sight.
In place of eq.(\ref{eq:rkintegral}) the 
Fourier component of density in velocity space
is now
\begin{equation}
\rho_s({\bf k}) =  \int_0^S dz  e^{i A f^{-1} k_z \cdot z} 
\int d^2{\bf X}  e^{i {\bf K} \cdot {\bf X}}
\rho({\bf x}) e^{-i A k_z u_z({\bf x^\prime})} ~~~, 
\end{equation}
where line-of-sight and image plane transforms are now treated separately
and we use convention ${\bf k}=({\bf K},k_z)$.

To calculate the statistical average
$\langle\rho_s({\bf k})\rho_s^{*}({\bf k^\prime})\rangle$ for Fourier 
components of the 
density in velocity space we assume statistical homogeneity for the 
quantities in
real space, which entails that
\begin{equation}
\Xi_2 (k_z, k_z^\prime, {\bf r})=\langle \rho({\bf x}) 
\rho({\bf x^\prime}) 
e^{i A [k_z {\bf u(x)} \cdot {\bf \hat z}-
k_z^\prime {\bf u(x^\prime)} \cdot {\bf \hat z}]}
\rangle
\end{equation}
depends only on the difference ${\bf r \equiv x-x\prime}$.
Choosing new variables ${\bf r}$ and ${\bf x^+=(x+x\prime)}/2$ we obtain
\footnote{Depending on convenience, we choose one of the following transformations
of the double integral along a pair of lines-of-sight through the slice of
finite thickness, $z_+=(z+z^\prime)/2$, $z^-=z-z^\prime$:
\begin{eqnarray}
\int_0^S dz \int_0^S dz^\prime F(z,z^\prime)
&=& \int_0^S dz^- \int_{z^-/2}^{S-z^-/2} dz^+ [F(z^+,z^-) + F(z^+,-z^-)] \nonumber \\ 
&=& \int_0^{S/2} dz^+ \int_{-2z^+}^{2z^+} dz^- [F(z^+,z^-) + F(S-z^+,z^-)]
\label{eq:trick}
\end{eqnarray}
}
\begin{equation}
\langle\rho_s({\bf k})\rho_s^{*}({\bf k^\prime})\rangle=
2 \delta({\bf K}-{\bf K^\prime}) \int_0^S dr_z \int_{r_z/2}^{S-r_z/2} dz^+
\int d^2{\bf R} \Xi_2 (k_z, k_z^\prime, {\bf R},r_z)
e^{i{\bf K}{\bf R}}
e^{i A f^{-1} [k_z^- z^+ - k_z^+ r_z]}
\label{eq:nonortho}
\end{equation}
Here $k_z^+=(k_z+k_z^\prime)/2$ and $k_z^-=k_z-k_z^\prime$.

Assuming that density and velocity
fields in galactic coordinates are uncorrelated (cf. Appendix D)
and described by a
correlation $\xi(r)$ and structure (see eq.~(\ref{struc})) functions
we get
\begin{equation}
\Xi_2 (k_z, k_z^\prime, {\bf r}) = 
\xi(r) 
e^{-\frac{1}{2} A^2 \left[(k_z-k_z^\prime)^2 D_z(\infty)/2+k_z k_z^\prime
D_z({\bf r})\right] }
\label{eq:nonortho1}
\end{equation}
Important feature of eq.~(\ref{eq:nonortho1}) compared to the kernel
in the main text is the explicit
presence of the velocity dispersion $<u_z^2>=D_z(\infty)/2$ in the exponent,
since we have to treat $k_z$ and $k_z^\prime$ as separate. This means that
one cannot use pure power-law structure functions do describe velocity
turbulence, but the turbulence must have a maximum scale. This is perfectly
in line with the introduction of finite cloud size along the line-of-sight.
In the absence of additional physics the size $S$ also serves as cutoff scale
for the structure function $D(r) \sim D(\infty) r^m/(r^m+S^m)$, but 
we notice that if this law is 
adopted for $D_{LL}(r)$, z-component of the structure function in the case of
solenoidal turbulence is
\begin{eqnarray}
D_z({\bf r}) &=& D(\infty) \frac{r^m}{r^m+S^m} \left( 1+  
\frac{m/2}{1+(r/S)^m}(1-\cos^2\theta)\right), ~~~~~ 
\cos\theta={\bf r} \cdot {\bf \hat z }/r \\
D_z(\infty)&=&D(\infty)=C  S^m
\label{projectzz}
\end{eqnarray}

For clouds introduce in Appendix~F 3D spectrum in velocity space, but 
here use
eq.~(\ref{eq:nonortho}) directly to obtain 2D spectrum of intensity 
in velocity slice from eq.~(\ref{eq:inten}).
Expressing left-hand-side intensity correlation function
as the Fourier integral of $P_2({\bf K})$ (see eq.~(\ref{ppp}))
and right-hand-side
3D density correlation in velocity space as double Fourier integral of the
$\langle\rho_s({\bf k})\rho_s({\bf k^\prime})\rangle$ given
by eq.~(\ref{eq:nonortho}) we can carry out integration over $k_z$ and $k_z^\prime$
to obtain
\begin{eqnarray}
P_2({\bf K}) & \propto & \int d^2{\bf R} e^{i{\bf K}{\bf R}}
\int_0^{\delta V} dz_s^-  \int_{z_s^-/2}^{\delta V-z_s^-/2} dz_s^+ \times
\label{eq:monster}  \\ 
&&\int_0^S \frac{1}{[B_1({\bf r}) D_z({\bf r})] ^{\frac{1}{2}}} 
e^{-(f^{-1} r_z-z_s^-)^2/B({\bf r})} 
\int_{r_z/2}^{S-r_z/2} dz^+
e^{-(f^{-1} z^+-z_s^+)^2/2D_z({\bf r})} 
 \xi({\bf R},r_z) \nonumber
\end{eqnarray}
where $B_1({\bf r})  =  D(\infty)[1-D_z({\bf r})/2D(\infty)] $ only weakly
 depends on $r$
and $\sim D(\infty)$.
Since our $z_s$ coordinates are in velocity units, integration over them is
done directly through velocity slice of  $\delta V$ width.
As one expects, parameter $A$ drops out of equations. 
Although eq.~(\ref{eq:monster}) can be further evaluated, resulting error 
functions
are not very illuminating. We shall analyze the limit $f^{-1} \to 0$ instead, 
which
describes the case when coherent motions through the cloud can be neglected.
It is useful to change the order of integration over $z_s^+$ and $z_s^-$
according to (\ref{eq:trick}). Then
\begin{equation}
P_2({\bf K}) \propto \int d^2{\bf R} e^{i{\bf K}{\bf R}}
\int_0^S dr_z (S-r_z)
\int_0^{\frac{\delta V}{2}} dz_s^+ 
 \frac{ 
e^{-z_s^{+2}/B_1({\bf r})} + e^{-(\delta V- z_s^+)^2/B_1({\bf r})}}
{B_1^{1/2}({\bf r})}
{\rm erf} \left[\frac{z_s^+}{(D_z/2)^{1/2}} \right]
\xi({\bf R},r_z) 
\label{eq:f=0}
\end{equation}
The structure of this expression looks somewhat different from 
eq.~(\ref{eq:slice}) 
mainly because here we have first performed integration over $k_z$, rather
than $z_s$. This form is convenient to discuss variation of the result with
the slice thickness. We can now distinguish three regimes.
If the velocity slice is {\it thick}, namely larger than velocity 
dispersion on the
scale of the cloud $\delta V \gg B_1^{1/2} \sim D^{1/2}(\infty)$, then 
an integration over the whole line (i.e. up to infinity) is appropriate
\begin{equation}
P_2({\bf K}) \propto
\int d^2{\bf R}
e^{i{\bf K}{\bf R}}
\int_0^S dr_z (S-r_z)
\tan^{-1} \left[ \left( B_1/2D_z \right)^{1/2} \right]
 \xi({\bf R},r_z) 
\end{equation}
The velocity effects disappear in this regime, with only
minor residuals through the almost constant function $\tan^{-1}$.

When the slice is not very thick, $\delta V \ll D^{1/2}(\infty)$,
exponential terms in eq.~(\ref{eq:f=0}) are close to unity and one returns
to the situation discussed in the paper. Namely, the slice is {\it thin}
when $\delta V < D_z^{1/2}(|{\bf K}|^{-1}) $. Then the error function in eq.~(\ref{eq:f=0})
can be expanded in series and in the leading order 
\begin{equation}
P_2({\bf K})|_t  \propto 
\int d^2{\bf R} e^{i{\bf K}{\bf R}}
\int_0^S dr_z (S-r_z)
B_1^{-1/2}({\bf r}) \frac{\xi({\bf R},r_z)}{D_z^{1/2}({\bf r})^{1/2}} 
\label{pcloud}
\end{equation}
This recovers the eq.~(\ref{eq:thinhighK}) with modification for the finite
size of the cloud\footnote{The presence of $r_z$ in eq.~(\ref{pcloud})
results in a subdominant term that decreases faster than the main term
for large $|{\bf K}|$.} . As $B_1({\bf r})$ varies slowly with
${\bf r}$ it can be considered as a constant $\approx D(\infty)\approx CS^m$.
Indeed, it is easy notice (see eq.~(\ref{projectzz})) that
$D_z({\bf r})/CS^m={\cal f}({\bf r}/S)$ which plays the role of 
$D_z({\bf r}/\lambda)$ in eq.~(\ref{eq:thinhighK}). {\it Thus
the size of the cloud $S$ plays the role of the correlation scale $\lambda$.}
Also, the physically transparent condition 
$\delta V < D_z^{1/2}(|{\bf K}|^{-1}) $
exactly corresponds to the transition point between thin and thick slice
regimes
derived earlier in (\ref{eq:phystrans}). The fact that in a completely
different regime we obtained the same condition for the separation 
between thin and thick slices shows that this condition is universal
and the cancellation of $f$ in eq.~(\ref{eq:phystrans}) was not
accidental. This finding allows us to apply our theory to both individual
clouds and to external galaxies without caring too much about a regular
shear. The thick and thin slice asymptotics are given by
Table~1 with $S$ being used instead of $\lambda$.

The effects of the emitting region being finite is seen in the
thick slice regime. This regime corresponds to an
intermediate
value of ${\delta V}$, $D_z^{1/2}(|{\bf K}|^{-1}) < {\delta V} <  
D^{1/2}(\infty) $.
Obviously
enough, the integration over the full velocity dispersion provides the
statistics that depends only on the density field. Inversion of this
statistics provides the 3D density power spectrum as it is shown in
L95. 

The opportunity of determining density spectrum before dealing with
the velocity is very welcome. 
In other words, individual HI clouds present an excellent example for
studying the interstellar statistics. Our results here are directly
applicable to optically thin species in molecular clouds.

\section{1D Spectrum along Line of Sight}

Calculation of the one dimensional power spectrum along a line-of-sight
$P_1(k_z)$ in velocity space is very straightforward under assumptions
adopted in the paper. Indeed
\begin{equation}
P_1(k_z) = \int d{\bf K} P_s({\bf K}, k_z)
\end{equation}
or, using eq.~(\ref{eq:kolmogorov})
\begin{equation}
P_1(k_z) =  e^{-f^2 k_z^2 v_T^2} \int dz \, e^{i k_z z} \xi(z)
\exp\left[-\frac{(k_z\lambda)^2 \tilde D_z(z/\lambda)}{2} \right] ~~~,
\label{eq:1Dspek}
\end{equation}
Asymptotics for the density $P_{1\rho}$ part (described by a
correlation function $\xi(r) \propto r^{-\gamma}$) and velocity
term $P_{1v}$ (which corresponds to $\xi(r)=const $) are given in the
table:
\begin{table}[h]
\begin{displaymath}
\begin{array}{lcc} \hline\hline\\
& \multicolumn{1}{c}{k_z \lambda \ll 1};  &
\multicolumn{1}{c}{k_z \lambda \gg 1}
\\[2mm] \hline \\
\lambda^{-1} (\lambda/r_0)^{3+n} P_{1\rho}(k_z): &
  (k_z\lambda)^{n+2};
&   (k_z\lambda)^{2(n+2)/m} \\[2mm] 
\hline \\[3mm]
\lambda^{-1} P_{1v}(k_z): &
(k_z\lambda)^{1-m}; &
 (k_z\lambda)^{-2/m} \\[3mm] \hline
\end{array}
\end{displaymath}
\caption{Asymptotics of the  1D spectrum along the line of sight at scales
larger than gas sound speed, $k_z < 1/(f v_T)$.} 
\label{tab:1Dspk_asymp}
\end{table}

This results are valid for $\gamma < 1$, so $n < -2 $.
At velocity scales below gas sound speed $k_z > 1/(f v_T)$ the
line-of-sight power is suppressed.
Note that the expression for $P_{1\rho}$ term in the longwave regime
would correspond to the density spectrum along the line-of-sight 
in the absence of velocity mapping.
The asymptotics are illustrated by the full solution, available for 
$m=1$ case.
\begin{equation} 
P_{1}(k_z) \propto e^{-(k_z \lambda)^2 (f v_T/\lambda)^2} 
\cos \left[(n+2) \tan^{-1} (1/k_z\lambda) \right]
{\left[(k_z\lambda)^2 (1+(k_z\lambda)^2) \right]}^{(n+2)/2}
\end{equation}
To get $P_{1v}$ term one should substitute $-3$ instead of $n$.
This spectrum allows to distinguish the short and long-wave density
regime easily.

For all practical purposes in the
case of the Galaxy $\lambda$ is sufficiently large that we
shall be in $k_z \lambda\gg 1$ regime.
In this regime the scale when the contribution from 
$P_{1v}$ and $P_{1\rho}$ terms are equal is
\be
(k_z \lambda)^{2/m}=\lambda/r_0
\ee
and does not depend on $n$. However,  which scales are dominated by the
$P_{1v}$ term and which by the $P_{1\rho}$ term does depend on 
whether $n < -3 $ or $n > -3$
If the underlying spectrum is steep, $n < -3 $, the short scales
$k_z \lambda > (\lambda/r_0)^{m/2}$ are dominated by velocity term
while the intermediate scales $ 1 < k_z \lambda < (\lambda/r_0)^{m/2} $ are
determined by the density. 
The situation is reversed for shallow $n > -3 $ spectra.
As a consequence, the spectrum on small scales always has shallower
slope than on the intermediate ones.

We can incorporate the effect of finite thickness of the cloud 
as follows. Let us model the density field inside cloud
as $\rho_{cl}({\bf x})=\rho({\bf x})\cdot F({\bf x},{\bf S})$ where $\rho({\bf x})$
is statistically homogeneous random field, while $ F({\bf x}) $ describes
the shape of the cloud. We assume the cloud is centered at the origin
and has characteristic size S. This size, in principle, can be different in different
directions, thus the vector notation. Similarly to eq.~(\ref{eq:preroman})
we obtain for the power spectrum in velocity space
\begin{equation}
\langle \rho_s({\bf k}) \rho_s^{*}({\bf k}) \rangle = e^{-f^2 k_z^2 v_T^2}
\int d^3 {\bf r}
e^{i {\bf k} \cdot {\bf r}} \Xi({\bf k}, {\bf r}) 
\int d^3 {\bf x_+} F({\bf x_+}+{\bf r}/2,{\bf S}) F({\bf x_+}-{\bf r}/2,{\bf S})~.
\label{eq:towards1dcloud}
\end{equation}
where
 ${\bf x}_+=({\bf x}+{\bf x^\prime})/2$ (compare to eq.~(\ref{eq:preroman})).
Note, that 
different Fourier components are not formally orthogonal when we deal with finite
clouds. We did not write down the off-diagonal elements, but the diagonal ones, in contrast
to eq.~(\ref{eq:roman}) contain not a delta-finction, but a convolution of shape functions $F$.
This convolution is the easiest to evaluate in Fourier space
\begin{equation}
F^2({\bf r},{\bf S}) \equiv
\int d^3 {\bf x_+} F({\bf x_+}+{\bf r}/2,{\bf S}) F({\bf x_+}-{\bf r}/2,{\bf S})
\propto \int d^3{\bf k}\, F^2({\bf k S})\, e^{i {\bf k} \cdot {\bf r}}
\end{equation} 

1D power spectrum with finite cloud size taken into account is giving by a modified 
version of eq.~(\ref{eq:1Dspek})
\begin{equation}
P_1(k_z) =  e^{-f^2 k_z^2 v_T^2} \int dz \, e^{i k_z z} \xi(z)
\exp\left[-\frac{(k_z\lambda)^2 \tilde D_z(z/\lambda)}{2} \right] F^2(z,S_z)~~~,
\label{eq:1Dspekfinite}
\end{equation}
The modification is insignificant for high $k_z > S_z^{-1}$ since at small scales
shape function is constant and we can use asymptotics in Table~(\ref{tab:1Dspk_asymp}).
To prove this consider
a  spherical cloud of constant density, $F(r/S)=const, ~ r \le S/2; ~ F(r/S)=0, r > S/2$.
Then
\begin{equation}
F^2(z/S) \propto 1 - \frac{3}{2} |z/S| + \frac{1}{2} |z/S|^3, ~~~ |z/S| \le 1 
\end{equation}
This is somewhat different from a model with a finite thickness $S$
of the cloud
along the line-of-sight, infitie dimensions in the image plane.
In this case
\begin{equation}
F^2(z/S) \propto  1 - |z/S|, ~~~ |z/S| \le 1
\end{equation}
Note, that in both expressions above for $z\ll S$ only unity term matters
and for small scale turbulence eq.~(\ref{eq:towards1dcloud})
presents the 3D spectrum in velocity space.

The 1D statistics is complimentary to the 2D statistics that we dwelt upon
in the main text.  Applying 1D spectral analysis to data cubes it is possible
to determine whether the spectrum is long-wave or short-wave dominated.

\end{document}